\documentclass[11pt,twoside,a4paper]{amsart} 
\usepackage{amsfonts,amssymb,latexsym,xspace}  
\usepackage{amsmath,enumerate}  
\usepackage[english]{babel}
\usepackage{graphicx}
\usepackage{amsfonts,latexsym,epsf}
\numberwithin{equation}{section}
\usepackage{pstricks}
\usepackage{pst-plot}
\usepackage{pst-node}
\def\cal {\mathcal}

\def\ds {\displaystyle}

\def\hat {\widehat}
\def\tilde {\widetilde}

\def \real{{\mathbb R}}
\def \complex{{\mathbb C}}

\def \bar {\overline}
\def\HH{{\cal H}}
\def \marginpar#1{}

\def\Pr{\operatorname{\mathbb P}} 
\def\E {{\mathbb E}}
\def\Var{{\mathbb V}}

\def\pn {\par \noindent}

\newcommand {\ntop} [2] {\genfrac {}{} {0pt}{1}{#1}{#2}}

\title [ Euclidean  Algorithms are Gaussian ]
{\bf 
Euclidean  Algorithms are Gaussian} 

\author{ Viviane {\sc Baladi} and  Brigitte {\sc Vall\'ee}} 
\address{
 Viviane {\sc Baladi} : CNRS UMR 7586, 
Institut de Math\'e\-ma\-ti\-ques de Jussieu, F-75251 Paris, France}

\email{ baladi@math.jussieu.fr}
\address{
 Brigitte {\sc Vall\'ee}: CNRS UMR 6072,  GREYC, Universit\'e de Caen, F-14032 
Caen, France}

\email{
brigitte.vallee@info.unicaen.fr  }

\date{Revised,  April 2004}
\begin{document}

\begin{abstract}
\pn
We obtain a Central Limit Theorem  for
a general class of additive parameters (costs, observables) associated 
to  three standard  Euclidean algorithms, 
with optimal speed of convergence. We also provide
very precise asymptotic estimates and error terms for the
mean and variance of such parameters.  
For costs that are lattice
(including
the number of steps), we go further and establish
a Local Limit Theorem, with  optimal  speed of convergence. 
We  view an algorithm as a dynamical system
restricted to rational inputs, and 
combine tools  imported from dynamics, such as 
transfer  operators,  with various other techniques:
Dirichlet series, Perron's formula, 
quasi-powers theorems, and the saddle-point method.  
Such dynamical analyses had previously been used to perform
the average-case analysis of algorithms. For the present
(dynamical) analysis in distribution, we
require  estimates on  transfer operators when a parameter varies 
along  vertical    lines in the complex plane.   To prove them,
we adapt techniques introduced   recently  by  Dolgopyat 
in the context of continuous-time dynamics
\cite {Do}.   \end{abstract}

\maketitle

\section{Introduction} 

\pn   According to Knuth~\cite[p.~335]{Kn},
{``we might call Euclid's method the granddaddy of all
algorithms, because it is the oldest nontrivial algorithm that
has survived to the present day.''}  Indeed, Euclid's algorithm 
is currently a basic  building block of
computer algebra systems and multi-precision arithmetic
libraries, and,  in many such applications, 
most of the time is  spent in computing gcd's. However,   
the Euclidean algorithm has not yet been completely
analyzed, and it is the purpose of this paper to provide such an
analysis.  

\smallskip
\pn We shall state informally
our Central Limit Theorem (CLT) and Local
Limit Theorem (LLT) later in this section,
but first we  discuss Euclidean algorithms
and cost functions (equivalently: additive parameters), 
recalling previously known facts about 
their average-case and distributional analyses. 
Our results have been announced in \cite{BVVB}.

\medskip
\pn {\bf  Continued fraction expansions of real numbers.}
Every  $x\in ]0, 1]$  
admits a  finite or infinite ($CF$)-continued fraction  expansion of the form
\begin {equation}\label{cfx}
x=  \frac{1}{\ds{  m_1+
  \frac{1}{\ds{ m_2+
  \frac{1}{~\ds  \ddots +\frac  {1} 
{m_n+\ldots }~}        }}}} \, .
\end{equation} 
Ordinary continued 
fraction expansions   
can  be  viewed as 
trajectories of a one-dimensional dynamical system, the Gauss map
$T: [0, 1] \rightarrow [0, 1]$,
$$ {\ds T(x):=    \frac 1 x  -  \lfloor \frac 1 x \rfloor,
\ \ \hbox {for $x \not = 0$}, \ \  T(0)= 0 \, .}$$
(Here, $\lfloor x \rfloor$ is the
integer part of $x$.)
For an irrational $x$, the   trajectory 
${\cal T}(x) = (x, T(x), T^2(x), \ldots, T^n(x),\ldots)$  
never meets $0$ and is  encoded
by  the infinite sequence of {\it digits}
$(m_{1}(x), m_{2}(x), m_{3}(x), \ldots,
m_{n}(x), \ldots)$, defined by 
$  m_{i}(x) :=m(T^{i-1}(x))$,
with  $  m(x) :=\lfloor {\frac {1}{x }}\rfloor $.
It is usual to consider  the truncated 
trajectory ${\cal T}_n(x) := (x, T(x), \ldots, T^n(x))$, and let  
$n$ tend to $\infty$. 
If $x\ne 0$ is rational,  the trajectory  ${\cal T}(x)$ reaches $0$ 
in a finite number of steps, 
and this number, $P(x)$,   is called the 
 depth  of $x$. We set $P(x)=\infty$ for irrational $x$.
\smallskip
\pn 
Associate a  nonnegative real value $c(m)$ to each possible digit $m \ge 1$.
We may also  regard $c$ as a function on the set $\cal H$
of inverse branches of $T$.
We    set for each
$x$ and  each $n$
\begin{equation}\label{bir}
C_n(x):= \sum_{i= 1}^{\min (n, \,P(x))} c(m_{i}(x) ) \, .
\end {equation}
We shall 
refer to $c$ as a {\it digit-cost}
and to $C_n$ as the associated {\it total cost}
(sometimes, just {\it cost}),   since a variety
of costs, in the usual sense of computational complexity,  are 
usually expressible in this way.
\smallskip
\pn The total cost (\ref{bir}) is a Birkhoff sum, i.e.,
a   sum  over iterates of the dynamics $T$. 
Hence,  the functions $c(m)$ play
the r\^ole of ``observables'' in physics,   and  it is of interest  
to characterize   their  probabilistic 
behavior.    Here,  $T$ admits  
 a unique    invariant density $f_{1} (x) = (1/\log 2) 
 (1+x)^{-1}$,  the 
 Gauss density. When   the integrability condition, 
\begin{equation}\label{intt}
\hat \mu(c):=\sum_{h \in \cal H} c(h)\cdot  \int_{h([0,1])} f_{1}(x) 
dx =
\int_0^1 c(m(y) ) \, f_{1}(y) dy  < \infty  \, ,
\end{equation}
 is satisfied, then the Birkhoff ergodic theorem implies that  for 
Lebesgue almost  every $x$ in $[0,1]$  the average $C_n(x)/n$  
converges to $\hat \mu(c)$ as $n\to \infty$.
We discuss next how {\it transfer operators} give refinements
of this fact.

\medskip
\pn The density transformer, also known as the Perron-Frobenius 
operator, 
\begin{equation} \label {H}
\mathbf {H}_1 [f](x)=\sum_{h \in \cal H} |h'(x)| \cdot f \circ h(x) 
\end{equation}
 was introduced
early in the study
of continued fractions  (see e.g. L\'evy \cite {Le}, 
Khinchin \cite {Kh}, Kuzmin \cite{Ku}, 
Wirsing \cite {Wi},  Babenko \cite{Bab}, 
and Mayer \cite {Ma}).  
The density transformer is a  special case of a {\it transfer
operator.}
Very general 
transfer operators were introduced by Ruelle,
in connection with his thermodynamic formalism  (see e.g. \cite {Ru}).
We shall see that  ``weighted'' transfer operators appear naturally
in the probabilistic analysis of dynamics.

\smallskip
\pn  Fix a reference probability 
measure  on $[0,1]$, absolutely continuous with a smooth
density $f$, and denote by $\E[\cdot]$
the corresponding expectation.
To establish probabilistic results on total costs along truncated trajectories
as $n\to \infty$,  it is standard to use  
 the sequence of {\it moment generating functions,} i.e., the sequence of
expectations  $\E[\exp(w C_n)]$, for $w$ complex.
In probabilistic i.i.d. situations,  $\E[\exp(w C_n)]$
is the $n$-th power of some expectation. 
In our setting, a {\it quasi-powers} approximation
(with remainder term) 
can be obtained for $\E[\exp(w C_n)]$ after expressing it  in terms
of the $n$th iterate $ {\bf H}^n_{1, w}$
of  a  transfer operator. The transfer operator in question
is the following perturbation 
of the density transformer:
\begin{equation} \label {H1w}
    {\bf H}_{1, w}[f]=\sum_{h\in h} 
\exp[w c(h)] \cdot |h'| \cdot   f\circ h \, .
 \end{equation}
The density transformer  ${\bf H}_{1}= {\bf H}_{1, 0}$ (acting on ${\cal C}^1$
functions) has a dominant eigenvalue  $\lambda = 1$, and a {\it spectral gap:}
the rest of the spectrum lies in a disk of radius $<1$.
For a non constant digit-cost satisfying a moderate growth condition
(\ref{MMGG}), elementary  {\it perturbation theory} \cite {Ka}  implies
that  ${\bf H}_{1, w}$ inherits the spectral
gap when $w$ is near $0$. 
This gives the above-mentioned quasi-powers expansion and, together
with convexity  of the logarithm of the
dominant eigenvalue, leads to a proof that
the asymptotic distribution of the total cost is   Gaussian,
with (optimal) speed of convergence $O(1/\sqrt {n})$.  
This Central Limit Theorem [stated below
more precisely as {\bf Theorem~1}]  is quite well-known.
See for instance \cite 
{Co} or  \cite {Bro} for interval maps, and
\cite{AAD}  for a more abstract framework and  references to the pioneering paper of Nagaev.
It is convenient to base
the proof below  on
a compact and versatile statement,
the ``Quasi-Powers Theorem'' of Hwang \cite{Hw, Hw1, Hw2} 
[{\bf Theorem 0} below], which encapsulates the consequences of  the   L\'evy 
continuity theorem and the Berry-Esseen inequality.

\medskip
\pn{\bf Continued fractions of rational
numbers:  Average-case analysis of Euclidean algorithms.}  
There are variants of the 
standard continued fraction algorithm 
induced by variations of the standard division procedure.
See  \cite{Va8} for many examples and a classification
into the ``Fast'' and  the ``Slow'' Class.
In this paper, we  study three  algorithms of the Fast Class, the {\it standard,} {\it centered,} 
and
{\it odd} algorithms, specified  at the beginning of Section~ 2. 
An execution of a  Euclidean algorithm on the input $(u, v)$ formed with
two integers
$u$, $v$  such that $u/v = x$
gives rise to  a rational trajectory ${\cal T}(x)$ 
which ends at zero in $P(x)$ steps, and the  total cost  of the trajectory is then 
\begin{equation}\label{birrat }
\qquad  C(x):=   \sum_{i= 1}^{P(x)} c(m_{i}(x) )\,.
\end{equation}  The    reference parameter   is no longer the 
truncation degree $n$, but the  size~$N:= \max (u, v)$
of the   input  $(u, v)$. The reference probability measure
$\Pr_N$
is now the uniform discrete measure
on the (finite) set of inputs of size  $\le N$.

\smallskip
\pn 
As it  is often the case, the discrete problem
is  more difficult than  its continuous counterpart. However,
the average-case complexity of Euclidean algorithms is   
already well-understood, as we explain next.
For the  number of steps $P(u,v)$, which corresponds to 
the  trivial digit-cost  $c \equiv 1$, 
the  standard  Euclidean algorithm  was first  analyzed  
in the average-case
around 1969, independently  by Heilbronn \cite{Hei} and Dixon \cite {Di}. 
The centered algorithm was studied by Rieger \cite {Ri}. 

\smallskip 
\pn Consider now a general digit-cost $c$ of moderate growth and the associated 
total cost $C$ of the rational trajectories.
The  expectation $\E_N [C]$ is described by the partial
sums of the coefficients  of a  generating function $S(s)$
(a
common tool in the average-case study of 
algorithms \cite {Fl,FS})
where the
parameter $s$   ``marks'' 
the size $N$ of inputs.
As it  is usual in   number theory,
the generating  functions  $S(s)$ are Dirichlet series.
Recently,   Vall\'ee \cite {Va8} has
related $S(2s)$  to the {\it quasi-inverse}  $(I-{\bf H}_{s})^{-1}$ 
of another  perturbation ${\bf H}_s$ of the 
density   transformer,   together with its weighted 
version ${\bf H}_{s}^{(c)}$:
\begin {equation} \label {Hs}
{\bf H}_{s}[f] =\sum_{h \in \cal H} |h'|^s \cdot f\circ h \, , \qquad {\bf H}_{s}^{(c)}
:= \sum_{h \in \cal H}
c(h) \cdot |h'|^s \cdot f\circ h.
\end{equation}
Then,  spectral information
on ${\bf H}_{s}$ may be used to 
show that $(I-{\bf H}_{s})^{-1}$ is analytic
in the half-plane $\{\Re s > 1\}$, 
and  analytic on $\Re s =1$ except for a simple pole
at $s=1$. Under these conditions,  one can extract asymptotically  the
coefficients of $S(s)$  by means of Delange's
Tauberian theorems \cite {De,Te}. (Just like in the usual prime number
theorem, or in weighted dynamical prime number 
theorems, see e.g. Sections 6--7
of \cite{PP}.)
For costs  of moderate growth and 
Euclidean algorithms in the Fast Class,
this dynamical approach gives \cite {Va8}  that
the mean value   $\E_N[C]$  of  the
total cost of the rational trajectory  satisfies
$\E_N[C] \sim \hat \mu(c) \cdot \mu  \log N$. Here, 
 $\hat \mu (c)$ is  the asymptotic
mean value (\ref{intt}) of truncated real trajectories,
and $\mu$ equals $ 2/|\lambda'(1)|$, where $\lambda(s)$ is   
the dominating eigenvalue of ${\bf H}_s$.

\medskip
\pn {\bf Euclidean algorithms: Distributional analysis and   dynamical 
methods.  Main results.}
We 
have seen that, with respect to any cost of moderate growth,  rational 
trajectories behave in average  similarly to the way
truncated real trajectories behave 
almost everywhere. 
It is then natural to ask  whether  this analogy extends to  
distributions: Is it true that the  distribution of the total cost 
$C(x)$  on  rational  trajectories with an input 
$x$ whose  numerator and denominator   are less than $N$ 
is asymptotically Gaussian (when $N$ tends to $\infty$)? 
How to compare the  distribution  of some cost on truncated real trajectories  
and 
on rational trajectories?
This paper provides a precise  answer to  all these 
questions for 
three  different algorithms that all belong to the Fast Class. 

\smallskip
\pn 
Concerning the standard Euclidean algorithm and
the number of steps (i.e., the constant cost $c\equiv 1$),
Hensley  \cite {He}  has obtained
a Central Limit Theorem,  and a  Local Limit Theorem  with
speed of convergence $O((\log N)^{-1/24})$. 
In the present work, we apply dynamical methods for the first
time to the distributional analysis of discrete algorithms;
in this way, we improve  Hensley's result while extending it to a large class
of cost functionals and to several algorithms.

\smallskip
\pn 
Our strategy consists in describing the moment generating function
$\E_N[\exp( w C)]$ as a quasi-power. It turns out that
$\E_N[\exp (wC)]$ is related to the partial sums of the
coefficients in a bivariate series 
$S(s, w)$. This series is 
of Dirichlet type with respect to the variable $s$,  while
the extra parameter $w$ ``marks" the cost $c$, 
 and
we  require  uniform estimates 
with respect  to  $w$.   Tauberian  theorems  are  now insufficient, since they  
do not provide remainder terms: We   need a more precise ``extractor'' 
of  coefficients, and  the Perron formula is  well-suited to this  purpose.

\pn We establish below (\ref{R1}) a simple
relation between $S(2s,w)$ and the
quasi-inverse of a two-variable transfer operator
${\bf H}_{s,w}$ defined by 
$${\bf H}_{s,w}[f]=\sum_{h \in \cal H} \exp[wc(h)] \cdot |h'|^s \cdot 
f\circ h \, .  $$ Note that this operator is a simultaneous extension of 
the three operators  ${\bf H}_1, {\bf H}_{1, w}, {\bf H}_s$ defined in  (\ref{H}), (\ref{Hs}), and (\ref{H1w}), while 
${\bf H}_s^{(c)}$ is just its derivative with respect to $w$ at $w= 0$. 
In order to apply Perron's formula with a convenient
integration contour, it is thus useful to know 
that, in a half plane  ``perforated''  at $s= 1$, of the form $\{ \Re s \ge 1- \epsilon \, , |s-1|> \epsilon/2  \}$
for small $\epsilon >0$, 
a certain norm of the
quasi-inverse satisfies 
\begin{equation}\label{txi}
\|(I- {\bf H}_{{s, w}})^{-1}\|_s\le \max(1,|\Im s|^\xi)\, , 
\end{equation}
with $\xi<1$,  uniformly in $w$. 
Then, spectral properties of ${\bf H}_{s,w}$ inherited from
${\bf H}_1$ give the desired quasi-power expansion for 
$\bar \E_N[\exp( w C)]$,
for $w$ close to $0$, and  some $\bar \Pr_N$ close to
$ \Pr_N$.
Note that Hensley    in  \cite {He} has used a
transfer operator ${\bf H}_{s,0}$, but in an appreciably different way:
Hensley obtains distributional results on rational 
trajectories upon aproximating discrete measures on rationals by 
continuous measures. 
In particular, his approach avoids parameters $s$ of large imaginary
parts.

\smallskip
\pn 
Adapting powerful methods due to Dolgopyat \cite {Do}, we
show [{\bf Theorem 2} and Lemma 6 below] that 
the quasi-inverse satisfies the estimates (\ref{txi}) for large $|\Im s|$. 
Dolgopyat  was interested in the decay of correlations
for hyperbolic flows satisfying some uniform nonintegrability
condition ({\em UNI}). Later on, Pollicott 
and Sharp  used Dolgopyat's bounds   together with Perron's formula 
 to find error terms in  asymptotic 
estimates for geodesic flows on surfaces of
variable negative curvature; see e.g. \cite{PS}, where 
only univariate Dirichlet series with positive cofficients appear.    
To the best of our knowledge, the present paper is the first instance 
where these powerful tools are extended to dynamical systems
with infinitely many branches and applied to distributional analyses in   
discrete combinatorics.  

\medskip
\pn Let us now state informally our two main results about the three algorithms:

\smallskip
\pn {\bf Theorem 3.} Consider a nonzero digit-cost $c$ of moderate growth. 
We  show the following Central Limit Theorem:
the asymptotic distribution of the total 
cost $C(u/v)$ of an  execution of the algorithm on the
rational input $u/v$,  uniformly randomly  drawn from  
$\{(u,v)\, , u\le v\le N\}$, is  asymptotically Gaussian,
with best possible speed  of 
convergence, of order $O(1/\sqrt {\log N})$. 
We give expansions for the  
expectation and variance, which are asymptotically  proportional to $\log N$.
The   constants $\mu(c)$ and $ \delta^2(c)$ in the main terms 
of the expectation and the variance  are expressed in function of
the partial derivatives  at $(s, w)= (1, 0)$  of the
dominant eigenvalue  of  ${ \bf H}_{s, w}$, 
and alternatively in terms of  
$\mu=\mu(1)$, $\delta^2=\delta^2(1)$, and the constants 
$\hat \mu(c)$, $\hat \delta(c)$
from Theorem~1. In particular $\mu(c)$ admits
a closed form.

\smallskip
\pn {\bf Theorem 4.} A digit-cost is {\it lattice}
if  it is non zero  and there exists
$L\in \mathbb R^+_*$  so that 
$c/L$ is integer-valued.  The largest such $L$ is then called the
{\it span} of $c$.  For instance, any nonzero constant $c$
is lattice.  For  lattice costs of moderate growth,
we obtain  a Local Limit Theorem with  optimal speed
of convergence 
$O(1/\sqrt {\log N})$.   
This time, we use estimates for
$\bar \E_N[\exp( i\tau C)]$, where $\tau$
 varies in a compact set of the real line. They lead, with 
the saddle-point method,   to a   very natural and concise proof.

\medskip
\pn 
Three   special  instances of our results  for lattice costs are of major interest.

\smallskip
\pn 
 $(i)$   {\sl Digit-cost $c\equiv 1$.} 
For each of
our three 
algorithms the number of steps    is asymptotically  
Gaussian, with mean $\mu \log N$ 
(where $\mu$ admits a closed form)
and   variance $\delta^2\log N$,  in
the sense of the  CLT and the LLT with speed
$O(1/\sqrt{\log N})$.

\pn $(ii)$   {\sl Digit-cost $c= c_{m}=$ the characteristic function  
of a digit $m$.}  
The total number of occurrences of a fixed digit $m$ in a rational trajectory
for our three algorithms
is asymptotically Gaussian (CLT, LLT), with  mean
$\mu \cdot  \hat\mu(c_{m})  \log N$.  
The constant $\hat \mu(c_m)$ is explicit for the three 
algorithms. In the case of the standard Euclidean algorithm we recover (see 
\cite{Va8})
$$ \hat \mu(c_{m}) = \frac{1} {\log 2} \log \left( 1 +\frac {1}{m(m+2)} 
\right) \, . $$

\pn $(iii)$ {\sl Digit-cost $c=$  the binary length $\ell$  of the digit.}  
The binary encoding of a rational trajectory  is 
asymptotically Gaussian (CLT, LLT), 
with mean-value  $\mu \cdot \hat \mu(\ell) \log N$. 
The constant $\hat \mu(\ell)$, explicit for the three  algorithms,
is a  variant of the Khinchine  constant \cite {Kh}.
For the standard Euclidean algorithm, it is equal to (see \cite{Va8})
$$\hat  \mu(\ell) = \frac {1} { \log 2} \log\prod_{k= 0}^\infty (1+ \frac {1} 
{2^k})\, .$$

\medskip
\pn {\bf Plan.}  In Section 2,
after  describing our three  Euclidean algorithms and 
their underlying dynamical systems, we state and prove Theorem~1 
(using Theorem 0) and state
our main results, Theorems 2 (our version of
Dolgopyat's bounds), 3 (CLT), and 4 (LLT).
The first part of  Section~3 is  devoted 
to proving Theorem~2
and checking that its assumptions hold for our three algorithms.
In the last two subsections of Section~3, we obtain
quasiperiodicity results.   With Theorem~2, they entail  
expected properties for  the Dirichlet series.
In  Sections 4 and 5, we present  the proofs of 
Theorems 3 and~4, respectively.

\section {Dynamical  methods and  statement of results.}
\pn  After a description of the three 
Euclidean algorithms to be studied  and their associated
dynamical systems (\S 2.1), we introduce the 
weighted transfer operator in \S 2.2.   Section 2.3
explains the fundamental r\^ole 
this operator  plays in the distributional 
analysis of truncated real trajectories. Theorem~1 is stated there,
and proved using Hwang's Quasi-Power result (Theorem~0). 

\pn Next, we  turn to   rational inputs and Euclidean algorithms. 
We introduce  in \S 2.4 Dirichlet series of moment generating functions, 
which  we relate    to the quasi-inverse of  the weighted transfer operator. 
 We briefly explain  in  \S 2.5 how to apply  the Perron 
formula  to extract coefficients of Dirichlet series, stating also 
estimates \`a la Dolgopyat  (Theorem~2)
useful for the applicability of this formula.  Finally, in
\S 2.6  we  state our CLT
(Theorem~3) and our LLT for lattice costs (Theorem 4).

\subsection {Euclidean algorithms and their associated interval maps.} 
Three   Euclidean algorithms are to be analyzed; 
each of them is related to a Euclidean division.
Let $v \ge u \ge  1$ be integers.
The classical division, corresponding to
the standard Euclidean algorithm  $\cal G$,  $v= mu +r$ produces an integer
$m \ge 1$  and  an integer remainder $r$ such
that $0\le r < u$. The centered  division 
(centered algorithm $\cal K$) requires
$v \ge 2u$ and takes
the form $v = mu +s$, with integer $ s \in [-u/2, +u/2[$. Letting $s = \epsilon r$, with $\epsilon = \pm 1$
(and $\epsilon=+1$ if $s = 0$),
  it produces  an integer  
remainder $r$ such that
 $0\le r \le{ u/2}$,  and  an integer $m \ge 2$. The odd division 
(odd algorithm $\cal O$) produces an odd quotient: it is  of the form 
 $v= mu +  s$ with  $m$ odd and integer $s \in [-u, +u[$. Letting $s =  \epsilon r$, with $\epsilon = \pm 1$
(and $\epsilon=+1$ if $s = 0$), 
 it produces an integer  remainder $r$  with $0\le r  \le u$, and  an odd integer 
$m\ge 1$.
In the three cases, the  divisions are defined  by  pairs $q= 
(m, \epsilon)$, which  are called the digits.  (See Figure 1.)

\pn Instead of the  integer pair $(u, v)$, we may consider the rational 
$u/v$, since  
the pair $(du, dv)$
produces the same sequence of  digits as $(u, v)$, up to multiplying
all  remainders $r$ by $d$.
Then, the division 
expressing  the pair $(u, v)$   as a function of $(r, u)$ 
is replaced by  a   linear fractional
 transformation (LFT)  $h$ that expresses the rational 
 $u/v$ as a function of $r/u$.  For  each algorithm,  the 
rational 
$u/v$ belongs to the interval   ${\cal I}'$ defined in Fig 1.
 
\smallskip
\pn To summarize,   on the input
$(u, v)$,  each   algorithm  performs a sequence of  admissible Euclidean   
divisions,  of the form $v= mu + \epsilon r$ with 
$r/u\in {\cal I}'\cup \{0\}$, and $(m, \epsilon) \in {\cal D}_1$. On the 
input $u/v$, it performs a sequence of LFT's from a
{\it generic} set $\cal H$  (depending on the algorithm) whose elements
$h_{[q]}$ are  indexed by
the admissible pairs $q= (m,\epsilon)$ of  ${\cal D}_1$ and are of the form 
$h_{[m, \epsilon]}(x) = 1/(m +\epsilon x)$.
The LFTs appearing in the
final step  belong to a subset ${\cal F}\subset \cal H$ and are  indexed by
the admissible pairs $q= (m,\epsilon)$ of  ${\cal D}_1\cap {\cal D}_2$.
  (Fig. 1).

\begin{figure}
\begin{small}
\hspace*{-2truecm}
\renewcommand{\arraystretch}{1.5} 
\renewcommand{\tabcolsep}{4pt}
\begin{tabular}{|l|l|l|l|}
\hline
\sl Algorithm &  $\cal G$ (standard)  &   $\cal K$ (centered) &  $\cal O$ (odd) 
 \\ \hline
{\sl Intervals}& ${\cal I}= [0, 1], \ \ \ {\cal I}'=]0, 1[$&${\cal I}=[0, 1/2], \ \ \  {\cal I}'=]0, 1/2]$&${\cal I}= [0, 1],
\ \ \  {\cal I}'=]0, 1]$
 \\ \hline
{\sl
 Generic conditions  }  &$m\ge 1$, $ \epsilon = +1$ &  $m\ge 2$, $\epsilon=\pm1$ &
$m\ge 1$ odd, $\epsilon=\pm1$
\\
{\sl $\cal D_1$ on  pairs $(m , \epsilon)$} &
 & if $m=2$ then $\epsilon=+ 1$& 
  if $m=1$ then  $\epsilon=+1$ 
\\
   \hline
\sl Final  cond. $\cal D_2$

& $  m\ge 2 $ 
& $  \epsilon = +1$ 
&  
$   \epsilon = +1$
		\\ \hline

\sl Graph of
&

\includegraphics[width=3.9cm]{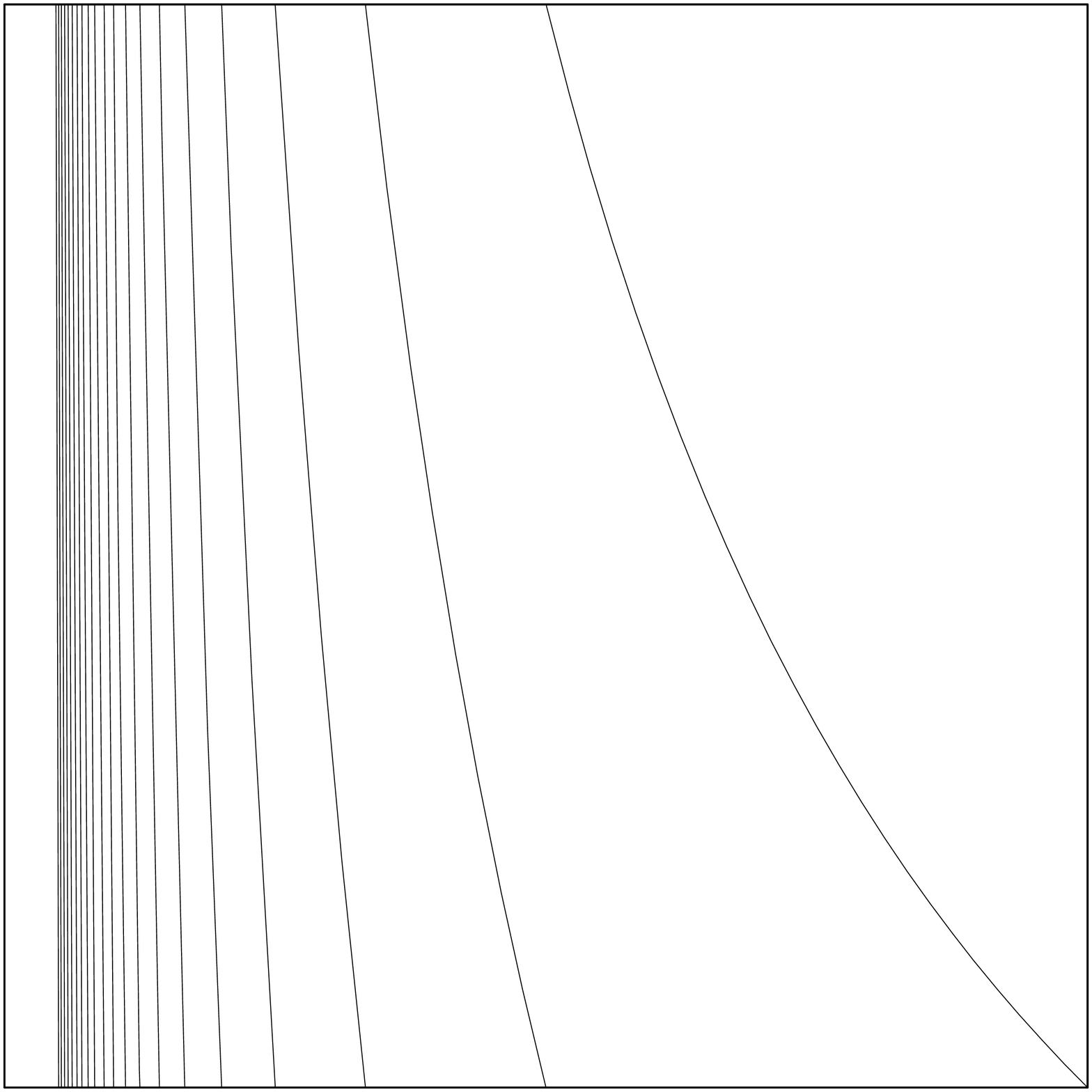}&
 \includegraphics[width=3.9cm]{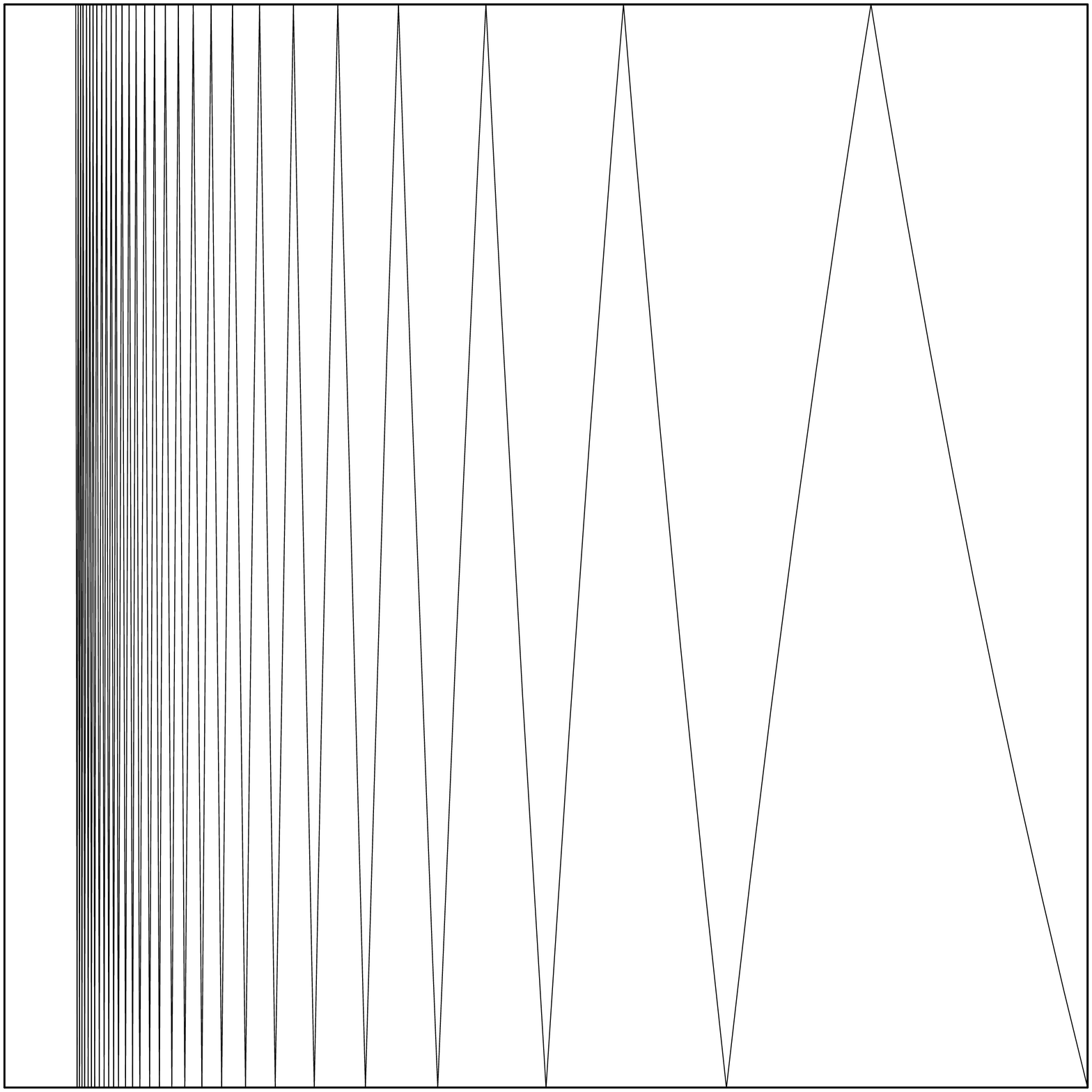}&
\includegraphics[width=3.9cm]{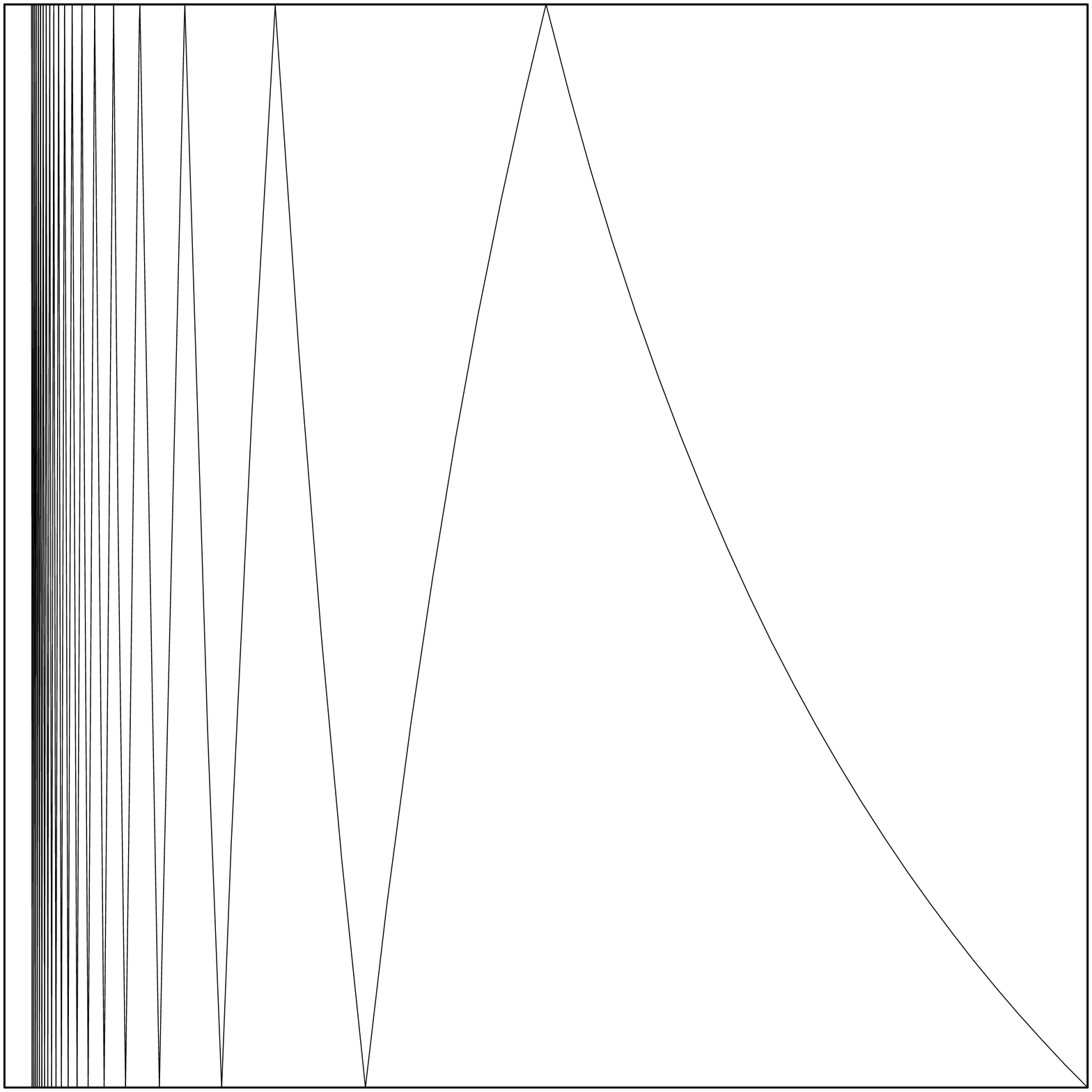}  \\
  \sl $T(x) =  | \frac 1 x  - A (  \frac 1 x ) |$&&&
\\ \hline
 \sl  Function $A(y)$ &    Integer part of $y$ &  Nearest integer  to $y$ &  Nearest odd  integer to $y$
\\
 \sl   &     &   i.e. $m$ s.t. $y-m\in [-1/2, +1/2[$ & i.e. $m$ odd s.t. $y-m\in [-1, +1[$
 \\ \hline
\sl Contraction ratio &$\rho= 1/\phi^2$ & $\rho=1/(\sqrt 2 +1)^2$ & $\rho=1/ 
\phi^2$
\\ \hline
 \sl Invariant density $f_1$ &    $\ds\frac {1} {\log 2} \frac {1} {1+x}$ & 
$\ds \frac {1} {\log 
\phi} [ \frac {1} {\phi +x} + \frac {1} {\phi^2 -x}] $   & $\ds  \frac {1} 
{\phi-1 +x} + \frac 
{1} 
{\phi^2  -x}$ 
\\ \hline
 \sl  Entropy & $\ds 
\frac {\pi^2} {6 \log 2}$ &  $\ds \frac {\pi^2} {6 \log \phi }$ & $\ds \frac 
{\pi^2} {9 \log \phi }$
\\ \hline
\end{tabular}
\end{small}

\caption{The  three  Euclidean   systems
($\phi = (1+ \sqrt 5)/2$).}
\label {les3}
\end{figure}

\smallskip
\pn Each algorithm  applied to a rational $u/v$   builds  a 
specific continued  fraction 
\begin {equation}
\frac{u}{v}=  \frac{1}{\ds{  m_1+
  \frac{\epsilon_1}{\ds{  m_2+
  \frac{\epsilon_2}{~\ds  \ddots +\frac  {\epsilon_{P-1}} 
{ m_P }~}        }}}} \, ,
\end{equation} 
of depth  $P$,
decomposing  also $u/v$ as
\begin {equation} \label{dec}
(u/v) =  h_1\circ h_2\circ \ldots \circ h_P (0) = h(0) \, , 
\end{equation}
 with 
$h_{i} \in \HH, 1 \le i \le P-1$, and  $h_P \in {\cal F}$.
We are 
interested in various costs related to an execution of the algorithm. The 
most basic one 
is the   number of steps $P(u, v)$.   
In general, given a digit-cost function $c$ on the set $\HH$,
 we consider  additive cost functions $C$ 
of the form  
\begin{equation}
\label{digcost1}
 C(u,v) := \sum_{i = 1}^{P(u, v)} c(h_{i}) \, .
\end{equation}

\pn We next see how to associate to each algorithm
a {\it dynamical system of the interval,} 
$T: {\cal I} \rightarrow {\cal I}$.   
The interval ${\cal I}$  is defined in Figure 1. The map 
$T$ extends the map defined on rationals by 
$T(u/v)= r/u$, where $r$ is the remainder of the Euclidean division  on $(u, v)$. 
We get 
$$ T(x) :=  \left | \frac 1 x  - A \biggl ( \frac 1 x \biggr  ) \right | \, , 
\quad x \ne 0 \, , 
\quad T(0)=0 
\,  ,
$$
$A$ depends on the algorithm   and is defined in 
Figure 1.
It is easy to see that the maps 
associated to the three
algorithms belong to the following class:
  
\smallskip
\pn 
{\bf Definition.} [Piecewise  complete maps of the interval] {\sl  A map  $T: 
{\cal I} \rightarrow {\cal I}$ 
is  piecewise complete if  there exist  a  (finite or countable) set $\cal 
Q$,  
whose elements are called digits,  
and  a   partition $\{{\cal I}_q \}_{q\in\cal Q}$ (modulo a countable set) of
the interval $\cal I$  into  open subintervals  ${\cal I}_q$ such that 
the  restriction of $T$ to  ${\cal I}_q$  extends to a a bijective  mapping   of 
class 
$\cal C^2 $ from the closure of  ${\cal I}_q$ to $\cal I$. }

\smallskip
\pn 
We may consider the 
set ${\cal H}= \{h_{[q]} \}$ of branches of the inverse function $T^{-1}$
of a general piecewise complete map
$T$, naturally indexed by the set $\cal Q$. The set of the inverse branches 
of 
the iterate $T^k$ is  ${\cal H}^k$;
its elements are of the form  $ h_{[q_1]}\circ h_{[q_2]}
\circ\cdots\circ h_{[q_k]}$ where $k$ is called the
{\it depth} of the branch.
Setting $\HH^0=\{\rm{Id}\}$,
the  set ${\cal H}^\star:=  \cup_{k\ge 0} {\cal H}^k$  is the semi-group 
generated by 
${\cal H}$.  
Each  interval  $h ({\cal I})$  for $h$ of depth $k$ is called a 
{\it fundamental interval of depth $k$.}

\subsection {Transfer operators of interval maps.}
\pn  The maps $({\cal I}, T)$ associated to  the three Euclidean dynamical systems
belong in fact to
a subclass of piecewise complete mappings :

\medskip
 \pn{\bf Definition}\,  [Good class]. {\sl A piecewise complete interval map 
$({\cal I}, T)$ 
belongs to the good class if:

\smallskip
 $(i)$  $T$ is  piecewise uniformly expanding, i.e.,     
there are $C $ and $\hat \rho < 1$  so that $|h'(x)|\le C \hat\rho^n$ 
for every inverse branch $h$ of $T^n$,  all $n$ and all $x \in {\cal I}$. 
The infimum of such $\rho$  is called the  contraction ratio, and satisfies
\begin {equation}  \label {rho1}
\rho =  \limsup_{n \to \infty}
\left ( \max \{ |h'(x)|; h \in 
{\cal H}^n, x \in {\cal I} \} \right) ^{1/n} \, .
\end{equation}

\smallskip $(ii)$  
 There is $ \hat K>0$, called the distortion constant, so that every inverse 
branch
$h$ of $T$ satisfies $|h''(x)|\le  \hat K |h'(x)|$ for all $x\in {\cal I}$.

\smallskip $(iii)$ There is $\sigma_0<1$ such that
$\sum_{h \in \HH}\sup |h'|^\sigma <\infty$ for all
real $\sigma > \sigma_0$.}

\medskip
\pn (Note that  maps in the good class are automatically
topologically mixing.)

\pn The distortion condition in $(ii)$ 
follows from the $\cal C^2$ assumption when there are finitely many branches, 
and is
just Renyi's condition  
otherwise.

\smallskip
\pn To check that maps associated to our algorithms
are in the good class, for $\sigma_0= 1/2$, use $|h'_{[m, \epsilon]}|=O(m^{-2})$.
(See also  Figure~1, and \cite{Sch} for proofs.)

\smallskip
\pn 	If  ${\cal I}$ is endowed 
with an initial probability density $g_0$
with respect to Lebesgue measure,  
 $T$  acts on it   and transforms it into a new density $g_1$.
The operator ${\bf H}$ such that $g_1={\bf H} [g_{0}]$  is called the density transformer, or the 
Perron-Frobenius 
operator  (acting now
on $L^1$ functions, soon we shall restrict its domain).
  An application of the change of variable formula gives
$${\bf H}[f](x):= \sum_{h \in \HH} |h'(x)| \  f \circ h 
(x) \,  .$$
It  is useful to deal with a more general 
operator, the  transfer operator ${\mathbf  H}_s$
which depends on a complex parameter $s$:
 $$
{\bf H}_{s}[f](x):= \sum_{h \in \HH} 
|h'(x)|^s\ \cdot  f \circ h 
(x)\, .
$$
(Note that  ${\bf H}_1= {\bf H}$.)
If $\sigma :=\Re s>\sigma_0$,  
then  ${\mathbf  H}_s$ acts boundedly on
the Banach space $\cal C^1(\cal I)$ of $\cal C^1$ functions on $\cal I$ endowed 
with the norm
$$
\|f\|_{1, 1}= \|f\|_{0} + \|f\|_{1}, \qquad  \hbox  {with \ \ } 
 \|f\|_{0}:= \sup|f|,  \|f\|_{1}:= \sup|f'|.$$
  
 \smallskip
\pn
We consider  nonnegative cost functions  satisfying the 
following  condition:

\smallskip
\pn {\bf  Definition} [Condition ${\cal {MG}}$\, --Moderate growth--]. {\sl  
Let $\HH$ be the inverse branches of a map
in the good class. A digit-cost $c:\cal H\to \real^+$ 
is of {\it moderate growth} if it is non identically zero and  if the series
\begin{equation}\label{MMGG}
\sum_{ h \in {\mathcal H}} 
  {\exp [w c(h)]} \cdot |h'(x)|^{s} 
\end{equation} 
converges when $(\Re s, \Re w)$ belongs to a   real neighborhood 
 of $(1, 0)$, or, equivalently to  $\Sigma_0\times W_0$  with 
$ \Sigma_0=]\hat \sigma_{0}, + \infty]$, for $\sigma_0 \le \hat \sigma_0 < 1$,
and $W_0=]- \infty , \nu_{0}[$ for $\nu_0 > 0$.
}

\smallskip
\pn {\bf Definition}\, [Class $\cal {GMG}$] {\sl  A triple  $({\cal I}, T, c)$  formed with 
an  interval map
$T:\cal I \to \cal I$ of the good class,  and a  digit-cost $c$ of 
moderate  growth will be said 
to be of ${\cal {GMG}}$-type.}

\medskip
\pn We extend the digit-cost to  a cost function, also denoted $c$, on 
${\cal H}^{\star}$   by 
$$ c (h_1\circ h_2\circ \ldots \circ h_k):= \sum_{i = 1}^k c(h_{i}) \, .$$
We can now   define  the weighted  composition 
operator which depends on two 
 (complex) parameters $s$ and $w$, 
\begin {equation} \label {transfer}
 {\mathbf H}_{s, w} [f](x) :=\sum_{h \in \cal H}   \exp [w c(h)] \cdot  
|h'(x)|^s \cdot  
f\circ h(x)\, .  
\end {equation}
The additive property of costs and the multiplicative
property of the derivatives entail
$$ {\mathbf H}_{s, w}^n [f](x) :=\sum_{h \in \cal H^n}   \exp [w c(h)] \cdot  
|h'(x)|^s \cdot  f\circ h(x), $$
 and   the quasi-inverse $(I-{\bf H}_{s, w})^{-1}$ can be written 
(formally) as 
\begin {equation} 
 (I-{\bf H}_{s, w})^{-1} [f](x):=
\sum_{h \in \cal H^\star}   \exp [w c(h)] \cdot  |h'(x)|^s \cdot  f\circ h(x)\, .  
\end {equation}

\medskip
\pn 
We  recall next  some well-known spectral properties of the  
transfer operator ${\bf H}_{s, w}$.
Endow the Banach space $\cal C^1({\cal I})$
with the norm $\|\cdot\|_{1, 1}$.
 It is known that
for  $(\Re s, \Re w)\in \Sigma_0 \times  W_0$ the operator
${\bf H}_{s, w}$ is bounded
but not compact acting on $\cal C^1(\cal I)$
[see e.g. \cite{Bal, Bro}]; however, it is quasi-compact.
We  recall the definition of
{\it quasi-compactness} for a bounded operator   ${\bf L}$ on a Banach space:
Denote by  $ {\rm Sp}\, {\bf L}$ the spectrum of  ${\bf L}$,   by  $R({\bf L})$ 
its  spectral radius, 
and  by $R_{e}({\bf L})$   its {\it essential
spectral radius,} i.e.,  the smallest 
$r\ge 0$ such that any  $\lambda \in{\rm Sp}({\bf L})$  
with modulus $|\lambda| >r$ is an isolated 
eigenvalue of finite multiplicity. 
An operator  ${\bf L}$ is {\it quasi-compact} if 
$R_e({\bf L}) < R({\bf L})$ holds.

\pn We denote the partial
derivatives of first and second order  of a function
$F(s,w)$ at $(a,b)$ by
$F'_w(a,b)$, $F'_s(a,b)$, $F''_{w^2}(a,b)$,
 $F''_{s^2}(a,b)$,  $F''_{ws}(a,b)$.

\medskip
\pn
{\bf  Proposition 0} [Classical spectral properties of transfer operators].  {\sl Let 
${\bf H}_{s, w}$ be the transfer operator 
(\ref{transfer}) associated to  a 
${\cal {GMG}}$ triple $(\cal I, T, c)$ with contraction constant
$\rho$.  Denote by $R(s,w)$ its spectral radius and $R_e(s,w)$ its essential 
spectral radius. Let $\Sigma_0$, $W_0$ be the real
sets from  (\ref{MMGG}).   When $w= 0$, we omit the second index  in the 
operator and  its associated objects. 

\smallskip
 $(1)$ {\rm [Quasi-compactness.] }Let $\rho < \hat \rho < 1$.
If $(\sigma=\Re s, \nu=\Re w)\in \Sigma_0 \times W_0$, 
then  ${\bf H}_{s, w}$ acts boundedly on $\cal C^1(\cal I)$.
Then  $R(s, w) \le   R(\Re s, \Re w)$ and
$R_e(s, w) \le \hat \rho \cdot  R(\Re s, \Re w)$,
in particular  ${\bf H}_{s, w}$ is (uniformly) quasi-compact for real $(s,w)$.

\smallskip
$(2)$ {\rm [Unique dominant eigenvalue.] } For real    $(\sigma, \nu) \in 
\Sigma_0\times W_0$,   ${\bf H}_{\sigma, \nu}$  has a unique  eigenvalue
$\lambda(\sigma, \nu)$ of maximal modulus, which is real and simple, the {\it
dominant eigenvalue.} 
The associated eigenfunction $f_{\sigma, \nu}$ is strictly positive, and
the associated eigenvector $\hat \mu_{\sigma, \nu}$ of the adjoint operator 
$ {\bf H}_{\sigma, \nu}^*$  is a positive Radon measure.    With the 
normalization conditions,  $\hat 
\mu_{\sigma, \nu}[1]=1$ and
$ \hat \mu_{\sigma, \nu}[f_{\sigma, \nu}]  =1$,  the measure 
$\mu_{\sigma, \nu}:=f_{\sigma, \nu}  \hat \mu_{\sigma, \nu}$
is a probability measure.   In particular, 
$\hat \mu_{1}$ is Lebesgue measure, with $ \lambda(1) = 1$.
  
\smallskip
$(3)$ {\rm  [Spectral gap.] }  For real parameters   $(\sigma, \nu) \in 
\Sigma_0\times W_0$, there is a spectral gap, i.e.,
the {\it subdominant spectral radius} 
$r_{\sigma, \nu}\ge R_e (\sigma, \nu)$ defined  by
$ 
r_{\sigma, \nu}:= \sup \{ |\lambda|;  \lambda \in {\rm Sp}({\bf H}_{\sigma,  \nu}), 
\lambda \not = \lambda(\sigma, \nu) \} 
$,
satisfies  $r_{\sigma, \nu}<   \lambda(\sigma, \nu)$. 

\smallskip
$(4)$ {\rm  [Analyticity in compact sets.] } The operator ${\bf H}_{s, w}$ depends  
analytically  on $(s, w)$  for $(\Re s, \Re w)  \in  \Sigma_0 \times  W_0$. 
Thus,  $\lambda(\sigma, \nu)^{\pm1}$,  $f_{\sigma, \nu}^{\pm 1}$, 
and $f'_{\sigma, \nu}$  depend analytically  on 
$(\sigma, \nu) \in  \Sigma_0 \times  W_0$, 
and are  uniformly bounded in any compact subset.

\smallskip
$(5)$ {\rm [Analyticity in a neighborhood of $(1,0)$.] } 
If $(s, w)$ is complex near $(1, 0)$ then
$\lambda(s, w)^{\pm1}$,  $f_{s, w}^{\pm1}$, and $f'_{s, w}$ are well-defined and 
analytic;
moreover,   for any  $\theta$, with $ r_1 < \theta < 1$,  one has  
$ r_{1, w}/|\lambda(1,w)|   \le \theta$.

\smallskip
$(6)$ {\rm [Derivatives of the pressure.] }  For $(\sigma, \nu) \in 
\Sigma_0\times W_0$, define the {\it pressure function} 
$\Lambda (\sigma, \nu)=\log \lambda(\sigma,\nu)$.

\pn $(6.a)$ 
$\Lambda'(1)$ is the opposite of the Kolmogorov entropy of the 
dynamical system  $(T, \mu_1)$. Also,
 $\Lambda' _w(1,0)$ is the $\mu_1$--average of the cost:
$$ \Lambda'(1)  = 
- \int_{{\cal I}} \log | T'(x)|\,   f_{1}(x) \, dx < 0 \, ,  \qquad 
\Lambda' _w(1,0)= \sum_{h\in \HH} c(h) \int_{h(\cal I)} f_1(x)\, dx.$$

\pn $(6.b)$ 
 If $c$ is not constant,  the  second derivative $\Lambda''_{w^2} (1, 0)$ is strictly positive. 

\smallskip
$(7)$ {\rm [Function  $w \mapsto \sigma(w)$.] } There is a complex neighborhood 
$\cal W$ of $0$ and
a unique function $ \sigma:\cal W \to \complex$ such that $\lambda(\sigma(w), w) 
= 1$,
this function is  analytic, and  
$\sigma(0) = 1$.  
 }

\medskip
\pn  {\bf Proof.} 
We refer to  \cite {Bal, Bro, Va7} except for   claim  (7) (which 
follows from $\lambda'(1)\ne0$
and the implicit function theorem)  and  for  claim (6):

\pn $(6.a)$ Taking the derivatives at   $(1, 0)$ of 
${\bf H}_{s,w} [f_{s,w}] = \lambda (s,w) f_{s,w}$ (with respect to $s$ or $w$), integrating on ${\cal I}$
with respect to $ \hat \mu_{1, 0}$ (equal to the Lebesgue measure), 
and using 
that
${\bf H}_{1}^*$ preserves $\hat \mu_{1}$,
gives the expressions as integrals.
To finish, apply Rohlin's formula.

\pn $(6.b)$ Convexity of the pressure is an old theme, see, e.g.,
\cite{Ru} and also  \cite{PP}, Chapter 4, Prop. 10 in  \cite {Va1}, 
Prop. 3.8 in \cite {CM}, Proposition 6.1 in \cite {Bro}. We  adapt  here the last work.   It is stated in the context of functions with
 bounded variation. Due to our  strong Markov assumption, we may work in ${\cal C}^1({\cal I})$. Since $f_{1}$ is a strictly positive
${\cal C}^1$ function and   ${\bf H}_{1}[c]$
 belongs to  ${\cal C}^1$,  we may transfer Broise's proof  to our ${\cal C}^1$ context:  it shows that  $\Lambda'' _{w^2} (1, 0)$ is zero if and only if there exists a constant $K$
and a function $u \in {\cal C}^1 ({\cal I})$ for which, for all $h \in {\cal H}$, the equality 
$ c(h) = u -u \circ h + K$ holds. Using the fixed point of each branch $h$ proves that $c$ is constant.\qed

\bigskip
\pn    
 We have already remarked that  the three Euclidean dynamical systems
  belong to the good class.  A sufficient condition for the cost $c$ to be of moderate growth  
is 
$c(m, \epsilon):= c(h_{[m,\epsilon]}) = O(\log m)$.
Note that,  for an inverse branch of depth $k$, of the form  $ h_{[q_1]}
\circ\cdots\circ h_{[q_k]}$, the interval $h ({\cal I})$ 
gathers all real $x$ for which the $k$ first digits  of the $CF$ 
expansion are $(q_{1}, q_{2}, \ldots, q_{k})$. 
Furthermore,  each inverse branch of any depth 
is a linear  fractional transformation 
$h(x)=(ax+b)/(cx+d)$, with $a$, $b$, $c$, $d$ coprime integers,
with determinant $ad-bc=\pm 1$,   and
denominator $D[h]$  related to  $|h'|$ through:
$$D[h](x): = |cx+d| =  |\det h|^{1/2}  \,   |h'(x)| ^{-1/2}=  |h'(x)| ^{-1/2} \, .
$$
Therefore,  the  transfer   operator  can be  alternatively  defined by 
\begin {equation} \label {Ruop}
 {\mathbf H}_{s, w} [f](x)  = \sum_{h \in \cal H}
\exp [w c(h)]  { \frac{1} { D[h](x)^{2s}}}\  f (  h  (x)) \, . 
\end {equation}
The above reformulation will be useful in \S2.4.

\pn We associate to each of the three algorithms and
its final set ${\cal F}\subset {\cal  H}$, 
a   transfer operator  
\begin {equation}\label {Ruop2}
  {\mathbf F}_{s, w} [f](x) := \sum_{h \in {\cal F}}
\exp [w c(h)]  { \frac{1}{ D[h](x)^{2s}}}\  f (  h  (x))
=     {\mathbf H}_{s, w} [f \cdot 1_{\cup_{ h \in \cal F} h(\cal I)}](x). 
\end{equation}
It is easy to see that $\mathbf F_{s,w}$  acts boundedly on $\cal C^1({\cal I})$
for $(\Re s, \Re w)\in  \Sigma_0 \times  W_0$, and to generalize the relevant
statements of Proposition~0 to this operator.

\subsection {Transfer operators and real trajectories.}  
We consider   one of our three 
algorithms, or more generally any  triple of ${\cal {GMG}}$ type, 
with a non-
constant cost $c$. The interval  ${\cal I}$ is endowed with  a probability 
measure  with smooth
density $f$ and the set of endpoints of fundamental intervals (rational points for Euclidean algorithms)
can be neglected. We  are  interested in the asymptotic behavior of the distribution 
of   
$ 
C_n(x) := \sum_{i= 1}^n  c(h_i) 
$ when the truncation 
degree $n$ tends to $\infty$.
As already mentioned,  a very convenient tool  is the  L\'evy moment generating 
function  of the cost, 
$\E [\exp (w C_n) ] =
 \sum_{h \in \HH^n} \exp [w c(h)] \cdot \int_{h({\cal I})} f(y) \, dy$.
The  change of variables $y= h(u)$ gives 
\begin{equation} \label {Mn}
\E [\exp (w C_n) ]= \int_{\cal I} 
\sum_{h \in \HH^n} \exp [w c(h)] \cdot |h'(u)|  \cdot f\circ h(u)\,  du 
=  
\int_{\cal I} {\mathbf H}_{1, w}^n[f](u) \, du  \, .
\end{equation}
The above relation is 
fundamental  for analysing costs  on truncated real trajectories, as we explain 
next.
 
\pn 
By Proposition 0,     for any $\theta$ with  $r_1< \theta < 1$, 
 there is a  small complex neighborhood $\cal W$ of 
$0$, so that,  for  $w \in {\cal W}$, 
the operator   ${\mathbf H}_{1, w}$ splits as 
 $
{\bf H}_{{1, w}} =  \lambda(1, w) \,  {\bf P}_{1,  w} +   {\bf N}_{1,   w}
$,
where ${\bf P}_{1, w}$ is the projector 
for the eigenvalue $\lambda(1,w)$ and 
$R({\bf N}_{1,   w})\le \theta \lambda(1,w)$. 
Therefore,  
$$ 
 {\bf H}_{{1, w}}^n[f](u) =  \lambda(1, w)^n \,  {\bf P}_{1,  w}[f](u)  +   
{\bf N}_{1,  w}^n[f](u) \, , \, \forall n \ge 1\, ,
$$
which entails
$$ \E [\exp (w C_n) ]= \left (  
\lambda(1, w)^n  \int_{\cal I} {\bf P}_{1, w}[f](u)\, du  \right) \left (1 + 
O(\theta^n )\right ) \, ,
$$
with a uniform $O$-term for $w\in {\cal W}$. 
In other words the moment generating function  behaves as a 
``quasi-power,''   and we may apply the following  result:

\medskip
\pn 
{\bf Theorem 0. }[Hwang's Quasi-Power Theorem] \cite {Hw, Hw1, Hw2}
{\sl Assume that the moment generating functions
for a sequence of functions $ \widehat  C_N$ on probability spaces  
$(\widehat \Omega_{N}, \hat\Pr_N)$ 
are  analytic in a   complex  neighborhood ${\cal W}$ of 
zero, where 
\begin{equation}\label{QP}
 \E_N [\exp(w \widehat C_N)]  = \exp [\beta_{N}  U(w) + V(w)]  \left (1 + 
O(\kappa_{N}^{-1}) \right) \, ,
\end{equation}
with $\beta_{N}$, $ \kappa_{N} \to \infty $ as $N \to \infty$, and $U(w)$, 
$V(w)$
analytic  on ${\cal W}$. Assume  
$U''(0) \not = 0$. Then, the mean and 
the variance satisfy
$$
 \E_N [\widehat C_N]= \beta_N U'(0) + V'(0) + O(\kappa_N^{-1}) \, , 
$$
$$ \Var_{N}[ \widehat C_N] = \beta_{N} U''(0) + V''(0) + O(\kappa_{N}^{-1}) \, .
$$
Furthermore, the distribution of $\widehat C_N$ on $ \widehat \Omega_{N}$ is 
asymptotically Gaussian,  
with speed of convergence 
$O (\kappa_{N}^{-1} + \beta_{N}^{-1/2})$.   

\smallskip
\pn  For each fixed $k\ge 3$, there  is 
 a polynomial $P_{k}$
of degree at most $k$, with coefficients depending on the
derivatives of order at most $k$ at $0$ of $U$ and $V$,
so that the moment of order $k$ satisfies
\begin{equation}
\E_N [\widehat C_N^k]  = P_{k}(\beta_N)\ + O \left( 
\frac{\beta_N^{k-1}}{ \kappa_N} \right) \, ,
\end{equation}
 with a $O$--term uniform in $k$.}

\medskip
\pn {\bf Proof.} [Sketch.] This statement encapsulates a classical 
calculation analogous to the proof of the central limit theorem 
by characteristic functions.The speed of convergence then results from the 
Berry-Esseen inequalities. the moment estimates are consequences of the derivability
of analytic functions. \qed

\medskip
\pn For our application, we set  $\widehat \Omega_n =(\cal I,  f \, dx)$ for all 
$n$,
$\widehat C_n=C_n$, $ \beta_n= n$,  $\kappa_n= \theta^{-n}$. The
function $U$ is the pressure function $w \mapsto \Lambda(1, w)$,
 and 
$V(w)=  \log  \left( \int_{\cal I} {\bf P}_{1,w}[f](u) \, du \right)$.
Since $c$ is not constant,  the function $ \Lambda(1,w) $ is 
absolutely convex at $w=0$,
 (see Proposition~0) and  $U'' (0)\ne 0$. 
Thus, using the formula for $\Lambda'_w(1,0)$
in Proposition~0, Theorem~0  entails  
the following Gaussian  asymptotic distribution result, which applies
in particular to our three Euclidean algorithms.

\bigskip
\pn {\bf Theorem 1. }{\sl  
For a  triple $(\cal I, T, c)$ of $\cal {GMG}$ type with
non-constant $c$ and any 
probability  $\Pr$ on $\cal I$ with a $\cal C^1$ density,
there are  $\hat \mu(c)> 0$ and $\hat \delta(c)> 0$  so that
for any $n$, and any $Y \in \real$
$$
 \Pr \left [x~\bigm|~
\frac {C_n(x) - \hat \mu(c) n}{  \hat  \delta (c)  \sqrt n} \le Y
\right]  =  
\frac {1} {\sqrt {2\pi} } 
\int_{- \infty}^Y e^{-y^2/2} \, dy + O \left (\frac {1}{\sqrt n}\right) \, . 
$$
Furthermore,  (recalling that $r_1$ is the subdominant spectral 
radius of the density transformer ${\bf H}$),  
for any  $\theta$ which satisfies  $r_1<\theta  <1$, one has : 
$$ 
\E\,  [C_n]= \hat \mu(c)  \cdot n+\hat \eta(c)+ O(\theta^{n}) \, , 
\quad
\Var\, [C_n] = \hat  \delta^2(c)\cdot   n  + \hat \delta_1(c) + 
O(\theta^{n})\, ,
$$
with $ \hat \mu(c)=\Lambda'_w (1,0) =\lambda'_w(1,0)$, 
$\hat \delta^2(c)= 
 \Lambda''_{w^2} (1,0)  =  \lambda''_{w^2}(1, 0)-\lambda'^2_w(1, 0)$.
\pn Finally, $\hat \mu (c)$ involves the  invariant density $f_1$, 
$$
\hat \mu (c) = \sum_{q \in {\cal Q}} c(q) \int_{{\cal I}_{q}}  f_{1}(x) \, dx \, .
$$}

\smallskip
\pn 
Note that $f_1$ is explicitly given in Figure~1 for $\cal G$, $\cal O$, 
and $\cal K$, so that $\hat \mu(c)$ is computable in these cases.

\subsection  {\bf Dirichlet generating functions and transfer operators.}   We  
restrict now our study, for each of our three algorithms,   to  (nonzero) 
rational inputs $x=(u/v)\in {\cal I}'$ and a
cost $c$ of moderate growth.  The intervals  ${\cal I}'$  are defined in Figure 1. 
We  consider the sets
$$ \tilde  \Omega := \{(u,v)\in \mathbb N_\star ^ 2 \, ,  \frac u v \in {\cal I}' \} \, , 
\qquad 
\Omega := \{(u, v) \in \tilde \Omega \, ; {\rm  gcd}\, (u,v)=1    \}\, , 
 $$
and we endow the sets 
$$ \tilde  \Omega_N := \{(u,v)\in \tilde \Omega ;    v\le N\} \, ,  \qquad
\Omega_N := \{(u, v) \in \Omega; v\le N\}    \}\, , 
 $$
with uniform 
probabilities  $\widetilde \Pr_{N}$ and $\Pr_{N}$, respectively. 
For the moment we only consider $\Omega_N$.

\smallskip
\pn 
To  study  the distribution of the total cost $C(u, v)$  (\ref  {digcost1})
associated  to some digit-cost $c$  and restricted to $\Omega_N$ we use its 
{\it moment generating function:}  
\begin{equation} \label {EN}
\E_{N} [\exp (wC)]:=  \frac {\Phi_{w}(N)} {\Phi_{0}(N)} \, ,
\end{equation} 
where $\Phi_{w}(N)=\Phi_{c,w}(N)$ is
the cumulative value of  $\exp[wC]$    on $\Omega_{N}$, 
\begin{equation} 
\Phi_w (N):= \sum_{(u, v) \in \Omega_{N}} \exp [w C(u, v)] 
\, ,\quad \Phi_0(N) = |\Omega_N| \, .
\end{equation} 
 Extending the principles defined in  \cite{Va5, Va7, Va8}, we    replace the  sequence of  moment  
generating  
functions by a  single bivariate {\it Dirichlet  series,} henceforth called 
the {\it Dirichlet-moment generating function:}
\begin{equation}\label {S} 
 S(s, w) := \sum_{(u, v)\in \Omega}
 \frac {1} {v^s}  \exp [w C(u, v)]
 = \sum_{n \ge 1} \frac {c_n(w)} {n^s} \, , 
\end{equation}
where  
$c_n(w):=\sum _{(u, v)\in \Omega_n\, , v=n} \exp [wC(u, v)]$.
  
\pn Since the partial sum  of the coefficients of the series   $S(s, w)$ 
satisfies
\begin{equation}\label{phiw}
\sum_{n \le N} c_{n}(w)=\Phi_w(N)\, ,
\end{equation}
to analyze
the  moment generating  function $\E_N [\exp(wC)]$ of the cost $C$ on 
$\Omega_{N}$, 
it suffices  to estimate
the functions $\Phi_w(N)$  (asymptotically in $N\to \infty$,
and uniformly in $w$ in a complex neighborhood of $0$).

\medskip
 \pn  As we previously 
did for  truncated real trajectories, we aim to relate 
the moment generating function of costs on rational trajectories
to the weighted transfer operator. An execution of  the 
Euclidean algorithm on a  input $(u, v)\in \Omega$,  performing  $P(u,v)$  
steps uniquely   decomposes  the rational $$ \frac{u}{v}=
  h_1\circ h_2\circ \ldots \circ h_P (0) = h(0),$$
with $h_{i} \in \HH, 1 \le i \le P-1$ and  $h_P \in {\cal F}$.
Thus, each Euclidean algorithm defines a bijection   
between  the 
sets $\Omega$ and ${\cal H}^\star \times {\cal F}$.
In view of (\ref{Ruop}, \ref{Ruop2}) and (\ref{S}),  the relations 
$$ 
v= D[h](0), \qquad C(u, v)= c(h)=\sum_{i=1}^{P(u,v)} c(h_i)
$$
 provide the desired
expression for the Dirichlet moment generating function $S(s, w)$ 
in terms of   the transfer operators  ${\mathbf H}_{s, w}$ and 
${\mathbf F}_{s, w}$:
\begin {equation} \label {R1}  
S(2s, w) =   {\mathbf F}_{s, w} \circ (I- {\mathbf H}_{s, w})^{-1}[1] (0) \, .
\end{equation}
Returning to $\tilde \Omega$,   we remark  that each element $(u', v')$ of $\tilde \Omega$ can be 
written  in a unique way as $(du,  dv)$ with $d= {\rm gcd}(u, v)$ and $(u, v)\in 
\Omega$, and, as already observed, an execution of the 
algorithm on $(u', v')$    leads to  the same  decomposition 
(\ref{dec}),  the two costs $C(u', v')$ and $ C(u, v)$ being equal.
 We may define
a Dirichlet moment generating function $\tilde S(s,w)$, and we get
\begin {equation} \label {R1tilde}   
\tilde S(2s, w):=   \sum_{(u, v)\in \tilde \Omega}
 \frac {1} {v^{2s}}  \exp [w C(u, v)] = \zeta(2s) \,  {\mathbf F}_{s, w} \circ (I- {\mathbf H}_{s, 
w})^{-1}[1] (0) \, .
\end{equation}
Using well-known properties of the Riemann zeta function $\zeta(s)$, all
our results for $\tilde \Omega_N$ will follow from those on $\Omega_N$.

\smallskip
\pn In view of (\ref{EN}, \ref{S}, \ref{phiw}),
the relations (\ref{R1}), and  (\ref{R1tilde})  connecting
the Dirichlet moment generating function 
with the transfer operator are the  analogues for rational trajectories  
of the relation (\ref{Mn}) for the truncated real 
trajectories.  In the case of rational trajectories, we have to  work with
the quasi-inverse and
extract the coefficients of  Dirichlet series: This is why the discrete 
problem  is more difficult to solve 
than the continuous problem. 

\subsection  {\bf Perron's  formula and Dolgopyat's estimates.}  
We  wish to evaluate the  sum $\Phi_{w}(N)$ of the first 
$N$ coefficients of the Dirichlet series $S(2s, w)$.
Our   first  main tool  towards this goal is the Perron formula. 
The
{\it Perron formula} of order two (see e.g. \cite {El2})
applied term by term to a Dirichlet series  
$F(s) =  \sum_{n \ge 1} a_n  n^{-s}$ and a vertical line 
$\Re s =D>0$  inside the domain of convergence of $F$ says that
\begin{equation} \label {Psigene}
  \Psi(T):=  \sum_{n \le T} a_n (T-n) 
= \frac {1} {2i \pi} \int_{D -i\infty}^{D +i \infty} F(s) \frac {T^{s+1}} {s(s 
+1)} \, ds \,  .
\end{equation}
Applying Perron's formula to the Dirichlet series $S(2s,w)$, we find 
 \begin{equation} \label {Psiw}
  \Psi_{w}(T):=  \sum_{n \le T} c_n(w) (T-n) 
= \frac {1} {2i \pi} \int_{D -i\infty}^{D +i \infty} 
 S(2s, w) \frac {T^{2s+1}} {s(2s +1)} \, ds \, .
\end{equation}
Thus, Perron's formula gives us information on
$\Psi_w(N)$, which is just a Ces\`aro sum of the   $\Phi_{w}(Q)$:
$$
 \Psi_{w}(N) =  \sum_{Q\le N} \sum_{n \le Q} c_{n}(w) =
 \sum_{Q\le N} \Phi_w(Q) \, .
$$

\smallskip
\pn    Proposition~0   and Relation  (\ref {R1}) show
that     $s \mapsto S(2s, w)$  has a possible pole
at $s= \sigma(w)$, where $\sigma(w)$ is the unique complex number  near $1$ for 
which  $\lambda(\sigma(w), w)=1$.
In particular, the  integral  
in (\ref {Psiw}) is well-defined only if $D > \sup_w (\Re\sigma(w))$
(note that $\sup_w (\Re\sigma(w))>1$ can be made arbitrarily close
to $1$ by taking $w$ close enough to zero).
To combine the Perron formula with Cauchy's residue theorem,
we wish to modify the integration contour $\Re s=D$ into a 
contour containing $\sigma (w)$ as a unique pole of $S(2s, w)$.
This is  possible and leads to a quasi-power expansion if
there is $\alpha>0$, such that 

\pn $(i)$ 
$S(2s,  w)$ admits $s= \sigma(w)$ as a unique pole in the
strip $ |\Re s - 1 |\le \alpha$.

\pn $(ii)$  In a ``perforated'' half-plane $\{ \Re s \ge 1- \alpha_0 \, , |s-1|> \alpha_0 /2 \}$,
for small $\alpha> \alpha_0 >0$, the estimates 
$|S(2s, w)|\le \max(1, (\Im s)^\xi)$ for $0<\xi <1$,
 hold,  uniformly in $w$ close to $0$. 

\smallskip 
\pn   Note that $(i)$ cannot be satisfied if the map $T$ is 
$C^2$ conjugated  with a piecewise affine  map, since, in this case
the Dirichlet series $S(s, 0)$ has  an infinite number of poles in the vertical  
$ \Re s = 1$.   Note also  that  bounds of the type  $(ii)$   are
extremely  difficult  to obtain for  general  Dirichlet series. By the above discussion, such bounds would follow from similar
estimates on the quasi-inverse of the operator, which 
are  closely related to those obtained by Dolgopyat \cite{Do}.
    In the spirit of Chernov \cite{Cher},  Dolgopyat \cite{Do}
introduced several ``uniform nonintegrability'' ({\em UNI}) conditions. They   
allowed him to control oscillatory integrals associated to
iterates of transfer operators ${\bf H}_{s}$ for $s=\sigma+it$, with
fixed $\sigma$ close to $1$, and prove exponential decay of correlations
for some flows.  We shall 
give a new formulation of Dolgopyat's strongest such   {\em UNI}  condition, which appeared
implicitly  in Section 5 of his paper \cite{Do} and turns out to be
satisfied by our three algorithms.
This condition is stated as an assumption on 
the  derivatives of the inverse 
branches of the dynamical system  in Section 3.2,  and  expresses that, in a sense,  the map $T$
is  quite different from a piecewise affine map.   
In Section~3, we shall  prove 
the following theorem, which is the  central functional analytic result of the 
paper.
In the statement, we use the following family of equivalent norms on 
$\cal C^1(\cal I)$:
\begin{equation}\label{tnorm}
\|f \|_{1,t}:=\sup|f|+ \frac{\sup | f'|}{ |t|}  \, , \quad t \ne 0 \, .
\end{equation}

\smallskip
\pn{\bf Theorem 2} [Dolgopyat-type estimates].
{\sl Let $({\cal I}, T, c)$ be a triple of ${\cal {GMG}}$-type,
with contraction ratio  $\rho<1$, and such that
Condition {\em UNI}   from \S3.2 holds. Let ${\bf H}_{s, w}$
be its weighted transfer operator (\ref{transfer}) acting on $\cal C^1(\cal I)$.

\pn For any $\xi$,  with $0< \xi< 1/5$,  there is a  (real)  neighborhood
$\Sigma_1=]1-\alpha, 1+\alpha[$ of $1$ (which depends only on $({\cal I}, T)$ and not on $c$),  a (real) neighborhood  $W_1$ of 
$0$, 
and there is   $M>0$  such that, 
  for all $s=\sigma+it$,  $w=\nu+i\tau$ 
with $(\sigma, \nu) \in  \Sigma_1 \times W_1$  and  $|t|\ge 1/ \rho^2$,    
 \begin{equation}\label {Dolgo2}
\|(I-{\bf H}_{s, w})^{-1}\|_{1,t} \le M\cdot |t|^\xi \, .
 \end {equation}}

\pn Note that we shall have to modify Dolgopyat's arguments since we  must  consider  dynamical 
systems which  possess an infinite number of branches
(see in particular Lemma ~1), and we 
work with  bivariate weighted  transfer operators
${\bf H}_{s, w}$ involving a cost function.  

\subsection  {\bf Statement of the Central and Local Limit Theorems.}
\pn  
We shall see in \S4.1 that Perron's  
Formula   (\ref{Psiw}) combined  with the fundamental relation
(\ref{R1}), together with the bounds \`a la Dolgopyat  (Theorem 2) provide  a
quasi-powers estimate for the  Ces\`aro sum
$\Psi_{w}(N)$.
It does not seem easy to transfer
this information  on $\Psi_{w}(N)$ to
estimates on $\Phi_{w}(N)$, because the
coefficients are complex. The way we overcome this is by
first proving  (Lemma~11 in Section 4.2) quasi-power estimates
for  the   moment generating function of some ``smoothed''
version  of  the  cost $C$, for which the transfer is possible
by standard methods. 
We are then able to  apply Theorem~0 to the
smoothed model, and show that the two models
are close enough in distribution so that the following holds:

\medskip
\pn {\bf Theorem 3.} [Central Limit Theorem for rational trajectories.] {\sl  For  a Euclidean 
algorithm   amongst ${\cal G}$, ${\cal K}$, ${\cal O}$,  there is $\gamma >0$, so that,  for any  
cost  $c$ of  moderate growth,  letting $\Lambda(s)$
be the function from Proposition~0:

\pn $(a)$ The 
distribution of  the total cost $C$ on $\Omega_{N}$  is asymptotically Gaussian,   
with speed of convergence  
$O  (1/\sqrt {\log N})$, i.e.,  there exist two constants $\mu(c)> 0$ and 
$\delta(c)> 0$ such that, for any $N$, and any $y \in \real$
\begin{eqnarray}
&&\nonumber \Pr_{N} \left [(u, v); 
\frac {C(u, v) -  \mu(c) \log N}{    \delta (c)  \sqrt {\log N}} \le y\right]\\
\nonumber &&  \qquad\qquad\qquad\qquad\qquad\qquad=  
\frac {1} {\sqrt {2\pi} } 
\int_{- \infty}^y e^{-x^2/2}\,  dx + O \left (\frac {1}{\sqrt {\log N}}\right) 
\, . 
\end{eqnarray}

\pn $(b)$ The
mean and the variance satisfy
$ \E_N [C]= \mu(c)  \log N  +  \eta(c)+ O({N}^{-\gamma})$, 
and
$ \Var_{N}[C] =  \delta^2(c) \log N  + \delta_1(c) +  O({N}^{-\gamma})
$.

\pn Generally, for each $k \ge 1$,  there is
 a polynomial $P_{k}$ of degree $k$ so that
$$ \E_{N} [C^k] = P_{k}(\log N)  + O  \left( \frac { (\log N)^{2k}} {N^{\gamma}} \right)\,  , $$
with a $O$-term uniform in $k$.

 \pn $(c)$  In the special case $c\equiv 1$, denoting $\mu:=\mu(1)$, 
$\delta^2:=\delta^2(1)$, we have
\begin{equation} 
\nonumber
\mu =\frac {2}{|\Lambda'(1)|}   > 0,  \quad 
\delta^2=  \frac {2 |\Lambda''(1)|} {|\Lambda'(1)^3|}  > 0 \, .
\end{equation}

\pn In the general case, 
$$  
\mu(c) = \mu \cdot \hat \mu (c) \, , \qquad   
\delta^2(c)= \hat  \mu^2 (c) \cdot \delta^2 + 
               \mu \cdot \hat \delta^2(c) + \mu^2 \hat \mu(c)  \cdot \chi(c) > 0 
\, ,
$$
where $\hat \mu (c) > 0$ and $\hat \delta^2(c)\ge 0$  are given
in Theorem~1, and 
$\chi(c)=\Lambda''_{sw}(1,0)$.

\pn  Claims (a), (b), and (c) also hold for $\tilde \Pr_{N}$  on 
$\tilde \Omega_{N}$.}

\medskip
\pn 
Note   that $2/\mu$ is the Kolmogorov entropy 
for $(T, f_1 dx)$. Also, $\gamma$    does not  depend on the
cost.
The constant $\chi(c)$ can be viewed as a covariance coefficient
between the number of steps  $P$ and the cost $c$.
Since there exists a 
closed form for  $f_1$ in  the three cases of 
interest (cf. Figure 1), the constants $\mu$, and  thus
$\mu(c)$ can be easily computed (see remark after Theorem~1 and  \cite{Va8}). 
The  constants  $\delta$, 
$\delta(c)$ are proven to be non zero in Proposition 1, Section 3.5. They  do not seem to admit a closed form. However, Lhote  has 
proved that they can be computed in polynomial time \cite {Lh}.

\medskip

\pn In  Section 5, restricting to  lattice
costs,  we  obtain bounds
for $\E_N [\exp(i\tau C)]$ with $\tau\in [-\pi,\pi]$, and combine
them
with saddle-point estimates to get:

\medskip
\pn {\bf Theorem 4. }[Local Limit Theorem for lattice costs.] {\sl   For any 
algorithm  among ${\cal G}$, ${\cal K}$, ${\cal O}$,  and
any lattice cost $c$  of span $L$ and of moderate growth,
 letting $\mu(c)>0$ and $\delta^2(c)>0$ be the constants  
from Theorem~3,  the following holds, 
$$
 \Pr_{N} \biggl [(u, v); -\frac L 2 < C(u,v)- 
 \mu(c) \log N -  \delta(c) x \sqrt {\log N} \le \frac L 2  \biggr  ]  $$
$$ = 
\frac {e^{-x^2/2}}  {\delta(c) \sqrt {2 \pi \log N} }  +
O   \left(  \frac {1} {\log N} \right),  
$$
 with a $O$ uniform for $x \in \mathbb R$. The same 
holds for $\tilde \Pr_{N}$ in $\tilde \Omega_{N}$.  }

\section { Property  {\em UNI} and Dolgopyat estimates.}
\pn  In this  section we shall first prove Theorem 2 and check (\S 3.5) that
its additional assumption {\em (UNI)} holds for our three Euclidean algorithms.
Then, we show that Theorem~2 together with aperiodicity results imply
useful estimates on $S(s,w)$.

\pn To prove Theorem~2,
we   use    ideas  due to  Dolgopyat \cite {Do}; however, we have to 
adapt them to our context, i.e., bivariate
weighted transfer operators $\mathbf{H}_{s,w}$ associated to 
triples $({\cal I}, T, c)$ 
of ${\cal {GMG}}$-type with infinitely many branches, as explained after the statement of Theorem 2
in Section 2.   
One of   the main ideas of Dolgopyat was to 
 deal first with the $L^2$-norm of some iterate 
 $ \widetilde  {\bf H}^n_{s, w}[f]$, with an index $n$ that depends on $t= \Im s$. 
Then, he made two 
transfers of estimates:  first from this $L^2$-bound  
into a bound  for the sup-norm, next  from this sup-norm-bound  
into the desired  bound  for the $(1, t)$-norm.
Following his strategy,
we establish    preliminary results in  Lemmata  1,  2 and 3.
(Note that Lemma~1 is new.)
\S 3.2 is central: our version of the {\em UNI} Condition is stated 
and shown to
entail the desired exponential estimate for the $L^2$-norm of the operator 
(Lemmata 4 and 5). 
The two   transfers are operated in \S 3.3: they lead to Theorem 2. 
 We then check in Lemma 6 (\S 3.4) that
its additional assumption {\em (UNI)} holds for our three Euclidean algorithms.
In \S 3.5, we     show  both an aperiodicity result on vertical $s$-lines and
$w$-lines together with a convexity result  (Lemma 7 and Proposition 1).
Finally, in \S3.6, we  return to the Dirichlet series $S(s, w)$  and   
obtain further estimates 
  in the compact
neighborhood of the real axis which is not covered
by Theorem 2 (Lemmata  8 and 9).

\subsection {Preparatory material.}   
\pn {\bf Transfer operators ${\bf H}_{s, w}$ and their normalizations 
$\tilde {\bf H}_{s, w}$.} 
Triples $(\cal I, T, c)$ of $\cal{GMG}$ type 
and their associated transfer operators ${\bf H}_{s, w}$, acting on
$\cal C^1(\cal I)$,  for $(\Re s, \Re w) \in \Sigma_0 \times W_0$ were 
introduced in \S~2.2.
We summarized in 
Proposition~0   well-known spectral properties that we shall need,
in particular the existence of a dominant eigenvalue $\lambda(\sigma, \nu)$, 
[with positive eigenfunction $f_{\sigma, \nu}$]
or $\lambda(s,w)$ for suitable real  $(\sigma,\nu) \in \Sigma_0 \times W_0$
[from (\ref{MMGG})]
and complex $(s,w)$ close to $(1,0)$.
Recall that  $\HH^{n}$ is  the set of  inverse 
branches of $T^n$,  $\HH^{\star}$ the set of all inverse branches
of any  depth, $\rho< 1$ the contraction constant  and $\hat K$
the distorsion constant. 

\smallskip
\pn 
Defining  $ K= \hat K/(1-\rho)$, it
is not difficult to check that, 
$$
|h''(x)|\le   K |h'(x)| \, , \quad \forall x\in {\cal I}\, , h \in \HH^{\star }  
\, .
$$  
The above   bounded distortion  property will play an important 
r\^ole. Note for further use that, 
since the ratio $h''(x) /h'(x)$ is the derivative of $\log  |h'(x)|$,  
setting $L:= e^K$, we get
\begin{equation}\label {L}\frac 1 L \le  \frac {|h'(x)|} {|h'(y)|}  \le L 
\qquad \hbox { for all 
$x, y \in {\cal I}$, for all $h \in \HH^\star$} \, .
\end{equation}

\pn It will be convenient to work
with the {\it normalized operators} $\tilde  {\bf H}_{s, w}$ defined by 
 \begin {equation} \label {normtransfer}
\widetilde  {\bf H}_{s, w} [f] =  \frac {1} {\lambda(\sigma, \nu) 
f_{\sigma, \nu}}\,{\bf H}_{s, w}  [f_{\sigma, \nu} \cdot f]\, ,
\quad s=\sigma+it,  \  w = \nu +i \tau \, .
\end{equation}
By construction, for  $(\sigma ,\nu) \in  \Sigma_0 \times  W_{0}$,  the operator 
$\widetilde{\bf  H}_{\sigma, \nu}$  acting on $\cal C^1({\cal I})$
has spectral radius equal to $1$,  and fixes the constant function $\equiv 1$.
Also, ${\rm Sp}\,  {\bf H}_{\sigma+it, \nu}=
\lambda(\sigma,\nu) {\rm Sp}\, \tilde{ \bf H}_{\sigma+ it,\nu}$.
Remark  next  the 
inequality  $|| \tilde {\mathbf H}_{s, w} [f] ||_0  \le 
|| f||_0 \, \tilde {\mathbf H}_{\sigma, \nu} [1] =  || f||_0$, 
which implies the useful  bound  
\begin {equation} \label {H1}
 ||\tilde  {\mathbf H}_{s, w}||_0 \le 1 \, .
\end{equation}
It is  easy to check that $\widetilde{\bf  H}_{\sigma, \nu}^*$ 
fixes the probability measure $\mu_{\sigma, \nu}=f_{\sigma, \nu} \cdot  
\hat \mu_{\sigma, \nu}$.

\smallskip 
\pn {\bf Remark about notations.} {\sl In 
the sequel, the notation $A(x) < \!\!<B(x)$ means:  $A $ is less than $B$ up to 
absolute multiplicative constants. This means that there exists some absolute  
constant 
$k$  such that for every $x$
of interest, $A(x) \le k B(x)$.  It is synonymous with $A(x) = 
O(B(x))$ with an absolute $O$-term.  The  
symbol ${\cal W}$ denotes a complex neighborhood of $0$ for the   variable 
$w$. If $J\subset \cal I$ is a union of intervals,
we denote by $|J|$ its Lebesgue measure.}

\bigskip 
\pn {\bf Relating $\tilde {\bf H}_{\sigma,\nu}$ and $\tilde {\bf H}_{1,0}$.}      
In order to exploit properties of Lebesgue measure,
which is fixed (only) by the dual of ${\bf H}_{1,0}$, Dolgopyat uses the 
following
property (see e.g.  last lines of p.~367 in  \cite{Do}):  When 
$({\cal I} , T) $ has finitely many branches, 
there is
$A_\sigma \to 1$ as $\sigma \to 1$ so that for positive $f\in \cal C^1(\cal I)$
\begin{equation}\label{finitely}
\tilde {\bf H}_{\sigma,0}  [ f ](x) \le A_\sigma\,  \tilde {\bf H}_{1,0} [f ](x) \, .
\end{equation}
The above inequality is not true in general when there are infinitely
many branches (it fails for the Gauss map).
The purpose of the following  lemma  is to extend (\ref{finitely})
to the case of infinitely many branches and bivariate operators,
comparing $\tilde {\bf H}_{\sigma, \nu}^n$, and  
$\mu_{\sigma, \nu}$,
to their analogues for $(\sigma, \nu) = (1, 0)$:

\medskip
\pn {\bf Lemma 1.} {\sl  For $(\sigma, \nu) \in  \Sigma_0 \times W_{0} $, denote 
$$ 
A_{\sigma, \nu} := \frac{\lambda(2\sigma-1 , 2\nu) ^{1/2} } 
 {\lambda(\sigma , \nu)} \, .
$$
Let  ${\cal L}$ be  a compact subset of $\Sigma_0 \times  W_{0} $.  
For   ${\cal J} \subset  \HH^{k}$, denote by $J= \cup_{h \in {\cal J}} h({\cal 
I})$.   
Then,  for $(\sigma, \nu)  \in {\cal L}$, 
$$ \mu_{\sigma, \nu} [J]   < \!\!<  A_{\sigma, \nu}^k  \   |J|^{1/2} \, ,
$$
furthermore,   for any $f \in {\cal C}^1 ({\cal I})$,  for 
any integer $k\ge 1$,  
\begin{equation} \label{2Le2}
|| \tilde {\bf H}_{\sigma, \nu}^{k}[f]||_{0}^2  < \!\!<  
   A_{\sigma, \nu}^{2k}\  || \tilde {\bf H}_{1, 0}^{k} [|f|^2]||_{0} \, .
\end{equation}
The absolute constants involved only depend on ${\cal L}$.}

\medskip
\pn The function  $A_{\sigma, \nu}$  depends  continuously on  
$(\sigma, \nu)$ 
and $A_{1,0}=1$.  
   
\medskip
\pn {\bf Proof.}
The equality $\mu_{\sigma, \nu} [f] = \mu_{\sigma, \nu} 
[\tilde {\mathbf  H}_{\sigma, \nu}^k[f]]$, when applied to the 
characteristic function of some fundamental interval $h({\cal I})$ of 
depth $k$ proves that 
$$ \mu_{\sigma, \nu} [h({\cal I})]  < \!\!< \frac { \exp [\nu c(h)]} 
{ \lambda({\sigma, \nu})^k}
\int _{\cal I}  |h'(x)|^\sigma d \mu_{\sigma, \nu} (x) \, .
$$
Moreover, by the  bounded distortion property (\ref{L}),  
the ratios (two by two) of the    three  
quantities $a(h)$, $b(h)$,  $c(h)$, 
$$ a(h):=  \int_{{\cal I}} 
|h'(x)|^\sigma  \,  d\mu_{\sigma, \nu}(x) \, ; 
\quad b(h):=  |h({\cal I})|^\sigma \, ; \quad c(h):=   \int_{{\cal I}} 
|h'(x)|^\sigma \, dx \, ,
$$
admit  upper  and lower bounds 
that do not depend on $h$, and are uniform  for $(\sigma, \nu) \in {\cal L}$.  
Then,   summing the inequalities
$$ 
\mu_{\sigma, \nu} [h({\cal I})]  < \!\!< 
\frac { \exp [\nu c(h)]} { \lambda({\sigma, \nu})^k} |h({\cal I})|^\sigma \, , 
$$
over  $ {\cal J} \subset  \HH^{k}$, and  applying the  Cauchy-Schwarz 
inequality, one gets
$$ 
\mu_{\sigma, \nu} [J]   \le  \frac {1}  { \lambda({\sigma, 
\nu})^k} \left (\sum_{h \in \HH^{k}} \exp [2 \nu c(h)] \cdot  |h({\cal 
I})|^{2\sigma-1} 
\right) ^{1/2}  \left (\sum_{h \in {\cal J}}  |h({\cal I})|
\right) ^{1/2} \, .
$$  
Then,  dominant spectral properties, 
together with bounded distortion,   entail the inequality 
$$ \sum_{h \in \HH^{k}} \exp [2 \nu c(h)] \cdot |h({\cal I})|^{2\sigma-1}  
< \!\!<  \lambda({2\sigma-1, 2 \nu })^k \, , $$ 
and, finally, the relation $|J|= 
\sum_{h \in {\cal J}}|h({\cal I})|$ provides the first claim. 

\smallskip
\pn 
 Consider now    $f \in \cal C^1({\cal I})$. The relation  
$$|(\tilde {\bf H}_{\sigma, \nu})^k[f](x)|
 < \!\!<  \frac {1} {\lambda(\sigma, \nu)^k}
\sum_{h \in {\mathcal H}^{k}}  \exp[\nu c(h)] \cdot |h'(x)|^\sigma 
\cdot  |f\circ h(x)| \, ,
$$
is valid  if $(\sigma, \nu)$ belongs to   ${\cal L}$, and, 
by the  Cauchy-Schwarz  inequality,
\begin{eqnarray}
\nonumber
\left (\sum_{h \in {\mathcal H}^{k}} \exp[\nu c(h)] \cdot |h'(x)|^\sigma 
\cdot |f\circ 
h(x)|\right)^2&& \\
\nonumber
\quad\le 
\left (\sum_{h \in {\mathcal H^{k}} } \exp[2 \nu c(h)] \cdot  |h'(x)|^{2 
\sigma-1}\right) &\cdot& \left (\sum_{h \in {\mathcal H}^{k}}  
   |h'(x)| \cdot |f|^2\circ h(x) \right) \, .  
\end{eqnarray}
The second factor is  exactly  ${\bf H}_{1, 0}^{k} [|f|^2](x)$, which is less 
than 
$\tilde  {\bf H}_{1, 0}^{k} [|f|^2](x)$ (up to absolute multiplicative  
constants). 
Thanks to dominant spectral properties, 
the first factor  is  easily related to  
$ \lambda(2 \sigma-1, 2\nu)^{k}$.
\qed

\bigskip
\pn  {\bf Lasota--Yorke bounds.} 
The following lemma describes how $\widetilde{\bf H}_{s, w}$ acts 
with respect to the quasi-norm $ ||.||_{1}$ when $s$ varies over a vertical
line:

\medskip
\pn 
{\bf Lemma 2.} 
{\sl For every compact  subset   ${\cal L}$  of $\Sigma_0 \times {  W}_0$, 
there is  $C >0$, so that for
all $(s,w)$ with $(\Re s, \Re w) \in {\cal L}$, and all $f\in \cal  C^1({\cal 
I})$,
$$
||\widetilde  {\bf H}^n_{s, w} f   ||_{1} \le   
C  \left (|s| \,   ||f||_{0} + \rho^n \,  ||f||_{1}  \right)
\, , \quad \forall n \ge 1  \, . 
$$}

\medskip
\pn 
{\bf Proof.}
The quantity $\widetilde  {\bf H}^n_{s, w} [f]$ can be written as a 
sum  over $h \in \cal H^n$ of terms
$$
\frac { \exp [w c(h)]}{\lambda(\sigma, \nu)^n}r_{h}(x) \qquad 
\hbox {with} \qquad  r_{h} := |h'|^{s}  \cdot\frac  {1 } 
{f_{\sigma, \nu} } \cdot (f_{\sigma, \nu} f) \circ h \, .
$$
The Leibniz sum for the derivative of $r_{h}$ contains three  terms.
We can bound the first for all $s$  using the distortion assumption since
$$
|s| |h''| | |h'|^{s-1}| \le | s| \bar K  | |h'|^{s} |=  |s|  K   
|h'|^{\sigma}  \, . 
$$
Compactness of ${\cal L}$ and continuity of $(\sigma, \nu) \mapsto 
f_{\sigma, \nu}$, and  $(\sigma, \nu) \mapsto  f'_{\sigma, \nu} $ 
 imply  that  
the second term  may be controlled by
$$
\frac  {|f'_{\sigma, w} | }{ f_{\sigma, w}^2}
  \le C_{{\cal L}}  \frac  {1}  {f_{\sigma, w}}\, ,
$$
for some $C_{\cal L} > 0$.
Finally the last term can be estimated using
$$
 | (f_{\sigma, w}\cdot f) ' \circ h |  |h'| \le \rho^n  [ 
 | f_{\sigma, w}' \cdot  f|  \circ h 
+   (f_{\sigma, w} \cdot |f'|)  \circ h ] \, .
$$
We can ensure
$$
\bar K |s| + C_{{\cal L}}+ \rho^n C_{{\cal L}}
\le C({{\cal L}}) |s| \, ,
$$  
so that  the derivative $(\widetilde  {\bf H}^n_{s, w} [f])'$ 
satisfies 
$$|| (\widetilde  {\bf H}^n_{s, w} [f])'||_{0}  \le 
C  \left (|s| \,   ||\tilde {\bf H}^n_{\sigma, \nu} [f]||_{0} +
 \rho^n \,  ||\tilde {\bf H}^n_{\sigma, \nu} [|f'|]||_{0}  \right) \, . 
$$
The final result follows from (\ref{H1}). 
\qed

 \bigskip
\pn {\bf First use of the  $(1, t)$-norm. } 
In the bound from Lemma 2 for the derivative of  $\widetilde{\bf H}^n_{s, w} f 
(x)$, 
there appear  two terms, one which contains a factor $|s|$,  the other
a decreasing exponential in $n$. In order to suppress the 
effect of the factor $|s|$,  Dolgopyat   uses the family of  equivalent  norms
$$\|f\|_{1,t}:= \|f\|_{0} + \frac {1}{|t|} \|f\|_{1}
= \sup|f|+  \frac {1}{|t|} \sup|f'|,  t \not = 0,$$ which appear
in the statement of Theorem 2.
With this norm and  Lemma 2,  together with   (\ref{H1}),  we obtain 
the first (easy) result: 

\medskip
\pn
{\bf Lemma 3.} {\sl For any $t_{1}>0$,   for every compact subset  ${\cal L}$  
of $\Sigma_0 \times   W_0 $,
there is  $M_0 >0$  so that for  all $n\ge 1$,
all $(s,w)$ for which $(\Re s, \Re w) \in {\cal L}$  
and  $ |\Im s | \ge t_{1}$  we have 
$||\widetilde  {\bf H}^n_{s, w} ||_{1, \Im s} \le M_0$.}

\subsection {{\em UNI}  Condition  and  $L^2$-estimates.} 
Assuming {\em UNI, } Dolgopyat   first  proves that there is
$\gamma < 1$ so that
 $$  \int_{{\cal I}} |\widetilde {\bf H}_{s}^{n_{0}}[f](x)|^2 dx
 \le \gamma^{n_{0}} ||f||_{1, t} \, , 
$$
 for all large $t$ and $n_{0} = O(\log |t|)$.
 In this subsection, we extend this result to 
the bivariate operator ${\bf H}_{s, w}$,  when the number 
of branches of  $T$ is possibly infinite.

\smallskip 
\pn  Writing $s= \sigma +it$, $w = \nu +i \tau$,
$|\tilde {\bf H}_{s, w}^{n}[f](x)|^2$ can be expressed as: 
\begin{equation}\label{Snsum}
\frac{1}{\lambda(\sigma, \nu)^{2n}} 
\sum_{(h, k) \in 
{\cal H}^n\times \HH^n}  \exp [w c(h) + \bar w c(k)] \cdot  \exp [it \Psi_{h, k} 
(x)]  
\cdot  R_{h, k}(x)  \, , 
\end{equation}
with
\begin{equation} \label {Psihk}  \Psi_{h, k} (x) :=\log \frac {|h'(x)|} 
{|k'(x)|} \, , 
\end{equation}
\begin{equation} \label {Rhk}\qquad R_{h, k}(x)  =
 |h'(x)|^\sigma |k'(x)|^\sigma   \frac {1} {f^2_{\sigma,\nu}(x)} 
 (f\cdot f_{\sigma, \nu})\circ h(x)\cdot ( \bar f \cdot  f_{\sigma, \nu})\circ k(x) 
\, . 
\end{equation}
Using   $f = \Re f + i\Im f$, the term $R_{h, k}$ decomposes 
into four terms, each
of which has the form 
\begin{equation} 
\label {rhk}\qquad r_{h, k}(x)  = e^{i\omega}
 |h'(x)|^\sigma |k'(x)|^\sigma   \frac {1} {f^2_{\sigma,\nu}(x)} 
 (g \cdot  f_{\sigma, \nu})\circ h(x) ( \ell \cdot 
 f_{\sigma, \nu})\circ k(x) \, , 
\end{equation}
for  two real functions $ g, \ell  \in \{\Re f, \Im f \}$ and 
$\exp{i\omega} \in \{ \pm 1, \pm i \}$.

\medskip
\pn  The functions $\Psi_{h, k}$ play an important r\^ole here:  the sum 
(\ref{Snsum}) will be  split  into two parts,  according  to  their properties.  
The first sum will gather the pairs for which the 
derivative $ |\Psi_{{h, k}}'(x)|$ has a ``small'' lower bound, and 
condition {\em UNI}  will precisely
require that there are not ``too many'' such pairs $(h, k)$, providing 
a convenient  bound  for the corresponding integral $I_{n}^-$ (Lemma 4).
The second sum will gather the  other pairs $(h, k)$, for which the  derivative
$ |\Psi_{{h, k}}'(x)|$ has a ``large'' lower bound. In this 
case, the Van Der Corput Lemma  on oscillatory
integrals will be applicable (Lemma 5), giving
a bound for the corresponding integral $I_{n}^+$.   

\medskip
\pn Let us introduce some notations needed for
our formulation of {\em UNI.}  
For two inverse branches $h$ et $k$  of same depth, we introduce a ``distance:''
\begin{equation} \label {delta}
\Delta (h, k) =  \inf_{x \in {\cal I}} |\Psi'_{h, k} (x)| =  
\inf_{x \in {\cal I}}\left| \frac {h''}{h'}(x) - \frac  {k''}{k'}(x)\right| \, .
\end{equation} 
For  $h$  in ${\cal H}^{n}$, and $\eta > 0$,  we denote 
\begin {equation} \label {J}
J(h, \eta) := \bigcup_{ k\in\cal H^{n} , \Delta(h, k) \le  \eta}
k({\cal I}) \, .
\end{equation}
Property {\em UNI(a)}  
expresses that the   Lebesgue  measure  of $J(h, \delta)$
is $<\!\!< \delta$ when  $\delta$ is scaled similarly 
to the  maximal length of  fundamental intervals of depth $n$. 
For any $\hat \rho > \rho$, this length  
is $O(\hat \rho^n)$ (up to absolute constants) and 
plays the role of a reference scale.  
This is a reformulation
of the {\em UNI} condition implicit in Dolgopyat's Section 5 \cite{Do}, which
we (finally) state:

\medskip 
\pn {\bf {\em UNI} Condition.}  {\sl   A dynamical system of the good class, 
with contraction ratio $\rho$,  fulfills  the UNI Condition if each inverse
branch of $T$ extends to a ${\cal C}^3$ function and
\pn $(a)$ For any $a$   ($0<a  <  1$) we have 
$
|J(h, \rho ^{an})|  <\!\!<  \rho^{an}\, ,
\forall n \, , \forall h \in {\cal H}^n \, . $
\pn $(b)$ 
$
Q:= \sup \{ |\Psi''_{h, k}(x) | ; n \ge 1 \, , h,k \in \HH^n, x  \in \cal I\} < 
\infty 
$. 
}

\medskip
\pn {\bf Remarks.} { Note first  that {\em UNI} does not involve the 
cost: this
is because $c$  is constant on the monotonicity intervals of $T$.   

\pn For  dynamical systems 
with affine branches, all the  $ \Delta(h, k)$ are zero, and, for  any $\eta>0$ and any $h \in {\cal H}^\star$,  the interval $J(h, \eta)$  equals ${\cal I}$. Then  dynamical systems 
with affine branches cannot satisfy {\em UNI}. We will see in Proposition 1, Section 3.5, that  this  is the same  when the map $T$ is  conjugated with a piecewise affine map. 

\pn Condition $(b)$ follows from the existence of $\tilde Q<\infty$ so that
 \begin {equation} \label {UNIb}
|h'''(x)| \le \tilde Q |h'(x)| \, , \forall n\ge 1\, , \forall  h\in \HH^n \, .
\end{equation}
It suffices to check (\ref{UNIb}) for $n= 1$ (similarly as for the distortion
condition). (Note that this condition  is  always satisfied if there are finitely 
many $\cal C^3$
inverse branches).

 \bigskip
 \pn {\bf  Study of the $L^2$-norm: the close pairs.}

 \smallskip
 \pn 
 {\bf Lemma 4.  } {\sl  Recall $A_{\sigma, \nu}$
from Lemma 1. Suppose that Condition UNI(a) holds.  
For any compact subset ${\cal L}$ of $\Sigma_0 \times W_0$,   for all  
$(\sigma=\Re s, \nu=\Re w)  \in {\cal L}$, for all $n$, for all 
$a$, with $0<a < 1$, 
 the integral  $I_{n}^-$ of the  sum (\ref{Snsum}) restricted
to pairs $(h, k)\in {\cal H}^{n}\times{\cal H}^{n} $  for which 
$\Delta(h, k) \le \rho^{an}$  satisfies 
$$  
|I_{n}^-|  =  
|I_{n}^-(s, w, f, a)|<\!\!<  \left (\rho^{a/2} A_{\sigma, \nu} \right)^n   
||f||_{0}^2 \, .
$$}

\smallskip
\pn {\bf Proof.} Up to a 
positive constant that only depends on $(\sigma, \nu)$ (and is uniform on the
compact subset ${\cal L}$), $|I_{n}^-|$  is 
less than     
$$  
\frac{||f||_0^2 } {\lambda(\sigma, \nu)^{2n} } 
\sum_{
\ntop{ (h, k)\in {\cal H}^n \times {\cal H}^n   }
{ \Delta(h, k) \le \rho^{an}           }}
\exp [\nu  (c(h) +c(k))] \cdot 
\int_{{\cal I}} |h'(x)|^\sigma |k'(x)|^\sigma \, dx \, .
$$
First,  using  the bounded distortion property, 
for all pairs $(h, k) \in {\cal H}^\star\times {\cal H}^\star$, up to 
multiplicative  absolute constants,  one has
$$  
\int_{{\cal I}} |h'(x)|^\sigma |k'(x)|^\sigma 
dx  < \!\! < 
{ \left (\int_{{\cal I}} |h'(x)|^\sigma dx \right) \cdot 
\left (\int_{{\cal I}} |k'(x)|^\sigma dx \right)} \, .
$$
Then, as in the beginning of the proof of Lemma 1, with the same bounded 
distortion  property and  using that
$\mu_{\sigma, \nu}$ is an invariant  probability for the 
normalized operator,   the ratios (two by two) of the   
four   quantities $a(h)$,  $b(h)$,  $c(h)$,  $d(h)$, 
$$ 
a(h):= \frac {\exp[\nu c(h)]} {\lambda(\sigma, \nu)^n} \int_{{\cal I}}  
|h'(x)|^\sigma  \, dx; 
\quad b(h):=  \frac { \exp[\nu c(h)]} {\lambda(\sigma, \nu)^n}  \int_{{\cal I}} 
|h'(x)|^\sigma 
d\mu_{\sigma, \nu }(x)\, ;
$$
$$ 
c(h) := \mu_{\sigma, \nu}[h({\cal I})] \, ; 
\quad d(h):= 
\frac {\exp[\nu c(h)]} {\lambda(\sigma, \nu)^n} |h({\cal I})|^\sigma \, ;
$$
admit upper and lower bounds 
which do not depend on $h$ and are uniform  when $(\sigma, \nu)$ varies in a
compact set. 
Up to a multiplicative constant, it is then sufficient to study 
the sum
$$ 
\sum_{h \in {\cal H}^{n}} \mu_{\sigma, \nu}[h({\cal I})]  
\left( \sum_{ \ntop {k \in  {\cal H}^n}{ \Delta(h, k) \le \rho^{an}} }
\mu_{\sigma, \nu}[ k({\cal I})] \right ) = 
\sum_{h \in {\cal H}^{n}} \mu_{\sigma, \nu}[h({\cal I})] \ 
\mu_{\sigma, \nu}[  J (h, \rho^{an})] \, .
$$
Now,  the  first relation  of Lemma 1, 
$ \mu_{\sigma, \nu} [J] \le C_{\sigma, \nu}  A_{\sigma, \nu}^n  \    |J| 
^{1/2}$, 
which  holds for any  
subset $J$  that is a union of fundamental intervals of depth $n$,   is 
 applied to   $  J (h, \rho^{an})$.  {\em UNI (a)} 
provides an evaluation of its Lebesgue 
measure,   and, finally,  
$ |I_n^{-} |<\!\!<  \left (\rho^{a/2} A_{\sigma, \nu} \right)^n   ||f||_{0}^2$. 
\qed

\bigskip
\pn {\bf 
 Study of the $L^2$-norm: Application of  the Van der Corput 
Lemma.} Consider now the  integral  $I_{n}^+$ of the   sum   relative to 
pairs $(h, k)$ which were not treated by Lemma~4: 

  \medskip
 \pn 
 {\bf Lemma 5.  } {\sl  Suppose that Condition  {\em UNI}(b) holds.  Letting
$\lceil x\rceil$ denote the smallest integer $\ge x$, set
\begin{equation} \label {n0}
n_{0}=n_0(t)= \left  \lceil  \frac {1}{|\log  \rho|}  {\log |t|} \right \rceil  
\, .    
\end{equation}
Then,  for any compact  subset ${\cal L}$ of $\Sigma_0 \times W_0$,   for any
$(\sigma=\Re s, \nu=\Re w)  \in {\cal L}$, and $|t|=|\Im s| \ge 1/ \rho^2$, 
for any   $0<a<1/2$,   the integral  $I_{n_0}^{+}$ of the  sum (\ref{Snsum})
for $n=n_0$,
restricted to  $(h, k)\in {\cal H}^{n_0}\times {\cal H}^{n_0} $  with
$\Delta(h, k) \ge \rho^{a n_0}$,  satisfies 
$$  
|I_{n_{0}}^{+}| =   |I_{n_{0}}^{+}(s, w, f, a)|  <\!\!<  \rho^{(1-2a)n_{0}} 
||f||_{1, t}^2 \, .
$$}

\medskip
\pn 
{\bf Proof.} We  start with a general integer $n$ and bound  
\begin{equation} \label {In+}
 |I_n^+|  \le  \frac{1} 
{\lambda(\sigma, \nu)^{2n} } \sum_{  
\ntop {(h, k)\in {\cal H}^n\times {\cal H}^n} { \Delta(h, k) \ge \rho^{an}}}
\exp [\nu (c(h) +c(k))] \cdot  \left |\hat I(h, k) \right | \, , 
\end{equation}
where the integral 
$ 
\hat I(h, k):=  \int_{{\cal I}} \exp [ it \Psi_{h, k} (x)] \, R_{h, k}(x) dx 
$   
involves  
$ \Psi_{h, k}$,  $R_{h, k}$  defined in (\ref {Psihk}), (\ref{Rhk}), and 
decomposes 
into four integrals of the form    
$$   
I(h, k):=  \int_{{\cal I}} \exp [ it \Psi_{h, k} (x)] \, r_{h, k}(x) dx \, , 
$$
with 
$r_{h, k}$  defined in  (\ref{rhk}). We shall
apply the following lemma to each oscillatory integral $I(h, k)$:

\smallskip
\pn {\bf Van der Corput Lemma} (See e.g. \cite {St}).  {\sl For each interval 
${\cal I}$
and every $Q > 0$, there is $C(Q)$, so that for all 
$t \in \real$, $\Psi \in \cal C^2 ({\cal I})$ with
$
|\Psi''(x)| \le Q$  ,   
$|\Psi'(x)|\ge \Delta$ with $ |t|^{-1} \le \Delta\le 1 $,
and  $r \in \cal  C^1 ({\cal I})$    with
$ ||r||_{0}\le R \, , \quad ||r||_{1, 1}\le R  D $,  
the integral
$
I (t)= \int_{{\cal I}} \exp [ it \Psi (x)] \, r(x) \,  dx 
$ 
satisfies
$$ 
| I (t)|\le R \, C (Q) \, \left [ \frac { D +1}{ |t|  \Delta} + 
  \frac {1}{ |t| \Delta ^2}\right ] \,  .
$$}

\smallskip
\pn   Consider  $(t, n)$ with $  {1}/ { |t|} \le 
\rho^{an}$.   Setting 
 \begin {equation} \label {Mhk}
M(h,   k):= \sup_{x \in {\cal I}}  |h'(x)|^\sigma |k'(x)|^\sigma   
  \frac {1} {f^2_{\sigma, \nu}(x)} 
  f_{\sigma, \nu}\circ h(x)  f_{\sigma, \nu}\circ k(x) \, ,
\end{equation}  the norm 
$||r_{h, k}||_{0}$ satisfies 
$$ 
||r_{h, k}||_{0} \le M(h, k) || g  ||_{0}\  ||\ell  ||_{0} \le  M(h, k) || 
g ||_{1, t} \ || \ell ||_{1, t} \, .
$$   
The arguments used in  the proof   of Lemma 2  for the function $r_{h}$ apply  
to  
the  function  $r_{h, k}$, and 
\begin{eqnarray}
\nonumber  ||r_{h, k}||_{1, 1} &< \!\!<&  M(h, k) \left 
[  ||g ||_0 \left (||\ell ||_0  
+  \rho^n \,  ||\ell ||_{1} \right) 
+ ||\ell  ||_0 \left  (||g ||_0  +  \rho^n 
\,  ||g ||_{1} \right)  \right] 
\\
\nonumber
  &<\!\!<&   M(h, k) || g  ||_{1, t}  || \ell  ||_{1, t} \left [ 1+ \rho^n  |t| 
\right ] \,  .
\end{eqnarray}
Then, by Property {\em UNI(b),} the Van Der Corput Lemma  can be  applied  to 
each integral 
$I(h, k)$, which  thus  satisfies
$$
|I(h, k)|  <\!\!<   M(h, k) \   ||g ||_{1, t} ||\ell ||_{1, t}\  \left   [ 
\frac {2+ |t |\rho^n}{|t |\rho^{an}} +  \frac {1}{|t| \rho^{2an}}\right ]\,  .
$$ 
Now, we choose  $n= n_{0}$  as in (\ref{n0}). 
Since $a <1/2$,  and $ \sqrt {|t|} \ge 1/\rho$ we have $n_0\ge 2$ and 
$\rho^{-an_0} \le \rho ^{-(n_0-1)} \le |t|$,  so that the Van der Corput  Lemma may be applied.  The previous inequality 
becomes
\begin{equation} \label 
{Ihk}
|  I(h, k)  |<\!\!<  M(h, k) \    ||g ||_{1, t} ||\ell ||_{1, t} \ 
   \rho^{(1-2a) n_{0}} \, . 
\end{equation}
Returning to the integral  $\hat I(h, k)$,   
$$ | \hat I(h, k)|  <\!\!<  M(h, k) \  \rho^{(1-2a) n_{0}} \  ||f ||_{1, t}^2 \,  .
$$
Now, take  $x_0$ in ${\cal I}$. Then, from the bounded distortion 
property (\ref{L}), and the definition of $M(h, k)$ in (\ref{Mhk}),    we get
  $$
M(h, k) <\!\!< |h'(x_{0})|^\sigma |k'(x_{0})|^\sigma   
  \frac {1} {f^2_{\sigma, \nu}(x_{0})} 
  f_{\sigma, \nu}\circ h(x_{0})  f_{\sigma, \nu}\circ k(x_{0})   \, ,
$$ 
and   therefore
  \begin{equation} \label {Mhk1}
  \frac{1} {\lambda(\sigma, \nu)^{2n} }
\sum_{(h, k)\in {\cal  H}^{n}\times {\cal  H}^{n} } 
  \exp [\nu (c(h)+c(k))] \, M(h, k) 
  <\!\!<
  \left (\widetilde {\mathbf  H}_{\sigma, \nu}^n [1](x_{0})\right)^2 = 1 \, .
\end{equation} 
  From (\ref{Ihk}, \ref{Mhk1}, \ref{In+}), we finally obtain 
  $|I_{n_0}^+| <\!\!<     \rho ^{ (1-2a)n_{0}}\,  ||f||_{1, t}^2$.
  \qed

\bigskip
\pn {\bf  Study of the $L^2$-norm: the final result.}
 Consider the integer $n_{0}$  from (\ref {n0}) of  Lemma 5 
(for $|t|\ge 1/\rho^2$) and   some $a$  with $ (2/5) < a < (1/2)$. 
Then, since $a/2 > (1-2a) >0$,   there exists a   (real) neighborhood 
of $(\sigma, \nu) = (1, 0)$ on which
\begin {equation} \label {vois1}  
A_{\sigma, \nu} \cdot  \rho^{a/2} \le \rho^{1-2a} \qquad \hbox {for any \ \ } (\sigma, \nu) \in { W \times \Sigma}.
\end{equation}
 Then,  from   Lemmata~4 and~5, 
\begin{equation} \label {Dolgo}
   \int_{{\cal I}} |\tilde {\bf H}_{s, w}^{n_{0}}[f](x)|^2 dx <\!\!< 
    \rho^{ (1-2a) n_{0}}  ||f||_{1, t}^2 \, .
\end{equation}
   
 \subsection {End of proof of Theorem 2.}    We operate now 
the transfers between various norms.

  \pn{\bf  From the $L^{2}$-norm to the  sup-norm. } 
Since  the  normalized density transformer $\tilde {\bf H}_{1}$  is 
 quasi-compact with respect to  the $(1, 1)$-norm,  and fixes the  constant
 function $1$, it  satisfies
\begin{equation} \label {QC} 
  ||\tilde {\bf H}_{1}^{k} [|g|^2] ||_{0}  =   
\left (\int_{{\cal I}}  |g|^{2}(x) \,  dx  \right) + 
 O( r^{k}_1) ||g^2||_{1, 1} \, , 
\end{equation}
where $r_1$ is the  subdominant  spectral radius of ${\bf H}_{1}$.

\smallskip
\pn Consider an iterate   $\tilde {\bf H}_{s, w}^{n}$ with $n\ge n_0$. Then 
$$  ||\tilde {\bf H}_{s, w}^{n}[f]||_0^2 <\!\!< 
   ||\tilde {\bf H}_{\sigma, \nu}^{n-n_{0}} [g] ||_{0}^2 \qquad 
   \hbox {with} \qquad g= | \tilde {\bf H}_{s, w}^{n_{0}}[f]| \, .
$$
Now,   using (\ref{2Le2}) from Lemma 1  and  (\ref{QC}) with $k:= n-n_{0}$, 
together with the bound  (\ref {Dolgo})  for the $L^2$-norm  and   finally   
Lemma 2 
to  evaluate $||g^2||_{1, 1}$, one obtains 
$$||\tilde {\bf H}_{s, w}^{n}[f]||_{0}^2  \le A_{\sigma, w}^{2(n-n_{0})} 
  \left [ \rho^ {(1-2a)n_{0}} + r_1^{n-n_{0}}\,  |t| \right ] 
  ||f||_{1,t}^2 \, . 
$$
We  now choose $n= n_{1}$ as a function of $t$
 so that  the two terms  $\rho^ {(1-2a)n_{0}}$ 
and $r^{n-n_{0}}_1 |t| $ are  almost equal (with $n_{0}(t)$ defined in (\ref{n0})): 
\begin{equation} \label {n1}
  n_{1} = (1 +  \eta) n_{0} \qquad \hbox {with} \qquad
  \eta := 2(1-a) \frac { \log \rho}{\log r_1} >0 \, .
\end{equation}
 Choose now $d$ such that  $\ 0< \eta (5a - 2) < d < 1-2 a< 1/5$ (which is possible if $a$ is
of the form $a= 2/5 + \epsilon$, with  a small $\epsilon >0$).
 Recalling  (\ref{vois1}) where a first neighborhood was defined, 
and considering   a  (real) neighborhood $\Sigma \times {W} $ of $(1, 0)$   for 
which
\begin {equation} \label {Voisi} 
  \sup \left [\lambda (\sigma, \nu)^{1+ \eta} , 
A_{\sigma, \nu}^\eta\right]  
<  \rho ^{-  \eta (5a/2 -1) } < \rho^{- d/2}  \,  ,
\end{equation}
we finally obtain, for  $n_1(t)$ and $\eta$ defined in (\ref{n1})  
\begin{equation} \label {N0} 
  ||\tilde {\bf H}_{s, w}^{n_{1}} [f]||_{0} <\!\!< \rho^{n_{1} b}\,  
||f||_{1,t}, 
   \qquad \hbox {with} \qquad  b:= \frac {1-2a- d} {1 + \eta} \, .
\end{equation}

\bigskip
\pn {\bf From   the sup-norm  to the $||.||_{1, t}$-norm.}
 Applying   Lemma 2 twice
and  using  (\ref{N0}) yields  the 
inequality
\begin{eqnarray}
\nonumber
||\widetilde  {\bf H}^{2n_{1}}_{s, w} [f] ||_{1}
& <\!\! < &{ |s|} \,  || \widetilde  {\bf H}^{n_{1}}_{s, w}[f]||_{0} + 
\rho^{n_{1}}  \,  ||\widetilde  {\bf H}^{n_{1}}_{s, w}[f]||_{1} \\
\nonumber &  <\!\!<&
   { |s|}  \,   \rho^{ n_{1} b} ||f||_{1,t}
+ \rho^{n_{1}}  |t|  \left (   \frac { |s|} { |t|}   ||f||_{0} +
 \rho^{n_1}   \frac{||f||_{1}}{ |t|}  \right)\\
& < \!\!<&  |t | \rho^{ n_{1}b} 
 ||f||_{1, t} \, ,
\end{eqnarray}
which finally  entails  for $n_2=2n_1$ (and $n_1(t)$ as above)
\begin{equation} \label {N1t}
||\widetilde  {\bf H}^{n_{2}}_{s, w}||_{1, t}  < \!\!<  \rho^{ 
n_{2} b/2} \, .
\end {equation}

\bigskip
\pn {\bf The last step in Theorem 2.} For fixed $t$ with
$|t| > 1/\rho^2$,  any integer $n$ can be written $n = k n_{2}+ \ell$ with 
$\ell < n_{2}(t)$. Then
(\ref {N1t}) and   Lemma 3 entail
$$ ||\widetilde {\bf H}_{s,w}^{n} ||_{1, t } 
 \le M_0\,   ||\widetilde {\bf H}_{s,w}^{n_{2}} ||^k_{1, t }
 \le M_{0} \, 
\rho^{b k n_{2}/2}  \le M_0 \, \rho ^{bn/2}\,  \rho^{-b n_{2}/2} \, . 
$$ 
Since $bn_2/2 = bn_1 = (1-2a- d)n_0$, with $n_0$ defined in (\ref{n0}), 
we finally obtain 
$$ ||\widetilde {\bf H}_{s,w}^{n} ||_{1, t }\le   M \,  |t|^\xi \,   
\gamma^n \, , 
$$
$$ \hbox {with\ \ }
 \xi := 1-2a - d,  \qquad  b  := \frac {\xi} {1 + \eta}, \qquad  \gamma := \rho ^{b/2} <1,\qquad 
   M_1= \frac {M_0} {\rho^\xi} .$$
 Then $\xi$ is any value between $0$ and $1/5$.  
Therefore,  returning to the operator ${\bf H}_{s,w}$,  we  have 
shown
\begin{equation}\label {Dolgo1}
\| {\bf H}_{s, w}^{ n} \|_{1,t}\le  M_1\cdot  \gamma^n \cdot |t|^{\xi}  \cdot  
\lambda( \sigma, \nu)^n  \, , \quad  \forall n\, ,
\forall |t|\ge 1/\rho^2  \, .
\end {equation}
\smallskip
\pn   Finally,    
for any $(\sigma, \nu) \in   \Sigma \times W$ 
as  in (\ref {Voisi}), one has
$$ \gamma \lambda(\sigma, \nu) \le   \rho ^{ \frac {\xi }{2(1+ \eta)}}   
\cdot \rho ^{- \frac {\xi }{4(1+ \eta)}} =   \rho ^{ \frac {\xi }{4(1+ \eta)}} 
=  \hat \gamma <1 \, .
$$
This proves  Theorem 2 with $M := M_1/(1- \hat \gamma)$. \qed

\subsection { {\em UNI} Condition and  Euclidean dynamical systems.} 
To  apply   Theorem 2   to our three algorithms, we  prove that they 
satisfy  {\em UNI.} 
  
\smallskip
\pn   For two LFT's $h_{1}$
 and $h_{2}$, with
 $h_{i}(x)=  ({a_{i} x+ b_{i}}) /({c_{i} x+ d_{i}})$, we have  
$$ \Psi'_{h_1, h_2}(x)= \left| \frac {h''_{1}}{h'_{1}}(x) - \frac 
{h''_{2}}{h'_{2}}(x)\right|  =   
 \frac {|c_{1} d_{2} - c_{2} d_{1}|} {\left | ( c_{1} x+ d_{1}) 
 ( c_{2} x+ d_{2}) \right |} \, ,
$$
so that the distortion property gives
\begin{equation} \label{h1h2}
   \Delta(h_{1}, 
h_{2}):=  \left | \frac {c_1}{d_1}-  \frac {c_2}{d_2}\right| \cdot 
 \inf_{x \in 
{\cal I}} \left |  \frac {h'_1(x) h'_2(x)} {h'_1(0) h'_2(0)}\right |^{1/2} \ge 
\frac {1}{L}  \left | \frac {c_1}{d_1}-  \frac {c_2}{d_2}\right| 
\, .
\end{equation}
Hence $\Delta$ only depends on the difference of the quotients $c_i/d_i$ of the 
denominators of
the LFT's. We shall next show that this difference is 
the ``honest'' (ordinary) distance between the rationals 
$ h^*_1(0)$ and $ h^*_2(0)$, where $h^*$ is the {\it mirror LFT} of $h$, 
 defined by:
$$  h^*(x) =  \frac {a x+ c} {b x+ d} 
\qquad \hbox {if}\qquad  h(x) =  \frac {a x+ b} {c x+ d} \, .  
$$
This mirror operation appears in \cite {Sch}  where Schweiger  
relates it to the natural extension, and in \cite {BDV}, where the authors use 
the  
geometric notion of  ``folded'' and ``unfolded.''

\pn Clearly, the mirror map is  an involution  satisfying the
morphism  property $( {h \circ k} )^*= k^* \circ  h^*$. 
It is not difficult to see that if $h \in \HH^p$ is a LFT from a Euclidean 
dynamical system, associated
to  the sequence $\epsilon_0,(m_1, \epsilon_1), (m_2, \epsilon_2)\ldots (m_p, 
\epsilon_p)$,
the mirror LFT $ h^*$ corresponds to  
$ \epsilon_p, (m_p, \epsilon_{p-1})\ldots (m_{p-1}, 
\epsilon_{p-2}) \ldots (m_1,  \epsilon_0) $, i.e., 
the decomposition  involves the 
same ``digits'' as $h$, but in the inverse  order. 
   
\pn By (\ref{h1h2}), the distance $ \Delta(h_{1},  h_{2})$ between $h_{1}$ and $ 
h_{2}$ 
indeed  satisfies 
\begin{equation} \label {Deltal}
\Delta(h_{1}, h_{2}) \ge \frac 1 { L }\, |  h^*_{1}(0)-  h^*_{2}(0)| \, .
\end{equation}

\smallskip
\pn 
It is not difficult to check for each of our three Euclidean algorithms   that  
the set $\{  h^* , h \in \HH\}$  is the set of inverse branches   of  a
dynamical system $ ( {\cal I}^*,  T^*)$,
the {\it dual dynamical system,} which belongs to the good class, with the
same contraction ratio $\rho^*=\rho$ and a distortion constant $ L^*$, as we now explain: 
For the
Classical Euclidean algorithm ${\cal G}$,  since all the $\epsilon$ are equal to 
$1$ note that $( {\cal I}^*,  T^*)=({\cal I}, T)$. 
For the two others,  the equality $\rho=\rho^*$ follows from 
the  definition (\ref{rho1}) of  $\rho$, the 
distortion property 
for the dual system, and the fact that, in  both
systems the worse  branch,  for which  the contraction ratio 
is  attained on a fixed point, is a LFT $h$ with $h =  h^*$.  
The  three dual dynamical systems are described  in Figure 2.      

\begin{figure}
\begin{small}
\hspace*{-1truecm}
\renewcommand{\arraystretch}{2.5} 
\renewcommand{\tabcolsep}{4pt}
\begin{tabular}{|l|l|l|l|}
\hline
\sl Algorithm &  $\cal  G^*$ & $\cal K^*$ & $ \cal O^*$
\\ \hline
 \sl  Interval     
&  $[0, 1]$&    $ [\phi-2, \phi-1] $ &      $ [\phi-2, \phi]$
\\ \hline
\sl Generic  &  $m\ge 1\, ,  \epsilon=+1 $
	&    $m\ge 2 \, ,    \epsilon= \pm 1$
& $m\ge 1$ odd, $\epsilon = \pm 1$
	\\
\sl  conditions
& & if 	$m=2$  then $ \epsilon= +1$
&     if  $m=1$ then $ \epsilon=+1$
\\ \hline
         
	  \sl Graphs & \includegraphics[width=4cm]{AlgorithmeG.eps}&
    \includegraphics[width=4cm]{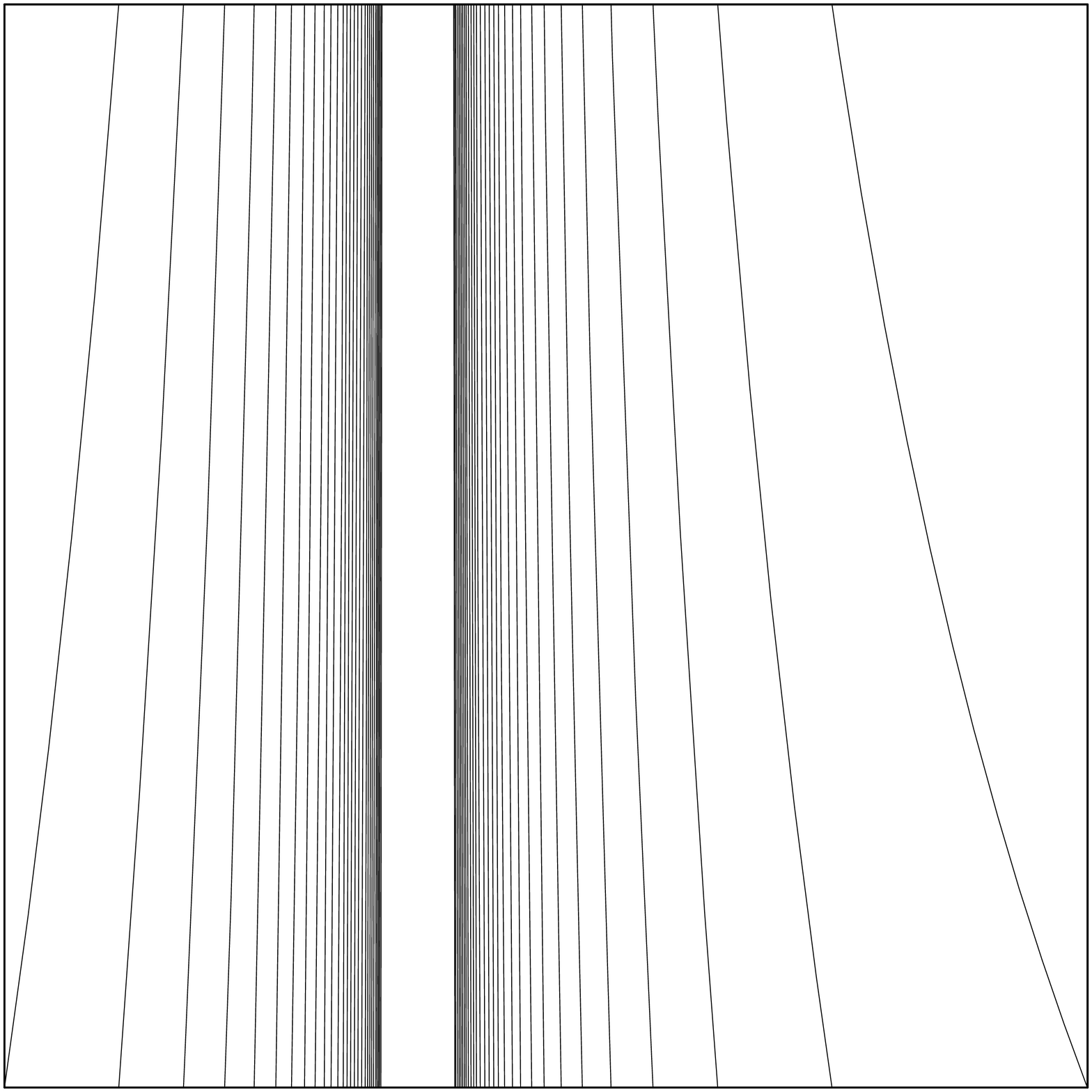}&
    \includegraphics[width=4cm]{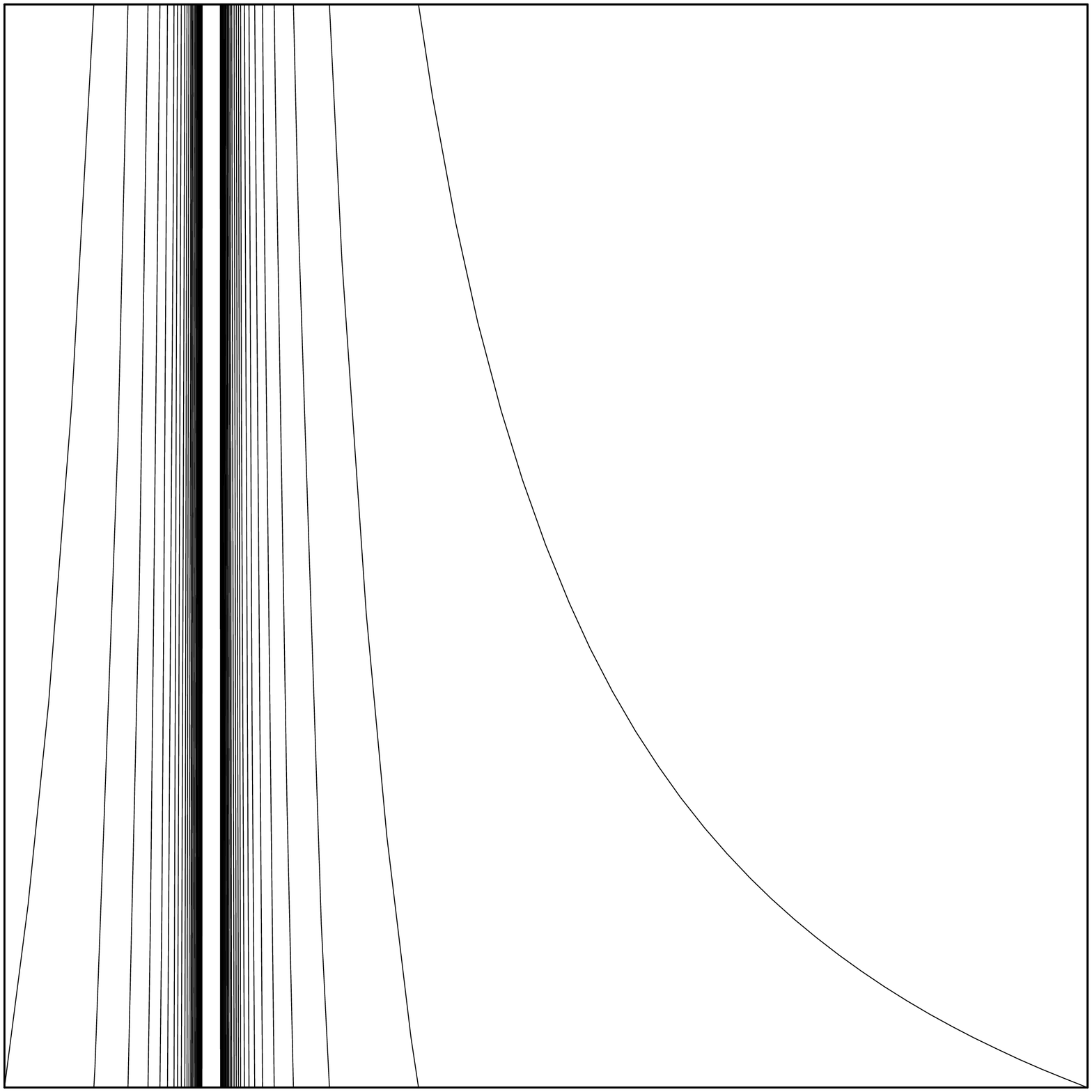}
    \\ \hline
        
\end{tabular}
\end{small}

\bigskip 
 \begin{center}

    \caption{ The three  dual Euclidean dynamical systems:  
Standard, Centered,  Odd. ($\phi={\frac {1+\sqrt 5} {2}}$)  }
\label {les3d}
\end{center}
\end{figure}

\smallskip
\pn 
Our next  goal is to check {\em UNI} for our three algorithms. 
Note  first that checking  {\em UNI (b)} amounts to
verifying (\ref{UNIb}), which is easy for the three algorithms.  For {\em UNI(a)}: 
\smallskip
\pn {\bf Lemma 6.} { \sl A Euclidean dynamical 
system of the good class  which admits a 
dual  system that belongs to the good class  satisfies  {\em UNI (a).}
In particular, the dynamical systems associated to the  Euclidean 
algorithms ${\cal G}, {\cal K}$, and ${\cal O}$ satisfy  {\em UNI.}} 
 
\smallskip
\pn{\bf Proof.}   Fix $0 < a < 1$ and $h \in \cal H^n$.
Denote by   $ J^* (h, \eta)$  the union of  the
intervals  $ k^*(  {\cal I}^*)$  for LFT's $k\in \cal H ^n$   
satisfying   $\Delta(h, k) \le \eta$. 
 First, we estimate $ |J^* (h, \eta)|$, second, we relate
$ | J ^*(h, \eta)|$ and $| J (h, \eta)|$. 

\pn By (\ref{Deltal}) if
$\Delta(h, k) \le \eta$ then  $ |h^*(0)- k^*(0)|\le L \eta$.   
Since  the 
length of a fundamental interval of depth $n$ in the dual system is
at most $\rho^{an}$ (up to  an absolute constant) we get 
\begin{equation} \label {UNI1}
   | J^* (h, \eta) | <\!\!< 2 L \eta +  2 \rho^{an} \,  .
\end{equation}
The fundamental   intervals $k({\cal I})$ and $ k^*( {\cal I}^*)$
(of depth $n$)
  have almost the same length:  Indeed, by the bounded distortion property, 
and since $( k^*)'(0) = k'(0)$ for all  $k \in \HH^{\star}$,
\begin{equation} \label {hh}
    \frac  1  {( {L  L^*})^{1/2}}\le \frac {|k({\cal I})|} {| k^*( {\cal 
I}^*) |}  =   \frac 
{|k({\cal I})|}{|k'(0)| }
\frac{|( k^*)'(0)| } {| k^*(  {\cal I}^*)|} \le ( {L  L^*})^{1/2} \, .
\end{equation}
Thus, since the intervals $k^*(\cal I^*)$ in $J^*(h, \eta)$
are disjoint,
 if $\eta \le   \rho^{an}$,  using (\ref {hh}, \ref{UNI1})
 \begin{equation} 
|J(h,\eta)| \le  ( {L  L^*})^{1/2} | J^*(h, \eta)| 
 <\!\!<   ( {L  L^*})^{1/2}( 2L\eta +  \rho^{an})  <\!\!<  \rho^{an} 
 \, . \hbox{\qed}
\end{equation}

\subsection { Condition {\em UNI}, aperiodicity and  absolute convexity.} 
Condition {\em UNI}  provides estimates for the $(1, t)$-norm of 
$(I- {\bf H}_{s, w})^{-1}$  when $|t|=| \Im  s|$ 
is  sufficiently large.  We   have  now to  consider  the case when $|t|$ 
is not large, and we explain how the {\em UNI} Condition intervenes in this context, via aperiodicity 
results. We must  also check that   variance constants $\delta(c)$ which will appear  in Theorems 3 and 4  are not zero, and
the {\em UNI} Condition intervenes in this context too, via 
 absolute  convexity results.

\smallskip 
\pn We first recall a classical   result.

\smallskip
\pn 
{\bf Lemma 7. } [Aperiodicity and absolute convexity.]  
{\sl  Let  $\mathbf H_{s,w}$ be the transfer operator associated to
$(\cal I, T, c)$ of $\cal{GMG}$ type.

\pn $(i)$ {\rm  [Aperiodicity.] } Denote 
by $R(s,w)$ its spectral radius, and consider a point   $(t_{0} ,\tau_0)\ne( 0,0)$.
The following are equivalent:

$(a)$  There exists 
$(\sigma, \nu) \in \Sigma_0 \times W_0$
for which $$ R(\sigma,\nu) \in {\rm Sp} \,  {\mathbf H}_{\sigma+it_0,\nu+i\tau_0} \, . $$

$(b)$  There  exists  $f \in \cal C^1(\cal I)$, $ |f| = 1$,
 so that,  for all $n$,  for all $h \in \HH^n$
\begin{equation}
|h'(x) |^{it_0} \cdot  \exp[i\tau_0 c(h)]  \cdot  f\circ h(x)=  f(x) \, ,
 \qquad  \forall x\in \cal I \, .
\end{equation}
 
\smallskip 
\pn  $(ii)$ {\rm  [Absolute convexity at $(1, 0)$]. } Consider  a point $(q, r) \not = (0, 0)$, the operator $\mathbf H_{1 +qw, rw}$ and its  pressure 
function $w \mapsto \Lambda(1+qw, rw)$. The following are equivalent 

$(a)$  The second derivative of  $w \mapsto \Lambda(1+qw, rw)$ is zero at $w = 0$

$(b)$ There  exists a function  $f \in \cal C^1(\cal I)$ strictly positive and a constant $\alpha>0$ , 
 so that,  for all $n$,  for all $h \in \HH^n$
\begin{equation}
 |h'(x) |^q \cdot \exp [ r c(h)] \cdot  f\circ h (x) = \alpha^n \cdot   f(x) \, ,
 \qquad  \forall x\in \cal I \, .
\end{equation}

}

\smallskip
\pn 
{\bf  Proof.}  $(i)$ See   for instance Prop. 6.2 
in \cite{PP}  or Prop.~9 in \cite{Va1}.
Since ${\rm Sp}\, {\bf H}_{\sigma+it_{0}, \nu+i\tau}=
\lambda(\sigma,\nu)\, {\rm Sp}\, \tilde{ \bf H}_{\sigma+ it,\nu}$
we may replace $R(\sigma+it,\nu+i\tau)$ and $R(\sigma,\nu)$
by the spectral radii of the corresponding normalized operators.
By Proposition~0, the spectral radius of $\tilde {\bf H}_{\sigma+ it, \nu+i\tau}$ 
is at most
 $1=R(\tilde {\bf H}_{\sigma,\nu})$,
while its essential spectral radius is at most  $\hat \rho < 1$.
 If $(a)$  holds, then  
$\tilde{ \bf H}_{\sigma+ it_0, \nu+i\tau_0}$ has an eigenvalue 
$\lambda=1$  with an eigenfunction
$f \in {\cal C}^1(I)$ with  $\max_I |f| = 1$. Suppose that this maximum is 
attained at $x_{0} \in \cal  I$. 
Then  the equality $\tilde{ \bf H}^n_{\sigma+ it_0, \nu+i\tau_0}[f](x_0) = f(x_0) $  can be written 
as $  \sum_{h\in \HH^n}  a_h b_h  = 1$ 
with 
$$  a_h := 
\frac {1} {f_{\sigma, \nu}(x_0)}  \exp[\nu c(h)]\cdot 
f_{\sigma, \nu}\circ h(x_0) \cdot |h'(x_0)|^{\sigma} 
$$
$$b_h:=   \frac  1 {f(x_0)} f\circ h(x_0)  \exp[i\tau_0 c(h) ] |h'(x_0)|^{it_0} $$
Normalization implies  that   $\sum_{h\in \HH^n}  a_h = 1$ 
while   each  factor $ b_h$
has modulus at most $1$.  Therefore,  each $b_h$ equals 1, and, for every $n\ge 1$, 
$$
f\circ h(x_0)= f(x_0) \exp[-i\tau_0 c(h) ] |h'(x_0)|^{-it_0}\, , \, \forall h\in \cal H^n 
\, .
$$
In particular, by density, $|f|$ is
the constant function $1$ and the above remarks also hold  for any $x 
\in {\cal I}$. 
Then, for all $x\in I$, all $n \ge 1$,   and all $h \in {\cal H}$, 
\begin{equation}\label{iterated}
\frac  { f(x)}{ f\circ h(x) } =  \exp[i\tau_0 c(h) ] |h'(x)|^{it_0}\, .
\end{equation}
Following the arguments backwards, we get the other implication. 

\smallskip
\pn $(ii)$ We apply to the operator ${\bf H}_{1+qw, rw}$ the results of \cite {Bro}.  In Proposition 6.1, Broise states her results  in the context of functions with
 bounded variation. Due to our  strong Markov assumption, we may work in ${\cal C}^1({\cal I})$, and the cost  of interest is  $D:= -q \log |T'| + r c$. Since $f_{1}$ is a strictly positive
${\cal C}^1$ function and   ${\bf H}_{1}[D]$
 belongs to  ${\cal C}^1$,  we may transfer Broise's proof  to our ${\cal C}^1$ context
:  it   shows   that Condition $(ii)(a)$ is equivalent to the following: there exist   $u \in {\cal C}^1 (I)$
and a constant $K$ for which, for any $h \in {\cal H}$, one has $ q \log |h'| +r c(h) = u - u \circ h + K$. This last condition
is clearly equivalent to $(ii) (b)$ with $f:= \exp [u]$. Note  that 
 the fonction $u$ introduced by Broise which  involves the centered version $\bar D$ of $D$, i.e., $\bar D:= D - \mu(D)$  
$$ u := - \frac{1}{f_1} (I- {\bf H}_1)^{-1} \circ {\bf H}_1 [\bar D \cdot f_1] $$  actually belongs to 
${\cal C}^1 ({\cal I})$.

\qed 

\medskip
\pn   Conditions $(i)(b)$ and $(ii)(b)$ are of the same form and closely related to conjugaison with piecewise affine maps.  As  we next  see it, 
the  {\em UNI} condition entails that this conjugaison cannot occur.  Then, under the   {\em UNI} condition,   the variance constants will be always strictly positive. On the other hand,     the aperiodicity result, together with the {\em UNI} condition,  provides some 
useful  informations  about  the spectrum  of ${\bf H}_{s, w}$, notably for lattice costs. We recall the definition of 
 lattice costs: a cost $c$  is said to be lattice if it is not   zero, and  there exists $L >0$
for which $c/L$ is integer. The  largest  such $L$ is the span of the cost. 

\medskip
\pn   {\bf  Proposition 1.}  {\sl  Consider  a $\cal {GMG}$ system satisfying
{\em UNI(a)}. The following  holds:

\smallskip
\pn $(i)$ The mapping $T$ is not ${\cal C}^2$ conjugated with a piecewise affine map.

\smallskip
\pn  $(ii)$ The pressure function of the operator ${\bf H}_{1+qw, rw}$ is absolutely convex at $w = 0$ 
for all fixed  $r\in \real$ and $q \ne 0$,  
(i.e., its second derivative is strictly positive). In particular,  for $r = 0$, one has $\Lambda''(1) >0$.

\smallskip
\pn $(iii)$  For  any $t \ne 0$,  $1$ does not belong to ${\rm Sp} \, {\bf H}_{1+it, 0}$. Furthermore, if $c$ is 
lattice with span $L$, for all $t$, and all $\tau$ not multiple of $2 \pi /L$,  
$1$ does not 
belong to ${\rm Sp} \, {\bf H}_{1+it,  \tau}$. }
	
\smallskip
\pn {\bf Proof.} $(i)$.  Suppose that $T$ is  ${\cal C}^2$ conjugated with a piecewise affine map. Then, there exists $f>0$   in ${\cal C}^1({\cal I})$ such that,  for any $n$, 
for each $h\in {\cal H}^n$, there is a constant $d(h)$  for which  $|h' (x)| f \circ h(x)  = d(h) f (x)$ for any $x \in {\cal I}$. 
Then taking the logarithm,
differentiating and putting $ \hat f:= \log f\in {\cal C}^1({\cal I})$, we get
$$
\Psi'_{h, k}(x) =   [h'(x) \hat f ' 
 \circ h (x) - k'(x)\hat f' \circ k(x) ] \, ,
\forall n\, , \quad \forall h\, , k \in {\cal H}^n \,  . $$ 
 Then,  for $\rho < \hat \rho <1$,    the distance $\Delta(h, k)$ satisfies
 $ \Delta(h, k) < \!\!< \hat  \rho^ n$ for any  $h$, $k \in {\cal H}^n$,   
 which contradicts  {\em UNI(a).}  

\smallskip
\pn $(ii)$ and $(iii)$.  It is clear  that  Condition $(ii)(b)$ of Lemma 7  with $q \ne 0$ or  Condition $(i)(b)$ of Lemma 7 with $t_0\ne 0$ entail that $T$  is ${\cal C}^2$ conjugated with a piecewise affine map. Then Condition $(ii)(a)$ of Lemma 7 (with $q \ne 0$) or 
Condition $(i)(a)$ of Lemma 7 (with $t_0 \ne 0$) cannot hold for   a $\cal {GMG}$ system satisfying
{\em UNI(a)}.  

\smallskip
\pn Finally,  assume {\em UNI (a)} and Condition  $(i)(a)$  of  Lemma 7 with $t_0 = 0$. Then  relations  of Condition $(i)(b)$
of Lemma 7  taken at
the fixed points $x_h$ of $h$ imply that  $\exp [i \tau_0 c(h)] = 1$ for all $h \in {\cal H}$.
This is only  possible if $c$ is lattice  of span $L$ and  $\tau_0$ is a multiple  of $2 \pi /L$. \qed

\subsection{ Final  bounds for the  Dirichlet series $S(s, w)$. }  
 With  Relation (\ref{R1}) 
which relates the Dirichlet series $S(2s, w)$  to the 
quasi-inverse of the operator ${\bf  H}_{s, w}$, we obtain now  the 
expected properties for the Dirichlet series $S(s, w)$.  The first result is relative to the case 
when $w$ is near 0, and will be useful in Section 4, while the second result 
is relative to the case 
when $w$ belongs to a compact vertical segment, and will be useful in Section 5.

 \smallskip 
  \pn {\bf Lemma 8.} {\sl   Consider one  of the three algorithms ${\cal G}, {\cal K}, {\cal O}$, and a 
cost $c$ of moderate growth.  For any $\xi$ with $0< \xi< 1/5$,  there is $\alpha_0 >0$ and,   
for all  $\hat \alpha_0$, with $0<\hat \alpha_0 < \alpha_0$,   there  are  
a  (complex)  neighborhood  ${\cal W}'$ of $ 0$   and  a constant $M''$ such that
for any $w \in \cal W'$, the following holds:

\pn $(i)$ 
$\Re \sigma (w) >1-(\alpha_0-\hat \alpha_0)$.

\pn $(ii)$ The  (meromorphic)
functions $s \mapsto S(2s, w)$, $s \mapsto  \tilde S(2s, w)$
have  a single   pole at
$s=\sigma(w)$ in the strip $|\Re s-1|\le \alpha_0$,  and this
pole is simple.

\pn $(iii)$ 
$\max \biggl ( |S(2s, w) | ,
 |\tilde  S(2s, w) | \biggr) \le  {M''} \max (1, |t|^\xi)\, ,\forall s,  \,  \Re s = 1 \pm \alpha_0 $.}

   \smallskip 
 \pn  {\bf Proof.}   
Let $\cal W$ be a complex neighborhood of $0$ in which
$\sigma(w)$ from Proposition~0$(7)$ is well-defined.
 Each vertical line $\Re s = \sigma$  is split into  three:

\smallskip
 \pn {\sl   Near the real axis.}   
 For $(s,w)$ in a  (complex) neighborhood ${\cal A}$ of $(1, 0)$,
Proposition~0$(5)$ gives  a  decomposition  
 $  {\bf H}_{{s, w}} =  \lambda(s, w) \,  {\bf P}_{s,  w} +   {\bf N}_{s, w} $
where ${\bf P}_{s,w}$ is rank-one and the spectral
radius of $ {\bf N}_{s, w} $ is $\le \theta$. It is easy to see that
the  $(1, 1)$-norm of  $(I -{\bf N}_{s,  w})^{-1}$ is 
bounded by some $\hat M_{2}$ on $\cal A$.
Since $\sigma(0)=1$,  taking
a smaller  neighborhood  ${\cal W}_{2}\subset \cal W$ of $0$, there are
 $\alpha_{2}>0$ and $t_2>0$ so that  the following set is contained in
$\cal A$
$$ {\cal A}_{2}:= \{ (s, w);  w \in {\cal W}_{2} , 
|\Re s -  1|  \le \alpha_{2},  \, \, |\Im s| \le t_{2} \}  \, .
$$
For $(s, w)\in {\cal A}_{2}$, the quasi-inverse of ${\bf H}_{s,  w}$ 
satisfies 
\begin {equation} \label {qinv}
   (I -{\bf H}_{s,  w})^{-1} =   \frac {\lambda(s, w)} {1- \lambda(s, w)}\,  
{\bf P}_{s,   w} +   (I -{\bf N}_{s,  w})^{-1} \, .
\end{equation}
It has as only singularities in
$\cal A_2$ a  simple pole at each point $(s= \sigma (w), w)$,   
 with residue  the nonzero operator
  \begin {equation}\label {R}
 {\bf R}(w) :=  \frac {-1} {\lambda'_{s}(\sigma(w), w)} \,    
{\bf P}_{\sigma(w), w}\, .
\end{equation}
  Next, note that since $ \lambda'_s(1,0)\ne 0$    we have
$\ell:= \inf_{{\cal A}_2}\bigl | \frac {\lambda(s, w) -1} {s- \sigma(w)}\bigr |> 0$.
Fix $\alpha$ with $0 < \alpha \le \alpha_2$.
Up to taking a smaller $\cal W_2$, we have 
$|\Re \sigma(w)-1|\le \ell \alpha/2$ for
$ w \in {\cal W}_{2} $.
Thus for $ \Re s = 1 \pm \alpha$, $|t| \le t_{2}$,
and $w \in \cal W_2$,  the dominant  eigenvalue satisfies 
$$|\lambda(s, w)-1|  \ge   \ell 
|s - \sigma(w)| \ge\ell |\Re s - \Re \sigma(w)| \ge 
\frac{\ell \alpha }{2}\, ,
$$
and, by
(\ref{qinv})   we have 
 $ || (I- {\bf H}_{s, w})^{-1} ||_{1, 1} \le  { M_2}/{\alpha} 
$.
  
  \smallskip
  \pn  {\sl  Compact region.}    Suppose that 
$t \ne 0$.   We first prove that 
$d(1,  {\rm Sp}\, {\bf H}_{1+it,0}) >0$.   
Fixing $\hat \rho > \rho$, the spectrum  ${\rm Sp}\, {\bf H}_{1+it,0}$ 
decomposes into two parts   ${\cal S}_t^- = {\rm Sp}\, {\bf H}_{1+it,0} \cap \{ |\lambda] \le \hat \rho\}$
and  ${\cal S}_t^+ = {\rm Sp}\, {\bf H}_{1+it,0} \cap \{ |\lambda] >\hat  \rho\}$. 
 Proposition~0$(1)$ implies that ${\cal S}_t^+$ is a
finite set of eigenvalues of finite multiplicity  and Proposition 1 
proves that $1$ does not belong to ${\cal S}_t^+$. Then  $d(1, {\cal S}_t^+ ) >0$. On the other hand, 
$d(1, {\cal S}_t^- )   \ge  1- \rho$, and finally   $d(1,  {\rm Sp}\, {\bf H}_{1+it,0}) >0$.  Then, 
by perturbation theory of finite parts of the spectrum,  there exist $\alpha_3>0$, $\beta>0$
 and a complex neighborhood 
${\cal W}_3$ of $ 0$ such that 
 the distance between $1$ and 
 the spectrum of ${\bf H}_{s, w}$ is at least $\beta$  on   
the compact set
  $$ {\cal A}_3:= \{ (s, w) ;\   w\in {\cal   W}_{3},  
|\Re s -  1| \le   \alpha_{3},  
 \  t_{2} \le |t| \le t_{0} \} \, ,
$$
where  $t_{0}= 1/\rho^2$ from Theorem 2. 
Thus  $ (s,w) \mapsto  (I -{\bf H}_{s,  w})^{-1}$ is analytic on  the compact 
set
${\cal A}_3$ and  its $(1, 1)$-norm  is bounded 
  by some $M_{3}$ there. 

\smallskip
\pn {\sl Domain $|\Im s|\ge 1/\rho^2$.} 
Consider  $\Sigma_1\times W_1$ from Theorem 2, for our 
fixed $\xi > 0$.   There exist $\alpha_{1}\in ]0, \sigma(0)-\sigma_0]$  
and a complex neighborhood ${\cal W}_1$ of $w = 0$ such that  any 
pair in 
$\{ (s, w) ;\   w\in {\cal W}_{1},  |\Re s -  1| \le   \alpha_1\}$
has its real part $(\sigma, \nu)$ in $\Sigma_1\times W_1$. 
  
 \smallskip
 \pn   Choose  first $\alpha_{0}:= \min (\alpha_{2}, \alpha_{3}, \alpha_1 )$,  
next
$ {\cal W}_4$ so that $\Re \sigma (w)>1 - (\alpha_0-\hat \alpha_0)$ 
on $ {\cal W}_4$,
and finally ${\cal W}' :=\bigcap_{j=1}^4 {\cal W}_j$. 
Taking  $M':= \max (M_2/\alpha_0, M_{3}, M_1)$,  
with $M_1$ as in Theorem 2, we obtain 
the claim 
 for $S(s, w)$.   For $\tilde S(s, w) $, we apply
(\ref{R1tilde}) and   use 
that $\zeta(s)$ is bounded
on vertical strips  near $\Re s = 2$. 
 \qed

\bigskip 
  \pn {\bf  Lemma 9.} {\sl  Consider one  of the three algorithms ${\cal G}, {\cal K}, {\cal O}$, and a 
 lattice cost $c$ of moderate growth with span $L$. Consider some $\xi $ with $0 < \xi < 1/5$, and
$\upsilon   >0$.   
Then,  there are $\gamma_1 >0$,   and  a constant $Q_3$  such that,  for any $ \tau \in \mathbb R, \, |\tau|\ \in [\upsilon, \pi/L] $, 

\pn $(i)$ $ s \mapsto S(2s, i\tau)$ and
$s \mapsto \tilde  S(2s, i \tau )$  are
analytic in the strip $|\Re s -1| \le \gamma_1$.

\pn $(ii)$
$\max \biggl ( |S(2s, i \tau ) | ,
 |\tilde  S(2s, i \tau ) | \biggr) \le  {Q_3} \max (1, |t|^\xi)\, ,\forall s,  \Re s = 1 \pm \gamma_1$.}

\smallskip
\pn {\bf Proof.} Fix $0< \xi < 1/5$. By {\em UNI},
Theorem 2  gives $\alpha>0$ and $Q_1$ so that for
 $|\Im s |\ge 1/\rho^2$, $|\Re s -1| \le \alpha$, and arbitrary real $\tau$, 
 \begin{equation}
\label{Dolgo3}
||(I-{\bf H}_{s, i\tau})^{-1}||_{1, t} \le Q_1\cdot |\Im s|^\xi \, ,
 \end {equation}

 \pn Suppose  that  $|t|\le 1/\rho^2$ and 
$|\tau|\in [\upsilon, \pi/L]$.  We first prove, as in Lemma 8,  that 
$d(1,  {\rm Sp}\, {\bf H}_{1+it, i \tau }) >0$.  
 Proposition~0$(1)$ implies that  the spectrum  ${\rm Sp}\, {\bf H}_{1+it,i\tau}$ 
decomposes into two parts   ${\cal S}_{t,\tau}^- = {\rm Sp}\, {\bf H}_{1+it,i \tau } \cap \{ |\lambda] \le \hat \rho\}$
and  ${\cal S}_{t,\tau}^+ = {\rm Sp}\, {\bf H}_{1+it,i \tau } \cap \{ |\lambda] > \hat \rho\}$,
where
${\cal S}_{t,\tau}^+$ is a finite
set of  eigenvalues of finite multiplicity. Proposition 1 
gives that $1$ does not belong to ${\cal S}_{t,\tau}^+$. Then  $d(1, {\cal S}_{t,\tau}^+ ) >0$. On the other hand, 
$d(1, {\cal S}_{t,\tau}^- )   \ge  1- \rho$, and finally   $d(1,  {\rm Sp}\, {\bf H}_{1+it,i \tau }) >0$. Then,  
by perturbation theory of finite parts of the spectrum, 
 there  are  $0<\gamma_1\le \alpha$ and  $\beta>0$ such that 
 the distance between $1$ and 
 the spectrum of ${\bf H}_{ \sigma  +it , i \tau }$ is at least $\beta$  on   
the compact set
  $ |\sigma -1|\ \le \gamma_1, \,   |t| \le 1/ \rho^2, \,   |\tau|\in [\upsilon, \pi/L]$. Then $s \mapsto 
(I- {\bf H}_{s, i \tau})^{-1} $ is analytic there, 
and there is $Q_2$   so that
\begin{equation}
\label{cpct}
||(I-{\bf H}_{1\pm \gamma_1+it, i\tau})^{-1}||_{1, 1}\le Q_2 \, ,
\forall |\tau|\in [\upsilon, \pi/L]\, , \, \forall  |t|\le 1/\rho^2\, . \hbox{\qed}
 \end {equation}

  \section {Limit Gaussian distribution for  costs of moderate growth.}
\pn 
In this section we  prove our Central Limit Theorem,
Theorem 3.  
We  first   explain our use of Perron's formula (\S  4.1).  
We  next introduce  in \S4.2 a smoothed  model endowed with 
probability   $\bar \Pr_N$.  For this smoothed model, Lemma 10  allows us
to deduce  from bounds  on $\bar {\Psi}_w$ the bounds on $\bar{\Phi}_w$
which entail quasi-power estimates.  Theorem 0
gives asymptotic normality for the smoothed model,
 with asymptotic estimates for its expectation and  variance  
(\S 4.3, in particular Lemma 12).
A  comparison   
 of the uniform and smoothed distributions of (Lemma 14 in \S 4.4) 
finally yields  Theorem 3.

\subsection {Using  Perron's formula.}   
Choose  $0<\xi <1/5$.  
Lemma 8 specifies $\alpha_0\in ]0, 1/2]$ and ${\cal W}'$. 
We  fix   $w \in {\cal W}'$, 
the Dirichlet series $S(s, w)$ being viewed as  functions of $s$.  
Consider  the strip  ${\cal S}(w)$   limited by 
the two  vertical lines $\Re s = 1- \alpha_0$ and $\Re s =  1+ \alpha_0$.  
By  Lemma 8,
 this strip  contains  $s= \sigma(w)\in \real$ as unique (simple)  pole  of 
 $ S(2s,w)$.  In the rectangle ${\cal U}(w)$ defined by the 
  strip ${\cal S}(w)$   and the two horizontal lines 
  $\Im s = \pm U$,  Cauchy's  residue theorem provides
  \begin{eqnarray}
\label{leresidu}
\frac {1} {2i \pi}  \int_{ {\cal U}(w)} 
 S(2s, w) \frac {T^{2s+1}} {s(2s +1)}\,  ds =  \frac {E(w)}{\sigma(w)(2 
\sigma(w) +1)}     {T^{2\sigma(w) +1}} \, , 
\end{eqnarray}
where  $E(w) = {\bf F}_{\sigma(w),w} \circ {\bf R}(w) [1](0)$,
with  ${\mathbf R}(w)$ the residue operator from (\ref {R}). 
Note in particular that
 \begin{equation}\label{residuenonzero}
E(0) =-\frac{1}{\lambda'(1)}
 {\bf F}_{1}{\bf P}_1 [1](0) =  -\frac{1}{\lambda'(1)}
 {\bf F}_{1} [f_1](0) \ne 0 \, .
\end{equation}

\smallskip
\pn 
We now let $U$ tend to $\infty$.  By Lemma 8, 
  the integral on the leftmost vertical line $\Re s = 1- \alpha_0$
  exists and  satisfies 
$$ 
\int_{ 1 -\alpha_0 -i\infty}^{1- \alpha_0  +i \infty}
 S(2s, w) \frac {T^{2s+1}} {s(2s +1)} \, ds = O \left ( T^ {3-2 \alpha_0 } \right) 
$$
 (with a $O$-term which is  uniform for $w\in {\cal W}'$
 and $T \to \infty$),   while the integrals on the horizontal lines of ${\cal 
U}(w)$ tend 
  to zero  for $U \to \infty$.  Finally,    Perron's formula (\ref{Psiw}) 
with  $D=1+\alpha_0$ gives the contribution from
 the rightmost vertical side, so that 
\begin{equation} \label {Psi}
   \Psi_{w} (T)
  =  \frac {E(w)}{\sigma(w)(2  \sigma(w) +1)}   {T^{2\sigma(w) +1}} \
\left[ 1 + O  \left ( T^{-2\hat \alpha_0} 
  \right) \right] \, ,
\end{equation}
with $E(0)\ne 0$ and a uniform $O$-term  with respect to~$w\in {\cal W}'$,
 as $T \to \infty$.

\subsection {Smoothed costs and transfer of estimates.} 
To  exploit the estimates (\ref{Psi}) on the Ces\`aro
sums $\Psi_{w}(N)$, we  introduce an auxiliary model, 
the {\it smoothed model.}

\pn 
 Associate to some  nonnegative  function  $T \mapsto \epsilon (T)$,  with 
 $\epsilon(T) \le 1$,  the probabilistic  models
 $(\bar \Omega_{N}(\epsilon), \bar \Pr_N(\epsilon))$  
as follows:  For any integer 
$N$, set $\bar \Omega_{N}(\epsilon)=\Omega_N$; next,
choose  uniformly
an integer $Q$  between $N-  \lfloor N \epsilon (N)\rfloor$ and $N$, and draw 
uniformly an element   $(u, v)$ of $\Omega_Q$.     
Slightly abusing language, we refer
to the function $C$ in the model  
$(\bar \Omega_{N}(\epsilon), \bar \Pr_N(\epsilon))$  
as the ``smoothed  cost.'' The
cumulative  value of $\exp[w C]$ for
$ \bar \Pr_N(\epsilon)) $
 is
 \begin{equation} \label {barphi}
  \bar \Phi_{w} (N):=  \frac {1} { \lfloor N \epsilon (N)\rfloor}
 \sum_{Q =  N- \lfloor N \epsilon (N)\rfloor }^{  N
 } \sum_{n \le Q} c_n (w) \, ,
\end{equation}
so that the moment generating function of the smoothed cost  is  just 
\begin{equation} \label {ratio} 
\bar \E_{N}[\exp (w  C)] =  \frac {\bar \Phi_{w}(N)} {\bar \Phi_{0}(N)} \, .
\end{equation}
Note that     $ \bar \Phi_{w} $
can be expressed as a function of $\Psi_{w}$, via
 \begin {equation} \label {trans1}
 \bar  \Phi_{w} (N) = \frac{1} {\lfloor N \epsilon (N) \rfloor} 
\left [ \Psi_{w} (N )- \Psi_{w}(N-  \lfloor N \epsilon (N) \rfloor)
\right ] \, . 
\end{equation}

\smallskip
\pn Now, to  transfer the bound  (\ref{Psi}) for $\Psi_w$  into a bound for 
$\bar  \Phi_{w}$,
we  shall appeal to  a   result  that is often used in  number theory contexts.
\medskip
\pn 
{\bf Lemma 10.} {\sl Let ${\cal W}$  be a complex neighborhood of $0$, and
let  $c_n(w)$ be a sequence of complex-valued functions on $\cal W$.
Assume that $\Psi_w(T) := \sum_{n \le T} c_n(w) (T-n)$ satisfies
$$\Psi_w(T) = F_w(T) \left [1 + O \left (G(T) \right) 
\right] \, ,   \, \, T\to \infty \, , $$ 
with a $O$-error term which is uniform for $w\in\cal W$, where
$F_w(T) = B(w) T ^{a(w)}  
$ and 
$B(w)$, $a(w)$  are bounded holomorphic functions   such that
$ \Re a(w) >1$, $B(w) \not = 0$ on ${\cal W}$.  Assume further
that $G(T)$  tends to $0$ for $T \to \infty$ and  is
of moderate variation, i.e., there exists $K$  so 
that  $ |G(cT)| \le K |G(T)|$ for any $c$  with $ 1/2 \le c \le 2$.
\pn
Then,    if $G(T)^{-1}=O(T)$  for $T \to \infty$, we have
\begin {equation} \label {wboundbis}
\frac {1}{\lfloor T  G(T)^{1/2}\rfloor}\left [ \Psi_{w} (T)
-\Psi_{w}\bigl(T-\lfloor T G(T)^{1/2}\rfloor  \bigr)\right ]  =    
F'_w(T) 
\left[ 1 + O  \bigl ( G(T)^{1/2} \bigr)\right]
\end{equation}
where the $O$-term is uniform with respect to~$w \in {\cal W}$.}

\smallskip
\pn 
{\bf Proof.} We first show (without using $G(T)^{-1}=O(T)$)
that for $T\to \infty$   
\begin {equation} \label  {wbound}
\frac {1}{T    }\left [ \Psi_{w} (T)
-\Psi_{w}\left(T-T G(T)^{1/2} \right) \right ]  =    
F'_w(T)\,   G(T)^{1/2} \,  
\left[ 1 + O  \left ( G(T)^{1/2} \right)\right] \, ,
\end{equation}
(without the integer parts)
with a uniform $O$-term  for  $w \in {\cal W}$. 
Consider some sequence $\epsilon (T)$ 
which tends to $0$.  The   estimate of $\Psi_w(T)$ and
the assumption on $G$  entail 
 \begin{eqnarray}
\nonumber&&\frac 1 {T \epsilon(T) } [\Psi_{w}(T)- 
 \Psi_{w}(T- T\epsilon(T))] 
\\
\nonumber&&\qquad= 
\frac {1} {T \epsilon(T) }   \left ( F_w(T)- F_w(T- T\epsilon(T) )  
\right) 
 + \frac {1} {T\epsilon(T) }  O \left ( F_w(T) G(T)\right )  
\\
 \nonumber &&\qquad = F'_w(T) \left [ 1+ 
O \left (T\epsilon(T) \frac{F''_w(T)}{F'_w(T)}, \frac{1} 
{T\epsilon(T)} \frac{F_w(T) G(T)} {F'_w(T)}
 \right)  \right] \, .
\end{eqnarray}
 Then our assumptions on $F_w(T)$ and 
 $a(w)$ imply
$$  F'_w(T) = 
 \Theta\left (T^{-1}  F_w(T)\right) ,\qquad    F''_w(T) =  \Theta\left (T^{-2}  
F_w(T)\right) \, ,
$$
with a uniform $\Theta$ [recall $A(T)=\Theta (B(T))$ as $T \to \infty$
means that there is an absolute constant 
$C>0$ so that
$A(T)\le C B(T)$ and  $A(T)\ge C B(T)$ as $T \to \infty$].  Therefore,
we obtain  (\ref{wbound}) by taking
$$ T \epsilon (T) := \left(  
\frac{F_w(T) G(T)}{F''_w(T)}\right)^{1/2}  =   \Theta \left ( T  
G(T)^{1/2}\right) \, .
$$

\pn To finish, remark that the difference  
between (\ref{wbound}) divided by $ G(T)^{1/2}$  
and (\ref{wboundbis}) can be split 
into two terms of order 
$$ F'_w(T)\cdot  O\left ( \frac {1} { T G(T)^{1/2}}\right)   
= F'_w(T)\cdot  O  \left ( G(T)^{1/2} \right) \, .
\hbox{\qed}$$

\medskip
\pn 
 We now apply   Lemma  10 to the smoothed costs.  From  (\ref{Psi}), 
(\ref{trans1}), upon 
setting $G(T) =   T^{- 2\hat \alpha_0}$ 
(recall $0<\hat \alpha_0 \le 1/2$), we find  
 $$ 
\bar \Phi_{w}(N)  =  \frac { E(w)}{\sigma (w)}  N^{2 \sigma(w)}   \left[ 
1 + O  \left ( N^{-\hat\alpha_0}  \right) \right] \, ,
$$
 where $E(0)\ne 0$
and the $O$-term
is uniform   (as $N$ tends to $\infty$) when $w$ varies in a sufficiently small   
neighborhood  of $0$.  For  $w= 0$, one has 
 \begin{equation} \label {CN0}  
\bar \Phi_{0}(N)=   E(0)  \,  N^{2 }   \left[ 1 + O \left ( N^{-\hat \alpha_0}
  \right) \right], \qquad \hbox {with} \ \ \
  E(0)= -\frac {{\bf F}_1 {\bf P}_1[1](0)}{ \lambda'(1)} \, .
\end{equation} 
Finally,  by (\ref{ratio}), we obtain:   
\smallskip
\pn {\bf Lemma 11.} 
[Quasi-powers for smoothed cost]
{\sl Let $0< \gamma_0< \alpha_0$ with $\alpha_0$ from  Lemma 8.
The    moment generating function  of the 
smoothed cost corresponding
to $\epsilon(N)=N^{-\gamma_0}$ satisfies
\begin{equation}
\label{qpowsmooth}\bar \E_{N}[\exp (w  C)] =  \frac { E(w)}{E(0) \sigma (w)}  
N^{2 (\sigma(w)-\sigma(0))}   \left[ 
1 + O \left ( N^{-\gamma_0} \right) \right] \, ,
\end{equation}
with $E(w)$ from (\ref{leresidu})
and a uniform $O$-term  when $N\to \infty$ and $w$ 
is near $0$.}

\subsection {Asymptotic Gaussian law for the   smoothed cost.}
In view of Lemma 11  which provides a quasi-power expression for the moment 
generating function of
the smoothed cost, 
we may apply  Theorem~0 to $\hat C_N=C|_{\Omega_N}$ and
$\hat \Omega_N=\Omega_N$, 
$\hat \Pr_N=\bar\Pr_N$ with
$ \beta_N:= \log N$, $\kappa_N= N^{-\gamma_0} $, and 
 \begin {equation} \label {UV}
U(w)=2 ( \sigma(w)-\sigma(0)), \qquad V(w)=  \log  \frac 
{E(w)}{E(0)\sigma(w)} \, .
\end{equation} 
We see from the above that function $\sigma(w)$ (which solves 
$\lambda(\sigma(w), w) = 1$) plays a central r\^ole.    
The next lemma expresses in particular  that 
$U''(0) = 2 \sigma''(0)>0$, so that Theorem~0 can be applied:

 \medskip
 \pn {\bf Lemma 12.}  [Computation of constants.] 
{\sl In the $\cal {GMG}$ setting we put $U(w)=2 ( \sigma(w)-\sigma(0))$. Then,
for the constant cost $c\equiv 1$ (recalling $\Lambda(s)=\log
\lambda(s)$ from Proposition~0), one has
\begin{equation} \label {mudelta}
\mu :=U'(0)=\frac {2} {|\lambda'(1)|} = \frac 
{2} {|\Lambda'(1)|} \, ,\qquad
\delta^2:= U''(0)=  \frac {2 \Lambda''(1) } 
{|\Lambda'(1)|^3}> 0 \, .
\end{equation}
More generally, recalling $\hat \mu (c)$ and $\hat \delta^2(c)$
from Theorem~1, and setting $\chi(c)=\Lambda''_{sw}(1,0)$,
we have 
\begin{eqnarray}
    \nonumber
\mu(c)&:=&U'(0)=\mu\cdot \hat \mu(c)\, , \\
\nonumber
\delta^2(c)&:=& U''(0)=  \hat \mu^2 (c) \cdot \delta^2+
\mu \cdot  \hat \delta^2(c) + \mu^2  \hat \mu (c) \cdot \chi(c)> 0 \, .
\end{eqnarray}
In particular,  $U(w)$ (or equivalently
$\sigma(w)$) is absolutely convex at $0$.}

\medskip 
\pn{\bf Proof.} Let us begin by the  case when   $c\equiv1$. Then 
$\sigma$ is defined     by  
$\Lambda(\sigma(w))={-w}$.
Therefore
$$ \sigma'(w) = - \frac {1} 
{\Lambda'(\sigma (w))}, \qquad 
 \sigma''(w)  =  
 \frac {\sigma'(w) \Lambda''(\sigma (w))}{\Lambda'^2(\sigma (w))} 
 ,  $$
and, recalling  that $\Lambda'(1)< 0$ (from Proposition~0) and $\Lambda''(1) >0$ (from Proposition~1), we get
$$ 
2\sigma'(0)= \frac 
{2} {|\Lambda'(1)|} \, , \qquad 
 2\sigma''(0)= \frac {2 \Lambda''(1) } 
{|\Lambda'(1)|^3}>0 \, .
$$
 
\pn Let us now study the general case. Taking the derivatives of 
the  relation  $\Lambda(\sigma(w), w)= 0$, one obtains
\begin{eqnarray}
\nonumber
0&=& \sigma'(w) \Lambda'_s(\sigma(w), w) +  \Lambda'_w(\sigma(w), w) \, , \\
\nonumber 0&=& \sigma''(w)\Lambda'_s(\sigma(w), w)  + \sigma'^2(w) 
\Lambda''_{s^2}(\sigma(w), w) \\
\nonumber &&\qquad\qquad\qquad\qquad\qquad
+2 \sigma'(w) \Lambda''_{sw}(\sigma(w), w) + 
\Lambda''_{w^2}(\sigma(w), w) \, .
\end{eqnarray}
Remark that
$(\sigma(0), 0)=(1, 0)$ and the derivatives with respect 
to $s$  satisfy
$$ \Lambda'_s(\sigma(w), w)|_{w= 0} = \Lambda'(1), \qquad  
\Lambda''_{s^2}(\sigma(w), w)|_{w= 0} =  \Lambda''(1) \, .
$$
Thus, setting $L(w):= \Lambda(1 + \sigma'(0) w, w)$, we find
$ \sigma''(0) = \frac{1}{|\Lambda'(1)|} L''(0)$.
\pn Since Proposition~1~ implies that $L''(0)>0$, 
we
get the  strict positivity of $\delta^2(c)$, as claimed.
Finally,   $U'(0)$ and $U''(0)$ are
\begin{equation}
    \label{muc}
    \mu(c)  =2\sigma'(0)=
  \frac {-2 \Lambda'_w(1,0)} {\Lambda'(1)} \, ,
\end{equation}
\begin{equation}
    \label {deltac}
    \delta^2(c)  =2\sigma''(0)=
\frac {2\Lambda'^2_w(1,0) \Lambda''(1)	}{|\Lambda'(1)|^3} + 
\frac {4 \Lambda'_w(1,0) \Lambda''_{sw}(1,0)}{|\Lambda'(1)|^ 2} +
 \frac {2\Lambda''_{w^2}(1,0)}{|\Lambda'(1)|} \, ,
\end{equation} 
and, using  ({\ref{mudelta}) as well as  Theorem~1, they may be
expressed as  functions of $\mu$, $\delta$, $\hat \mu(c)$, $\hat \delta(c)$,
and  $\chi(c)$. Note that   $\chi(c) =\hat \delta(c)=0$ for constant~$c$. 
\qed

\medskip
\pn 
By Lemma~12,  Theorem~0   applies and provides the following result:
\smallskip
\pn {\bf Lemma 13.} {\sl  Let $0<\gamma_0<\alpha_0$, with
$\alpha_0$ the constant in  Lemma 8.  
The  smoothed  
cost $C$ associated to $\epsilon(N)=N^{-\gamma_0}$
has an asymptotically Gaussian
distribution, with speed of convergence  $O(1 /\sqrt {\log N})$. 
Moreover, 
\begin{equation}\label {ENCb}
 \bar \E_N [ C]= U'(0)\log N  +  V'(0)+ O( N^{-\gamma_0}) \, , 
\end{equation}
 \begin{equation}\label {VNCb} 
\bar \Var_{N}[ C] =  U''(0)  \log N  +  V''(0) +  O( N^{-\gamma_0}) \, ,
\end{equation}
with  $V$ defined
from the residue function $E(w)$  in (\ref{leresidu}) through  (\ref{UV}), 
and $U(w)$, $U'(0)$, and $U''(0)>0$  as in Lemma 12.

\pn Moreover, for each fixed $k\ge 3$, there  is 
 a polynomial $ P_{k}$
of degree exactly $k$ with coefficients depending on the
derivatives of order at most $k$ at $0$ of $U$ and $V$,
so that the moment of order $k$ satisfies
\begin{equation}
\bar \E_N [ C^k] =  P_{k}(\log N)\ + O \left( \frac  
{(\log N)^{k-1}} {N^{\gamma_0}} \right) \, ,
\end{equation}
with a $O$-term uniform in $k$.}

 \subsection {Proof of Theorem 3.}
The next lemma will allow us to deduce Theorem 3 from
Lemmata 12 and 13:

  \medskip
 \pn {\bf Lemma  14.} {\sl    Suppose that 
 $\lim_{N\to \infty} \epsilon(N)=0$  with
$\epsilon(N)^ {-1}=O({N}/{\log N})$.  Then
the  distance between the distributions 
$\Pr_N$ and $\bar \Pr_N(\epsilon)$ on 
$\Omega_{N}$ is $O( \epsilon(N))$. The same holds
for the distance between $\widetilde \Pr_N$ and its
smoothed version.   }

 \medskip
 \pn {\bf Proof.}  
First recall that there is
$K> 0$ so that 
$$|\Omega_{N}| =
 K N^2 (1+O\bigl (  \frac {\log N}{N})\bigr ), \qquad {\rm Pr}_{N}(u, v) = \frac {1}{K  N^{2}} (1+ O\bigl( \frac  {\log N}{N})\bigr ), $$
for all $(u,v)\in \Omega_N$.
The estimate  provided  in \cite{Te} \S I.3.4 gives 
$K=3/\pi^ 2$ for $\cal G$ and $\cal O$,  and  $K=3/(4\pi^ 2) $  for  $\cal K$. Also,
$|\widetilde \Omega_{N}| =
\widetilde K N^2 (1+O(N^ {-1}))$. 

\pn Denote by $N':= N- \lfloor N \epsilon(N) \rfloor$. 
 Then $\Omega_{N}$ decomposes into the ordinary 
 subset  ${\cal O}_{N} := \Omega_{N'}$ 
 and the exceptional subset  ${\cal E}_{N}:= \Omega_{N}\setminus {\cal O}_{N}$, 
with
 $$\Pr_{N}({\cal E}_{N})   = O \left (  \frac {N^2-N'^2} {N^2}  
 \right)  = O( \epsilon(N)) \, .
$$ 
 Now,  the  probability  $ \bar \Pr_{N}(u, v)$ 
  satisfies,   for any $(u, v) \in {\cal  O}_{N}$,
$$ 
\bar \Pr_{N} (u, v) = \sum_{q= N'}^N   \Pr_{q} (u, v) \cdot \Pr [Q= 
q] = \frac {1} {1+ \lfloor N \epsilon(N) \rfloor} \sum_{q= N'}^N \frac 
{1}{|\Omega_{q}|} \,  .
$$
Then,  using  $N=O(N')$, the relations 
\begin{eqnarray}
K\sum_{q= N'}^N \frac 
{1}{|\Omega_q|}  &=&   \sum_{q= N'}^N  \frac {1}{ q^2} 
\biggl (1+O\biggl ( \frac {\log q}{q}\biggr )\biggr )=
  \biggl (1+O\biggl (\frac {\log N}{N'}\biggr )\biggr)
\left [\frac{1}{N'} - \frac {1}{N+1}\right ]   \cr
&=& 
\biggl (1+O\biggl (\frac{\log N}{N}\biggr )\biggr )
 \left ( 1 +{\lfloor N \epsilon(N) \rfloor} \right)  \frac {N} {N'}\, ,
\end{eqnarray}
give,  for any $(u, v) \in {\cal O}_{N}$, 
$$ \bar \Pr_{N}(u, v)  =     \Pr_{N}(u, v)\frac {N} {N'}
\left (1+O \left (\frac {\log N}{N}  \right) \right)
=   \Pr_{N}(u, v)\left (1+O(\epsilon(N))\right).$$
Finally, for any  $A\subset \Omega_{N}$,  the difference  
$ |\Pr_{N}(A) -  \bar \Pr_{N}(A)|$ is less than
$$  
\big|\Pr_{N}(A\cap {\cal O}_{N})-  \bar \Pr_{N}(A
\cap {\cal O}_{N})\big| + \big|\Pr_{N}(A\cap {\cal E}_{N}) - \bar \Pr_{N}(A
\cap {\cal E}_{N}) \big|=O(\epsilon(N)) \, . \hbox{\qed}  
$$

 \medskip
 \pn {\bf Proof of Theorem~3.}
For   $\epsilon (N) = N^{-\gamma_0}$,   Lemmata 13 and 14
imply that the
asymptotic distribution of  $C$ is Gaussian, with a speed of 
convergence in $O( 1/\sqrt {\log N})$. This proves 
Theorem 3$(a)$ for any $\gamma \le \gamma_0<\alpha_0$.   

\smallskip
\pn To get claim $(b)$ of Theorem 3,
we use the fact that for each of our three
algorithms, the  maximal number of steps $P(u,v)$  on $\Omega_N$ is bounded by
$K_0\log N$. This is a classical result  
due to the fact  that the  three  Euclidean divisions we consider are
of the form $v= mu + \epsilon r$ with $r \le v/2$.  
For $(u, v) \in \Omega_N$, all the quotients $m_i$ are
at most $N$, and the moderate growth assumption then  implies that 
$$\sup \{ C^k (u,v); (u, v) \in \Omega_N \} \le K (\log N)^ {2k } $$ 
for some $K$ and all $N$,  and finally,  Lemma~14 gives
$$
\big|  \E_N [C^k]-\bar \E_N [ C^k ]\big |  
 =  \, O\left (  (\log N)^{2k} \,   N ^{-\gamma_0}\right) . 
$$
 Applying Lemma 12, 
we get the expression for  $\E_N [C^k]$  and for the variance in Theorem~3$(b)$, 
  ending the proof of 
Theorem 3.
\qed

\section{Local Limit Theorem for lattice costs.}
\pn   This section is devoted to  proving Theorem 4. 
Consider an algorithm
with associated  dynamics in the good class  and satisfying
{\em UNI.} Let $c$ be a lattice cost, of span $L>0$, 
and of moderate growth. 
By Lemma 14, it suffices to consider the smoothed model
$\bar \Pr_N$
associated to $\epsilon(N)= N^{-\gamma}$, for 
$\gamma$ satisfying $0<\gamma<\alpha_0$
 ($\alpha_0$ given by Lemma 8) which is 
to be fixed later. By Lemma 11,
we have a quasi-power expression for $\bar\E_N[\exp(wC)]$
for small $|w|$.

\smallskip
\pn 
 We first consider costs $c$ with span $L= 1$, so that 
$$ \bar\E_{N}[e^{i\tau C}]  = \sum_{\ell\ge 0} \bar \Pr_N [C = \ell ] \  e^ {i \tau \ell} \, . $$
Our starting point is the relation 
$$ \bar  \Pr_{N} [  C(u,v) = \ell] = \frac {1} {2 \pi} 
  \int_{-\pi}^\pi e^{-i\tau \ell}  \cdot 
\bar\E_{N}[e^{i\tau C}] \, d\tau \,  .$$
Since we are looking for an LLT  result,   the convenient scale of the problem is $n:= \log N$, and 
we set 
$$\ell= p_x (n) :=   \mu(c) n +  \delta(c) x \sqrt {n}, \qquad q_x(n) := \lfloor p_x(n)\rceil \, .$$
(Here $\lfloor \cdot \rceil$ denotes the nearest integer function.)
We consider 
\begin {equation}
\label {starting}
 I_n:=  2 \pi  \sqrt {\log N} \cdot  \bar  \Pr_{N} [ C(u,v) = q_x(\log N)] $$
$$ \qquad \qquad \qquad  = \sqrt n 
 \int_{-\pi }^{+\pi} \exp [-i\tau  q_x (n)]  \cdot 
\bar\E_{N}[e^{i\tau C}] \, d\tau  \, .
\end{equation} 
Our strategy is to decompose  the integration interval
$[-\pi, +{\pi}]$ into a neighborhood 
of zero, $[-\upsilon, \upsilon]$,
and its complement $|\tau|\in (\upsilon, \pi]$. 
This gives rise to two integrals $I_n^{(0)}$ and 
$I_n^{(1)}$.

 \subsection {Estimate of  the integral  $I_n^{(0)}$.} 
 We begin with $\alpha_0$ and a neighborhood 
${\cal W}'$ as provided by Lemma 8.  We assume
that  $[-\upsilon, + \upsilon]$ is  contained in 
${\cal W}'$ and study 
$$   I_n^{(0)} :=   \sqrt n 
 \int_{-\upsilon }^{+\upsilon} \exp [-i\tau  q_x (n)]  \cdot 
\bar\E_{N}[e^{i\tau C}] \, d\tau \, .$$
Lemma 11 grants us a quasi-power expression for $\bar\E_N[\exp(i \tau
C)]$. Recalling that 
$\mu(c) = 2 \sigma'(0)$, we see that
the integrand can be written as a product $g_nf_n$, where
the first factor is
$$  g_n(\tau):=  \frac { E(i\tau)}{E(0) \sigma (i\tau)} \cdot   \left (1+ O  (  e^{-n\gamma}) \right) \cdot \exp [i \tau (p_x(n)-q_x(n))] \, ;$$
and the second factor is
$$ f_n(\tau):=  \exp [2 n (\sigma(i\tau)-1-i\tau\sigma'(0))]  \cdot  \exp [-i \tau  x\delta(c) \sqrt n]\, .$$ 
The function $ z\mapsto  \sigma(z)-1-z\sigma'(0)$ has a saddle point at $z= 0$, and 
the main  useful properties of $g_n(\tau), f_n(\tau)$  are
\begin {equation} \label {propfg}
|g_n(\tau)-1| <\!\!< |\tau| + \exp [-n \gamma] , \qquad  |f_n (\tau)| \le \exp [- n\tau^2 \delta_0] \, ,
\end{equation} 
 for $|\tau| \le \upsilon$, where $\upsilon$ is 
taken sufficiently small,  so that 
\begin {equation} \label {deltazero} 
\delta_0 := \inf  \{ |\Re  \sigma''(i\tau)| ; |\tau| \le \upsilon \} >0 \, .
\end{equation} 

\smallskip
 \pn We  first prove that
the dominant part of $I_n^{(0)}$  is 
$$  J_n^{(0)} :=  \sqrt n 
 \int_{-\tau_n}^{\tau_n}  \exp [-i\tau  q_x (n) ]  \cdot 
\bar\E_{N}[e^{i\tau C}] \, d\tau , \qquad \hbox {with}\ \  \tau_n := \left ( \frac {\log n}{ \delta_0 n }
\right ) ^ {1/2} \,  .$$
 Properties (\ref{propfg}) entail that   
$$ \sup \{ |g_n(\tau) f_n(\tau)|;  \  \tau_n \le |\tau| \le \upsilon \} <\!\!< \exp [-n \tau_n^2 \delta_0]  <\!\!< \frac 1 n \, ,$$
and this implies   that  $|I_n^{(0)}-J_n^{(0)}|$ \  is $O( 1/\sqrt n)$.

\smallskip
\pn 
We now  define   the following quantities, 
\begin{eqnarray}
\nonumber
J_n^{(1)}&:=& \sqrt {n} \int_{-\tau_n}^{\tau_n} 
\exp [-i\tau x \delta(c) \sqrt n ]  \cdot 
\exp [2n (\sigma(i\tau)-1-i\tau \sigma'(0))]    \, d\tau\, ,  \cr
  J_n^{(2)}&:= & 
\int_{-\tau_n\sqrt{n}} ^{\tau_n\sqrt{n}} 
\exp[ -i\vartheta \delta(c) x - \frac 1 2 \vartheta^2\delta^2(c)] \, d\vartheta\, , \cr
 J_n^{(3)}&:= &  \int_{-\infty} ^{\infty} 
\exp[ -i\vartheta \delta(c) x -  \frac 1 2 \vartheta^2\delta^2(c)]  d\vartheta=
\frac {\sqrt{2\pi}} {\delta(c) }  e^{-x^2 /2} \, , 
\end{eqnarray}
and will show that they satisfy  $|J_n^{(i)}- J_n^{(i+1)}| = O(1/\sqrt n)$ for $1 \le i \le 3$. 
Let us check this line by line.  

\pn {\sl From $J_n^{(0)}$ to $J_n^{(1)}$.}    Properties (\ref{propfg}) 
imply that  
$$  \frac {1} {\sqrt n} |J_n^{(1)} -J_n^{(0)}|  <\!\!< \int_{-\tau_n}^{\tau_n}  |\tau|  \exp [ - n \tau^2 \delta_0]  d\tau  
+ e^ {-n \gamma} \int_{-\tau_n}^{\tau_n}  \exp [ - n \tau^2 \delta_0]   d \tau \, ,$$
and the two  terms  are $O(1/n)$. 

\smallskip
\pn {\sl  From $J_n^{(1)}$ to $J_n^{(2)}$.} There is first a change of variable $ \vartheta = \tau \sqrt n$. 
Remark that 
$$\vartheta^3/\sqrt n  = \tau^3  n  = O\left (\frac{ (\log n)^{3/2} } {\sqrt n}\right) $$ 
tends to $0$  for $n \to \infty$.  
Furthermore, since $\delta^2(c)= 2 \sigma''(0)$, 
$$  2n \left[ \sigma (i\ \frac {\vartheta} {\sqrt n}) - \sigma(0) -  i\ \frac {\vartheta} {\sqrt n} \sigma'(0)  \right] 
= -   \vartheta^2  \frac{\delta^2(c)}{2}   +   O\left (\frac  { \vartheta^3}{  \sqrt n}  \right)  \, ,
$$
so that the difference between  the two integrals  satisfies 
$$ |J_n^{(1)}- J_n^{(2)}| \ <\!\!<  \frac 1 {\sqrt n}  \int_{-\tau_n \sqrt n }^{\tau_n \sqrt n} 
\vartheta^3 \exp [ -  \vartheta^2 \frac{\delta^2(c)}{2}] d \vartheta = O \left(  \frac 1{\sqrt n}  \right) \, . $$

\smallskip
\pn  {\sl   From $J_n^{(2)}$ to $J_n^{(3)}$. } The difference between the two integrals  satisfies 
$$ | J_n^{(2)}- J_n^{(3)}| \le \int_{|\tau| \ge \tau_n \sqrt n}   \exp [ -  \vartheta^2 \frac{\delta^2(c)}{2}]  d \vartheta
 =O\left( \frac {1}{\sqrt n}\right)\,  .$$ The last equality for   $J_n^{(3)}$ is obtained by ``completing" the square.

\medskip
\pn  Finally, the first part  $I_n^{(0)}$ of the integral   $I_n$ satisfies
\begin {equation} \label {In0}
 I_n^{(0)} = \sqrt {2 \pi} \,  \frac {e^{-x^2/2}}{ \delta(c) }+ 
O\left( \frac {1}{\sqrt n}\right) \, .
\end{equation}

\subsection {Estimate of  the integral   when $\tau$ is not close to 0.}  We   now  consider the integral (\ref{starting})
outside of the neighborhood of zero.

\medskip
\pn
{\bf  Lemma 15.} 
{\sl  Assume that $c$
is  a lattice cost of moderate growth with span $L$. For each $\upsilon > 0$, there are $Q < \infty$ and
$\gamma_0 >0$ so that,  for the smoothed cost corresponding
to $\epsilon(N^{-\gamma})$ with small enough $\gamma$,
we have 
$$ |\bar \E_{N}[\exp (i\tau C)] |\le Q  N^{-\gamma_0}
\, , \forall |\tau|\in [\upsilon, \pi/L] \, .$$}

\smallskip
\pn  {\bf Proof.}   We apply Lemma 9 with Perron's formula,  keeping
the notations of Lemma 9.  Consider, as in Section 4.1,  Perron's
formula  
(\ref{Psiw}) 
with  $D=1+\gamma_1$,  
$$  \Psi_{i \tau} (T)  = 
\int_{\Re s=1+\gamma_1}
 S(2s, i \tau ) \frac {T^{2s+1}} {s(2s +1)} \, ds$$
In the rectangle ${\cal U}$ defined by the  strip  $\Re s = 1 \pm \gamma_1$  and the horizontal lines 
 $\Im s = \pm U$,  the function $ s \mapsto S(2s, i \tau)$ is analytic (from Lemma 9)  and  Cauchy's  residue theorem gives
$$\int_{ {\cal U}} S(2s, w) \frac {T^{2s+1}} {s(2s +1)}\,  ds = 0 .$$
  Furthermore,   using once more Lemma 9, 
$$ 
\int_{\Re s=1-\gamma_1}
 S(2s, w) \frac {T^{2s+1}} {s(2s +1)} \, ds = 
O \left ( T^ {3-2 \gamma_1 } \right) \, .
$$   
 We  let $U$ tend to $\infty$.
The integrals on the horizontal lines of ${\cal U}$ tend 
to zero,    Perron's formula  gives  the right side, and finally
\begin{equation} 
 \Psi_{i \tau } (T) =    O(T^{3-2\gamma_1}) \, .
\end{equation}
Now, if $\gamma < 2\gamma_1$, setting $\gamma_0 =2\gamma_1-\gamma$, the smoothed 
quantity  $\bar \Phi_{ i \tau } (N)$ satisfies  
\begin{eqnarray}
\bar \Phi_{ i \tau } (N)  &=&\frac{1}{\lfloor N^{1-\gamma }\rfloor}
\bigl (\Psi_{i \tau} (N)-\Psi_{i \tau}(N-\lfloor N^{1-\gamma }\rfloor \bigr )\cr
&=& O(N^{\gamma-1}) O(N^{3-2\gamma_1})
=     O(N^{2-\gamma_0}) \, .
\end{eqnarray}
 Furthermore, with (\ref{CN0}),  
$\bar \Phi_{0} (N)   =\Theta (N^{2})\, $. Therefore
$$
\bar \E_{N}[\exp (i\tau  C)]=
\frac{\bar \Phi_{w} (N)}{\bar \Phi_0(N)}  =     O(N^{-\gamma_0})
\, . \hbox {\qed}
$$

\medskip
\pn We deduce from Lemma 15  that $I_n^{(1)}$ is $O(e^{-n\gamma_0})$. 

\subsection {End of proof of Theorem 4.} 
Finally, we take $\alpha_0$ and a neighborhood 
${\cal W}'$ as granted by Lemma 8.  We assume
that  $[-\upsilon, + \upsilon]$ is  contained in 
${\cal W}'$  and is sufficiently small  
to entail validity of (\ref{deltazero}). Then, $\upsilon$ is fixed and we choose $\gamma$
so that $\gamma$ is less than $2 \gamma_1$ (from Lemma 9) and less than $\alpha_0$ (from Lemma 8).  
 
\pn Now, with (\ref{In0}) and Lemma 11,   Theorem 4 is proven in the case of span $L= 1$. Remark that  if $c$ has span $L$, 
then $ d:= c/L$ has span 1, with $\mu(c)= L \mu(d)$ and $\delta(c) = L \delta(d)$. This proves Theorem 4 in the general case. \qed

\subsection  {\bf The nonlattice case.}
We say that a nonlattice digit-cost is {\it shifted lattice}
if  there exist 
$L\in \mathbb R^+_*$ and a number
$L_0$, with $L_0/L$ irrational,   so that  $(c-L_0)/L$ is integer-valued.  
The number  $L_0$ is called the {\it shift.}
The span of $c$ is the largest possible $L$. Then, the  digit  cost $c$ can be written as 
the sum of  two costs $c= L_0  + d$,  where $d$ is  lattice with span  $L$,  and
the total cost $C$ equals $c= L_0 P + D$. It is then possible to apply the previous result to $D$.
Furthermore, since the two $\Lambda$  functions relative to $c$ and $d$ are related by 
$ \Lambda_c (s, w) =  w L_0 + \Lambda_d(s, w)$, 
one has 
$$ \mu(d)= \mu(c) - \mu L_0,  \qquad \delta^2(d)= \delta^2(c)+ L_0^2 \delta^2 -  L_0 [ \chi (c) \mu^2 +
2 \frac{\mu(c) } {\mu} \delta^2] \, ,$$
and, finally:  

\smallskip
\pn {\sl For  the algorithms  ${\cal G}$, ${\cal K}$, ${\cal O}$ and for
any shifted lattice cost $c$ of span $L$ and  shift $L_0$,
which is of moderate growth, one has,  for any $x \in \mathbb R$
 $$
\Pr_{N} \left[ (u, v); -  \frac{ L} {2}  <  C(u,v)-L_0 \cdot P(u,v) - 
\mu(d) \log N -   \delta(d) x \sqrt {\log N}   \le \frac {L} {2}\right] $$
$$  \qquad \qquad \qquad \qquad   =  
\frac {e^{-x^2/2}}  {\delta(d) \sqrt {2 \pi  \log N}}  + O \left( \frac{1} {\log N}\right) 
\, ,
$$
with a $O$-term uniform in $x$.}

\medskip
\pn The case when $c\not \equiv 0$ is neither lattice nor shifted lattice
is essentially different, the noncompact situation
requiring Dolgopyat-type estimates in $w$.
The generic nonlattice situation  will
be considered in a  forthcoming paper.

\section {Conclusion.} 

\pn This article has presented a unified approach to  a large body of
results relative to Euclid's algorithm and its major (fast)
variants. We recapitulate here some of its consequences,
relations to the vast existing literature on the subject,
as well as extensions and open questions. It should be stressed that
most of the improvements can eventually be traced to the existence of
pole-free strips for Dirichlet series of cost parameters, a fact that
precisely devolves from our extension of Dolgopyat's estimates to continued
fraction systems. 

\smallskip
\pn First our methods lead to extremely precise estimates of the moments of
costs, and in particular the number of steps, a much studied
subject. Our estimates, when
specialized to the mean number of steps, yield
\begin{equation}\label{meannumstep}
\E_N[C]=\mu\log n+\eta+O(N^{-\gamma});
\end{equation}
see above Theorem~3, Parts~$(b)$ and~$(c)$. In the case of the
standard algorithm, this covers 
the original estimates of Dixon and Heilbronn in 1969--1970 (the main term), 
as well as Porter's 1975 improvement (the second term, $\eta$), while
providing the right shape of the error term ($O(N^{-\gamma})$),
for which Porter further showed that one could take $\gamma=\frac16-\epsilon$
in the case of the standard algorithm. 
We refer the reader to the accounts by Knuth~\cite{Kn} and
Finch~\cite{Finch03} 
for more material on this classical topic. Our formula~(\ref{meannumstep})
also extends Rieger's analyses (first published around
1980, see~\cite{Ri,Rieger80})
of the centered algorithm 
and the odd algorithm.
Note that, in our perspective, the second-order constant~$\eta$ comes out  as a
spectral quantity. It is an open problem to derive an explicit
form starting from our expressions 
(e.g., such a form involving $\zeta'(2)$
is known under the name of Porter's constant in the
standard case). 

\smallskip
\pn In sharp contrast to the mean value case, variances do not
seem to be amenable to elementary methods. The first-order
estimate has been given
by Hensley (1994) in the paper~\cite{He}
that first brought functional analysis
to the field. Our formula,
\begin{equation}\label{varstep}
\Var_N[C]=\delta^2\log n+\delta_1+O(N^{-\gamma}),
\end{equation}
stated in Theorem~3 can be viewed as a precise form of Hensley's 
estimate relative to the standard algorithm, one that also extends to
the odd and centered algorithms. (Incidentally, the quantity~$\delta$,
called Hensley's constant~\cite{Finch03}, is not recognized to be related to
classical constants of 
analysis, though it is now known to be polynomial-time computable 
thanks to a recent study of Lhote~\cite{Lh}; the nature of~$\delta_1$
is even more obscure.) Note that the complex-analytic
properties of the moment generating functions
provided by the functional-analytic methods furnish
similarly strong estimates for moments of arbitrarily high order
(our Theorem~3), a fact which also appears to be new.

\smallskip
\pn Regarding distributional results, several points are worthy of note. 
Dixon in his 1970 paper had already obtained exponential tail bounds
and these were further improved, albeit in a still fairly qualitative way by
Hensley in his 1994 study (see his Theorem~1 in~\cite{He}). Our approach gives 
much more precise information on the distribution, as we now explain.

\pn For simplicity, let us specialize once more 
the discussion to the number of steps. 
First, regarding the central region of the Local Limit Theorem
(Theorem~4), the nature of the error terms obtained ($O(N^{-\gamma_0})$)
and the fact that the saddle point method lends itself to the
derivation of full asymptotic expansions (see, e.g., Henrici's
book~\cite{Henrici77}) 
entail the existence of a \emph{full asymptotic expansion}
associated with the Gaussian approximation of Theorem~4, namely, for a
computable numeric sequence~$\{c_j\}$,
\begin{equation}\label{loca}
\Pr_N[\cdot] =\frac{e^{-x^2/2}}{\delta\sqrt{2\pi\log N}}
\left(1+\sum_{k=1}^r \frac{c_r}{(\log
N)^{r/2}}+O\left(\log^{-(r+1)/2}N\right)\right)
\end{equation}
(the argument of~$\Pr_N$ being that of Theorem~4).
This expansion is valid for~$x$ in any compact set of~${\mathbb R}$.

\pn Regarding large deviations,  Lemma~11
(quasi-powers for smoothed costs) implies that $\bar \E_N[\exp(wC)]$ is of
the form of a quasi-power, provided~$w$ stays in a 
small enough fixed neighbourhood of~0.
By a standard argument (originally due to Cram\'er
and adapted to the quasi-powers framework by Hwang~\cite{Hw,Hw1,Hw2})
and Lemma~14,
this implies the
existence of a large deviation rate function, that is, 
of a function~$I(y)$ such that, for the right tail corresponding
to~$y\ge0$, there holds
\begin{equation}\label{largedev}
\lim_{N\to\infty} \frac{1}{\log N} \log \Pr_N[C> (1+y)\mu\log N] =-I(y),
\end{equation}
with~$I(y)$
defined on an interval $[0,\delta]$ for some $\delta>0$. In other
words, the probability of exceeding the mean by $y\mu\log N$
is exponentially small in the scale of the problem, being of the form 
$\exp[-I(y)\log N]=N^{-I(y)}$. A simplified version
of~(\ref{largedev}) is then: the probability of
observing a value at~$z$ or more standard deviations from the mean is 
bounded by a quantity  that decays like
$e^{-C_1z^2}$ (Hensley's notations~\cite{He}). 
Analogous properties naturally hold for the left tail.
Cramer's technique of ``shifting the mean'' and use of the resulting
``shifted'' quasi-powers lead in fact to very precise quantitative versions
of~(\ref{largedev}) in the style of Hwang---thereby providing optimal
forms of tail inequalities \emph{\`a la} Dixon--Hensley.

\pn Similar properties hold for other costs measures, like the ones 
detailed in the introduction. Of particular interest is the statistics
of the number of digits assuming a particular value, for which 
estimates parallel to~(\ref{meannumstep}), (\ref{varstep}),
(\ref{loca}), and~(\ref{largedev})
are also seen to hold. For instance, the 
frequency of occurrence of digit~$m$ in the expansion
of a rational number has mean
\[
{} \sim \log_2\left(1+\frac{1}{m(m+2)}\right),
\]
and it exhibits Gaussian fluctuations.
It is again unknown to us, which of the involved
constant (beyond the mean value factor)
may be related to classical constants of analysis. The
spectral forms of our
Lemma~12 may at least provide a starting point for such
investigations.

\smallskip
\pn A major challenge is to derive distributional information that are
so to speak ``superlocal''. By this, we mean the problem of estimating
the behaviour of the number of steps and other cost measures
over fractions of denominator \emph{exactly} equal to~$N$, i.e.,
rationals belonging to $\Omega_N\setminus\Omega_{N-1}$. In view of what is known
in the average-case~\cite{Finch03,Kn}, we expect  arithmetical 
properties of~$N$ to come into the picture. The analytical difficulty
in this case stems from the fact that a further level of ``unsmoothing'' 
(how to go from~$\Omega_N$ to $\Omega_N\setminus\Omega_{N-1}$?) would be
required.

\medskip
\pn   {\bf Further works.} Our  
dynamical  approach  provides a   general framework,
where  it seems  possible  to  answer other  
questions about distributional analysis. For instance, all the questions that 
we solve for rational trajectories can be asked for the periodic trajectories.  A
periodic trajectory is  produced by a  quadratic number, and the  reference parameter
is related to   the length of the geodesics on the modular surface. In a forthcoming paper, 
and following the same principles as in \cite {PP, PS, Va6}, 
we prove that  the distribution of costs on periodic trajectories can be studied  in a similar way as here,
 replacing the  quasi--inverse $(I-{\bf H}_{s, w})^{-1}$ by $\det (I-{\bf H}_{s, w})$. 

\medskip
\pn  {\bf Open problems.}  We also ask  various  questions about Euclidean algorithms: 
for instance, what happens for other Euclidean algorithms of the Fast 
Class (in particular for the Binary algorithm  \cite {Br, Va5})? The extension of our results to cost functions that are still
``small'' but take into account the boolean cost 
(also known as the bit complexity) of each arithmetic
operation is on our agenda. Note that
an average-case analysis is already known to be possible via operator
techniques~\cite{AV, Va8}. On another register,
the extension to ``large'' costs is likely to lead
us to the realm of stable laws: see for instance Gouezel and Vardi's
works~\cite{Gouezel03,Vardi87} for occurrences of these laws in
continued fraction related matters.

\medskip
\pn {\bf Acknowledgements. }  
During his thesis, Herv\'e Daud\'e made experiments providing
evidence for the 
Gaussian limit property of
the number of steps.  Joint work with Eda Cesaratto \cite {CeVa}
involved an extensive use of the weighted transfer operator, 
and some of the ideas that we shared 
on that occasion proved very useful  for the present paper.  
We have had many stimulating discussions with Philippe Flajolet 
regarding the saddle-point method and the notion of smoothed costs.
Finally, we thank S\'ebastien Gou\"ezel who
found a mistake in a previous version.
This work has been supported in 
part by two {\sc CNRS Mathstic} grants
and by the European Union
under the Future and Emerging Technologies programme of the Fifth
Framework ({\sc Alcom-ft} Project IST-1999-14186).

\end{document}